\documentclass[a4paper]{article}   	
\pdfoutput=1 
\usepackage{geometry}                		
\usepackage{graphicx}				
\usepackage{amssymb,amsmath}

\allowdisplaybreaks[4]

\newcommand{\beq}{\begin{eqnarray}}
\newcommand{\eeq}{\end{eqnarray}}

\title{Quantum quench in the attractive regime of the  sine-Gordon model}
\author{Axel Cort\'{e}s Cubero and Dirk Schuricht\\[5mm]
\normalsize
Institute for Theoretical Physics, Center for Extreme Matter and Emergent Phenomena,\\ 
\normalsize Utrecht University, Princetonplein 5, 3584 CC Utrecht, the Netherlands}

\date{}							

\begin{document}
\maketitle

\begin{abstract}
We study the dynamics of the sine-Gordon model after a quantum quench into the attractive regime, where the spectrum consists of solitons, antisolitons and breathers. In particular, we analyse the time-dependent expectation value of the vertex operator, $\exp\left({\rm i}\beta\Phi/2\right)$, starting from an initial state in the ``squeezed state form" corresponding to integrable boundary conditions. Using an expansion in terms of exact form factors, we compute analytically the leading contributions to this expectation value at late times. We show that form factors containing breathers only contribute to the late-time dynamics if the initial state exhibits zero-momentum breather states. The leading terms at late times exponentially decay, and we compute the different decay rates. In addition, the late-time contributions from the zero-momentum breathers display oscillatory behaviour, with the oscillation frequency given by the breather mass renormalised by interaction effects. Using our result, we compute the low-energy contributions to the power spectrum of the vertex operator. The oscillatory terms in the expectation value are shown to produce smooth peaks in the power spectrum located near the values of the bare breather masses.
\end{abstract}

\section{Introduction}
The study of  quantum quenches in many-body systems has gathered significant attention in recent years. The quantum quench protocol consists on preparing a system to be in an eigenstate (typically the ground state) of a Hamiltonian, $\mathcal{H}_0$, then suddenly changing some parameter and evolving with a new Hamiltonian, $\mathcal{H}$, with respect to which the system is no longer in equilibrium. This is motivated by the realisation of closed quantum systems using cold atoms and ions~\cite{experimental}, which have provided an experimental probe for quantum non-equilibrium dynamics. More generally, one may consider the time evolution starting from any initial state, even ones that are not eigenstates of some initial Hamiltonian. 

Out-of-equilibrium dynamics are particularly interesting in integrable (1+1)-dimensional systems, where the additional conserved charges prevent thermalisation at long times. Expectation values of local operators are instead expected to be described by a generalised Gibbs ensemble (GGE) \cite{gge}, which takes into account the additional local (or quasilocal) conserved charges\footnote{The question of which conserved charges need to be included in the GGE of an integrable field theory is a matter of ongoing investigation. In both, discrete lattice models~\cite{quasilocallattice}, and field theories~\cite{quasilocalfield}, it has been established that strictly local conserved charges are not sufficient, and certain quasilocal charges must be included.}. Furthermore, there are powerful non-perturbative methods associated with integrability in quantum systems, so the non-equilibrium dynamics can be evaluated analytically in more detail than is possible in general interacting systems. To list a few examples, one can use conformal symmetry~\cite{cardycalabrese}, apply the recently introduced quench-action approach~\cite{quenchaction,repulsive}, or employ the knowledge of the complete spectrum to directly tackle the Lehmann representation for the time evolution of observables~\cite{repulsive,Lehmann,gritsev,fioretto,SE}. The latter approach relies on the knowledge of the matrix elements of the operators of interest, which, in the case of integrable field theories, can be obtained from the form-factor bootstrap~\cite{bootstrap,mussardo}. Furthermore, using form-factor perturbation theory~\cite{ffpt} the approach has recently been applied to study the effects of small integrability breaking terms~\cite{delfino,delfinotwo}. 

Our main objective in this paper is the computation of time-dependent expectation values of local operators at finite times after a quench in a specific integrable field theory, namely the attractive sine-Gordon model. Previously, a similar problem has been analysed in Ref.~\cite{repulsive}, where the sine-Gordon model (\ref{hamiltonian}) in the repulsive regime (where the spectrum consists of only solitons and antisolitons, but no bound states) was considered. In this reference, the expectation values of a vertex operators were computed at large (but finite) times, using two complementary approaches, namely the quench-action formalism and a linked-cluster expansion based on the exact form factors. It was seen that the expectation values decay exponentially in time, and the decay rate was computed. Similar results for a quench in the sine-Gordon model were obtained using a semiclassical approach \cite{semiclassical,semisemiclassical}, where at low energies, the quantum dynamics of solitons and antisolitons can be ignored. It was also argued in this reference that breathers (bound states of solitons and antisolitons) would not contribute at this semiclassical level, instead there effect would be purely quantum mechanical. A similar semiclassical approach has also been used to study quantum quenches in O(3)-symmetric models~\cite{semiclassicalothree}.

In this paper we extend the techniques developed in Ref.~\cite{repulsive} to the attractive regime of the sine-Gordon model. In this regime, the spectrum includes breathers as well as solitons and antisolitons. Specifically, we compute the expectation value of the vertex operator $\exp({\rm i}\beta\Phi/2)$ using a form-factor expansion. We notice that the contributions to this observable coming from form factors involving breathers are qualitatively different from those coming from soliton-antisoliton form factors. This is mainly because solitons are non-local excitations with respect to the vertex operator, while breathers are local. This results in the breather contributions being suppressed, in agreement with the semiclassical arguments put forward in Ref.~\cite{semiclassical}.

We find, however, that the new contributions to observables arising from breather form factors are not negligible if the initial state after the quench exhibits zero-momentum breathers corresponding to boundary bound states~\cite{gloshal}. Quenches from such an initial state have been previously considered \cite{gritsev}, where the first few terms in a form-factor expansion for the expectation value of the local vertex operator $\exp({\rm i}\beta\Phi)$ were studied. In the computation of Ref.~\cite{gritsev} the boundary bound states lead to permanent oscillatory contributions to the expectation values, which do not decay in time. These terms are seen to produce sharp poles in the power spectrum of this operator. This result is based on the assumption that the higher terms in the form-factor expansion are negligible. We find that this is, in general, not the case, as the higher order terms may diverge at long times.

Our main result, stated in Eq.~(\ref{finalresult2}), is the calculation of the time-dependent expectation value of the vertex operator $\exp({\rm i}\beta\Phi/2)$ after a quantum quench. We find that the leading terms at late times exhibit both, oscillations in time as well as exponential decay, in contrast to the repulsive regime where oscillatory behaviour only appears as sub-leading corrections. The oscillation frequency is not simply given by the breather mass but contains corrections due to interaction effects, which vanish in the semiclassical limit. 

This article is organised as follows: In the following section, we present a brief introduction to the sine-Gordon model, and discuss some of its potential realisations in solids or in cold atomic systems. In Section~\ref{initialstate} we present a discussion of the initial states we consider, and discuss their relation to integrable boundary conditions. In Section~\ref{linkedclusterexpansion} we present an overview of the analytical method used to compute the time evolution of expectation values. This method, which has been named ``linked-cluster expansion", consists in expanding the expectation values in terms of exact form factors, while carefully ensuring the cancellation of infrared divergences. We present our main result in Section~\ref{themainresult}, ie, the time-dependent expectation value of the vertex operator, $\exp({\rm i}\beta\Phi/2)$, in the simplest regime of the attractive sine-Gordon model where only one species of breather bound states exists. From the time evolution we compute the power spectrum in Section \ref{thepowerspectrum}. In contrast with the results of Ref.~\cite{gritsev}, the sharp delta-function peaks at the transition energies are broadened to Lorentzian-function peaks. We then discuss the relation between our results and the semiclassical approach of Ref.~\cite{semiclassical} in Section~\ref{semiclassicalmethods}. In Section~\ref{morethanone} we generalise our results for the time evolution to the regimes in the sine-Gordon model where more than one species of breathers exists. Finally, we present our conclusions in Section~\ref{conclusions}.  The explicit computation of the terms of the linked-cluster expansion, as well as a brief discussion of exact form-factor axioms and finite-volume regularisation techniques, are presented in the appendices.

\section{The attractive sine-Gordon model}
We consider the sine-Gordon model with Hamiltonian given by
\beq
\mathcal{H}=\frac{1}{16\pi}\int dx\left[\left(\partial_x \Phi\right)^2+\left(\partial_t\Phi\right)^2\right]-\lambda\int dx \cos\left(\beta \Phi\right),\label{hamiltonian}
\eeq
where we have set the velocity to one, $v=1$. In general the coupling constant lies in the range $0<\beta^2\le 1$, where the cosine term is relevant in the renormalisation-group sense and opens a gap $\Delta$ in the spectrum corresponding to the mass of solitons and antisolitons. In this paper we will specifically consider the so-called attractive regime, $\beta^2< 1/2$, where the spectrum consists of solitons and antisolitons with attractive interactions that form bound states, the so-called breathers. The number of species of breathers depends on the particular value of the coupling constant. We mostly focus on the simplest case, $1/3<\beta^2<1/2$, where there exists only one species of breathers. The case $\beta^2<1/3$, where there are many species of breathers, is briefly discussed in Section \ref{morethanone}. Quantum quenches in the repulsive regime $\beta^2>1/2$ have been studied in Ref.~\cite{repulsive}.

 \subsection{Realisations of the sine-Gordon model}
The sine-Gordon model arises as the low-energy description of a variety of systems in condensed-matter physics. Probably the most prominent realisation is provided by the Heisenberg chain with a field-induced gap~\cite{solids,benzoatetwo}
\beq
H=J\sum_{j=1}^L\left[ S_j^x S_{j+1}^x+S_{j}^yS_{j+1}^y+S_{j}^zS_{j+1}^z\right]+h_u\sum_{j=1}^LS_j^z+h_s\sum_{j=1}^L(-1)^jS_j^x.\label{benzoate}
\eeq
In the thermodynamic limit, the low energy dynamics of (\ref{benzoate}) is described by a quantum sine-Gordon model, typically in the attractive regime we are interested in. The vertex operator $\exp({\rm i}\beta\Phi/2)$ we are considering here corresponds to the bosonised form of the staggered transverse magnetisation, $S_j^+\sim(-1)^j\exp({\rm i}\beta\Phi/2)$, see Refs.~\cite{solids,repulsive} for more details.

The sine-Gordon model can also be used to describe the low-energy dynamics of  systems of interacting bosons~\cite{coldatoms}, which can be experimentally realised with ultra cold trapped atoms~\cite{optical}. Two such approaches have been outlined in Ref.~\cite{repulsive}. The first of these is to consider a single species of bosons in a periodic potential, where bosonisation~\cite{bosonizeboson} yields an effective sine-Gordon model with the cosine term originating from the periodic potential. The vertex operator then describes the leading oscillating term in the particle density, $\rho_\mathrm{osc}\sim\exp({\rm i}\beta\Phi/2)$. The second realisation in cold atomic systems is provided by a pair of coupled one-dimensional condensates~\cite{coupledcondensates} as can be realised experimentally using atom chips~\cite{atomchips}. Here the dynamics of the relative phase of the condensates is governed by the sine-Gordon model.

 \subsection{Particle spectrum and factorised scattering}
 The quantum sine-Gordon model (\ref{hamiltonian}) is integrable, and thus a factorisable scattering theory~\cite{mussardo,zamolodchikovsmatrix}. The energy and momentum of solitons and antisolitons can be parametrised by a rapidity, $\theta$, as $E=\Delta \cosh\theta$ and $p=\Delta\sinh\theta$, respectively, where $\Delta$ denotes the soliton mass (recall that we set $v=1$). The mass of the breather is given by 
\beq
\Delta_B=2\Delta \sin\frac{\pi\xi}{2},
\eeq
where the parameter $\xi$ is defined by
\beq
\xi=\frac{\beta^2}{1-\beta^2},
\eeq
and we assume $1/3<\beta^2<1/2$ for which only one type of breathers exists. We define particle creation and annihilation operators $Z^\dag_{a}(\theta)$ and $Z_a(\theta)$ with $a=\pm$ for solitons and antisolitons with rapidity $\theta$, as well as $B^\dag(\theta)$ and $B(\theta)$ for the breather respectively. Classically, solitons and antisolitons are field configurations connecting adjacent minima of the cosine potential, as is reflected by the topological charge 
\begin{equation}
Q=\frac{\beta}{2\pi}\int dx\,\partial_x\Phi(x)
\end{equation}
taking the values $\pm 1$. On the other hand, breathers correspond to bound states of solitons and antisolitons and are thus charge neutral, $Q=0$. The energy and momentum of the breathers are given by $E=\Delta_B\cosh\theta$ and $p=\Delta_B\sinh\theta$. The scattering matrix between solitons and antisolitons, $S_{a_1 a_2}^{b_1b_2}(\theta)$, defines their algebra as
\begin{eqnarray}
Z_{a_1}(\theta_1)Z_{a_2}(\theta_2)&=&
S_{a_1a_2}^{b_1b_2}(\theta_1-\theta_2)Z_{b_2}(\theta_2)Z_{b_1}(\theta_1),\nonumber\\
Z^\dagger_{a_1}(\theta_1)Z^\dagger_{a_2}(\theta_2)&=&
S_{a_1a_2}^{b_1b_2}(\theta_1-\theta_2)Z^\dagger_{b_2}(\theta_2)Z^\dagger_{b_1}(\theta_1),\nonumber\\
Z_{a_1}(\theta_1)Z^\dagger_{a_2}(\theta_2)&=&2\pi\delta(\theta_1-\theta_2)
\delta_{a_1,a_2}+
S_{a_2b_1}^{b_2a_1}(\theta_2-\theta_1)Z^\dagger_{b_2}(\theta_2)Z_{b_1}(\theta_1).
\end{eqnarray}
Explicit expressions for the soliton scattering matrix $S_{ab}^{cd}(\theta)$ are for completeness given in Appendix~\ref{aap:explicit}. Here we only state the relevant relations, starting with the Yang--Baxter equation,
\beq
S_{a_2a_3}^{b_2b_3}(\theta_2-\theta_3)S_{a_1b_3}^{b_1c_3}(\theta_1-\theta_3)S_{b_1b_2}^{c_1c_2}(\theta_1-\theta_2)=S_{a_1a_2}^{b_1b_2}(\theta_1-\theta_2)S_{b_1a_3}^{c_1b_3}(\theta_1-\theta_3)S_{b_2b_3}^{c_2c_3}(\theta_2-\theta_3).
\eeq
Furthermore, we have the unitarity and crossing conditions
\beq
S_{a_1a_2}^{c_1c_2}(\theta)S_{c_1c_2}^{b_1b_2}(-\theta)&=&\delta_{a_1}^{b_1}\delta_{a_2}^{b_2},\\
S_{ab}^{cd}({\rm i}\pi-\theta)&=&S_{\bar{c}b}^{\bar{a}d}(\theta)=S_{a\bar{d}}^{c\bar{b}}(\theta),\quad\bar{a}=-a,
\eeq
as well as the relations
\beq
\left(S_{ab}^{cd}(\theta)\right)^*=S_{ab}^{cd}(-\theta)\,\,\,\,\,{\rm for}\,\,\,\,\theta\in\mathbb{R},\,\,\,\,S_{ab}^{cd}(\theta)=S_{ba}^{dc}(\theta)=S_{cd}^{ab}(\theta)=S_{\bar{a}\bar{b}}^{\bar{c}\bar{d}}(\theta).
\eeq
The scattering matrix $S_{+-}^{-+}(\theta)=S^{+-}_{-+}(\theta)$ possesses a pole $\theta=\mathrm{i}\pi(1-\xi)$ corresponding to a soliton-antisoliton bound state with mass $\Delta_B^2=2\Delta^2[1+\cos(\pi(1-\xi))]=4\Delta^2\sin^2(\pi\xi/2)$, which is just the first breather state. The scattering of solitons and antisolitons off breathers is governed by the scattering matrix, $S_B(\theta)$, defined by
\beq
Z_a^\dag(\theta_1)B^\dag(\theta_2)=S_B(\theta_1-\theta_2)B^\dag(\theta_2)Z_a^\dag(\theta_1),
\eeq
which is diagonal in the index $a$. Similar relations hold with $Z_a^\dagger(\theta)$ and $B^\dagger(\theta)$ replaced by the corresponding annihilation operators. Finally, the scattering of breathers is described by the breather-breather scattering matrix, $S_{BB}(\theta)$, via
\begin{eqnarray}
B(\theta_1)B(\theta_2)&=&S_{BB}(\theta_1-\theta_2)B(\theta_2)B(\theta_1),\\
B^\dag(\theta_1)B^\dag(\theta_2)&=&S_{BB}(\theta_1-\theta_2)B^\dag(\theta_2)B^\dag(\theta_1),\\
B(\theta_1)B^\dag(\theta_2)&=&2\pi\delta(\theta_1-\theta_2)+S_{BB}(\theta_1-\theta_2)B(\theta_2)B^\dag(\theta_1).
\end{eqnarray}
These scattering matrices satisfy the unitarity and crossing conditions
\beq
&&S_{B}(\theta)S_B(-\theta)=1,\quad S_{BB}(\theta)S_{BB}(-\theta)=1,\\
&&S_{B}({\rm i}\pi-\theta)=S_{B}(\theta),\quad S_{BB}({\rm i}\pi-\theta)=S_{BB}(\theta).
\label{Scrossing}
\eeq

Using the analogously defined particle annihilation operators $Z_a(\theta)$ and $B(\theta)$, the ground state of the sine-Gordon model is defined by
\beq
Z_a(\theta)\vert 0\rangle=B(\theta)\vert 0\rangle=0.
\eeq
A complete basis of eigenstates is found by acting on the vacuum with particle creation operators,
\beq
\vert \theta_1,\dots,\theta_N,\phi_1,\dots,\phi_M\rangle_{a_1,\dots,a_N}=Z_{a_1}^\dag(\theta_1)\dots Z_{a_N}^\dag(\theta_N)B^\dag(\phi_1)\dots B^\dag(\phi_M)\vert0\rangle.
\eeq

\section{The initial state} \label{initialstate}
The determination of the exact initial state that corresponds to a given quantum quench protocol in an interacting theory is a difficult and unresolved problem, even in quenches of integrable field theories~\cite{ZFinitial}. The particle dynamics in interacting theories are usually not factorisable across the $t=0$ boundary~\cite{delfino,schuricht}, which suggests that it may not be possible to determine the initial state using the standard methods from integrability. One known exception~\cite{planar} is provided by the planar large-$N$ limit of the principal chiral sigma model, where factorisation across the $t=0$ boundary is maintained in the interacting theory. Apart from this, quenches starting from the ground state of an initial Hamiltonian have been analysed using perturbation theory in the quench parameter~\cite{delfino,delfinotwo} as well as other approximate methods~\cite{initialsinh,sinetcsa}. 

Here, however, we will not investigate the general properties of the initial state. Instead, we assume a simple initial state of the ``squeezed state form" and focus on the subsequent time evolution. Such states have been proposed by Fioretto and Mussardo~\cite{fioretto} as a natural starting point in the study of quantum quenches, given their simplicity. A second motivation to consider such initial states comes from the observation~\cite{cardycalabrese} that analytically continuing to imaginary times, the problem of computing observables after a quantum quench can be mapped to that of computing observables in a field theory with boundaries, the boundaries being identified with the initial state. Thus a natural starting point for the study of quenches in integrable field theories are provided by integrable boundary states~\cite{gloshal}. Motivated by this we thus assume the following initial state
\beq
\vert \Psi_0\rangle=\left(1+\frac{g}{2}B^\dag(0)\right)\exp\left[\int_0^\infty\frac{d\theta}{2\pi}K^{ab}(\theta)Z^\dag_{a}(-\theta)Z^\dag_b(\theta)+\int_0^\infty\frac{d\phi}{2\pi}K_B(\phi)B^\dag(-\phi)B^\dag(\phi)\right]\vert 0\rangle.\label{initial}
\eeq
We note that the breather particles contribute both as pairs with finite rapidities as well as individual, zero-momentum particles created by the operator $B^\dagger(0)$. For convenience, we label soliton and antisoliton rapidities by the letter $\theta$ and breather rapidities by $\phi$. The functions  $K^{ab}(\theta)$ and $K_B(\phi)$ are assumed to satisfy the boundary Yang--Baxter equation,
\beq
K^{a_1c_1}(\theta_1)K^{c_2c_3}(\theta_2)S_{c_2c_1}^{a_2c_4}(\theta_1+\theta_2)S_{c_3c_4}^{b_2b_1}(\theta_1-\theta_2)=K^{c_1b_1}(\theta_1)K^{c_2c_3}(\theta_2)S_{c_3c_1}^{b_2c_4}(\theta_1+\theta_2)S_{c_2c_4}^{a_2a_1}(\theta_1-\theta_2),
\label{BYBE}
\eeq
and the so-called ``cross-unitarity" conditions,
\beq
K^{ab}(\theta)&=&S_{cd}^{ab}(2\theta)K^{dc}(-\theta),\label{BCU1}\\
K_B(\phi)&=&S_{BB}(2\phi)K_B(-\phi).\label{BCU2}
\eeq
These conditions ensure that the exponential in the initial state (\ref{initial}) is well defined. In order to simplify the calculations, we will restrict ourselves to initial states with vanishing topological charge, which implies the conditions 
\beq
K^{++}(\theta)=K^{--}(\theta)=0.\label{dirichletlike}
\eeq
Furthermore, we require the initial state to be normalisable, which implies that $K^{ab}(\theta)$ and $K_B(\phi)$ have to decay to zero sufficiently fast at large rapidities. We note that the functions $K^{ab}(\theta)$ and $K_B(\theta)$ are not required to satisfy the boundary unitarity condition, hence (\ref{initial}) does not satisfy all the conditions required to be an integrable boundary state, as defined in Ref.~\cite{gloshal}. For later convenience, we also define the functions
\beq
G(\theta)=\vert K^{ab}(\theta)\vert^2=\vert K^{a\bar{a}}(\theta)\vert^2,\quad G_B(\phi)=\vert K_B(\phi)\vert^2.
\eeq
To summarise, for the purposes of this paper, we will simply assume an initial state of the form (\ref{initial}) with the functions $K^{ab}(\theta)$ and $K_B(\phi)$ and the parameter $g$ unspecified except for the requirements (\ref{BYBE})--(\ref{dirichletlike}). Our main results will be given in terms of these general functions. The question of which particular function correctly describes a given quantum quench is beyond the scope of this paper and will be left for future investigations. 

Still one can make further assumptions on the form of the functions $K^{ab}(\theta)$ and $K_B(\phi)$ by considering an initial state corresponding to Dirichlet boundary conditions\footnote{We note that in the attractive boundary sine-Gordon model the integrability implies non-trivial relations~\cite{gloshal} between the functions $K^{ab}(\theta)$, $K_B(\phi)$ and the parameter $g$, eg,  
${\rm Res}[K^{ab}(\theta),\theta={\rm i}\pi(1-\xi)]\sim {\rm i}g/2$. However, for our quench setup the relation between $K^{ab}(\theta)$ and $g$ is not required.} with $\Phi(t=0,x)=\Phi_0=0$. Such an initial state is compatible with the condition (\ref{dirichletlike}), thus we call states satisfying (\ref{dirichletlike}) also ``Dirichlet-like" initial states. Physically it can be identified with a quench in the mass parameter $\lambda$ from an infinite value $\lambda_0=\infty$ to a finite value at $t=0$. This implies an infinite-to-finite change in the soliton and breather masses. Such an initial state is, however, problematic since it introduces an infinite amount of energy density and the initial state is not normalisable. This problem manifests itself in the fact that the functions $K^{ab}_{\rm Dirichlet}(\theta)$ and $K_{B\,{\rm Dirichlet}}(\phi)$ tend to constant values at large rapidities instead of decaying to zero, as is required of a normalisable state. A prescription to obtain normalisable initial states from Dirichlet boundary conditions was proposed in Ref.~\cite{fioretto}, motivated on the similar approach in the study of quenches in conformal field theories~\cite{cardycalabrese}. The idea is to modify the functions $K^{ab}_{\rm Dirichlet}(\theta)$ and $K_{B\,{\rm Dirichlet}}(\phi)$ introducing an ``extrapolation time", $\tau_0>0$, such that
\beq
 K^{ab}(\theta)=e^{-2\tau_0 \Delta \cosh\theta}K^{ab}_{\rm Dirichlet}(\theta), \quad\,K_B(\phi)=e^{-2\tau_0\Delta_B\cosh\phi}K_{B\,{\rm Dirichlet}}(\phi).
 \label{extrapolationtime}
\eeq
As a result, all appearing integrals over the rapidities will be regularised and thus the initial state becomes normalisable. We will always assume that an extrapolation time has been introduced in this way. We note in passing that it has been shown~\cite{initialsinh}, however, that such a simple regularisation cannot accurately describe realistic quenches in massive integrable field theories unless the extrapolation time is taken to be rapidity dependent, ie, replacing the constant $\tau_0$ by a function $\tau_0(\theta)$.

Very recently, an approximation for the function $K_B(\phi)$ for mass quenches from $\Delta_0$ to $\Delta$ was proposed~\cite{sinetcsa}, which is given by
\beq
K_B(\phi)=K_\mathrm{free}(\phi)K_{B\,{\rm Dirichlet}}(\phi),\quad 
K_\mathrm{free}(\phi)=\frac{\cosh\phi-\sqrt{\left(\frac{\Delta_0}{\Delta}\right)^2+\sinh^2\phi}}{\cosh\phi+\sqrt{\left(\frac{\Delta_0}{\Delta}\right)^2+\sinh^2\phi}}.
\eeq
Note that the prefactor ensures the normalisability at large rapidities. The result was obtained by analytically continuing the corresponding function in the sinh-Gordon model and numerically checking the result using the truncated conformal space approach. However, since the sinh-Gordon model does not possess soliton-like particles, an approximate expression for the function $K^{ab}(\theta)$ cannot be derived in this way.

The time evolution starting from the initial state is then given by
\beq
\vert \Psi_t\rangle=e^{-{\rm i}t\mathcal{H}}\vert \Psi_0\rangle,
\eeq
resulting in the expectation value of an operator $\mathcal{O}$ as
\beq
\frac{\langle \Psi_t\vert \mathcal{O}\vert \Psi_t\rangle}{\langle \Psi_t\vert \Psi_t\rangle}=\frac{\langle \Psi_0\vert e^{{\rm i}t\mathcal{H}}\mathcal{O}e^{-{\rm i}t\mathcal{H}}\vert \Psi_0\rangle}{\langle \Psi_0\vert\Psi_0\rangle}.\label{expectation}
\eeq
We will  focus specifically on ``small quenches" where we assume that the functions $K^{ab}(\theta)$ and $K_B(\phi)$ as well as the parameter $g$ are small, such that we can restrict ourselves to the leading terms in an expansion in these formal parameters. This does not imply, however, that our calculations are limited to a specific order. In fact, the resummation of the long-time behaviour leading to the exponential decay in the final result (\ref{finalresult}) requires the analysis of higher-order terms. Our assumption of small quenches rather refers to the limitation of the calculation of the obtained decay rates (\ref{decayrates1})--(\ref{decayrates3}) to leading order in $K^{ab}(\theta)$ and $K_B(\phi)$.

In the next sections we will focus particularly on the vertex operator $\mathcal{O}=\exp\left({\rm i}\beta\Phi/2\right)$. This operator is chosen for its semi-locality properties with respect to the solitons and antisolitons, which simplify the computations (as seen in the form-factor axioms presented in Appendix~\ref{app:FFA}). The computation for a general vertex operator, $\exp\left({\rm i}\alpha\Phi\right)$, turns out to be significantly more involved and thus is beyond the scope of this paper.

\section{Linked-cluster expansion} \label{linkedclusterexpansion}
To compute expectation value (\ref{expectation}), we expand the states $\vert \Psi_0\rangle$ in terms of the eigenstates of the sine-Gordon Hamiltonian and compute each term in this expansion using exact form factors. These terms can have singularities that need to be regularised. It is expected that some singularities from the numerator and denominator cancel each other.

The denominator in (\ref{expectation}) can be formally expanded as
\beq
\langle \Psi_0\vert \Psi_0\rangle&=&\sum_{N,J=0}^{\infty}\int_0^\infty\frac{d\theta^\prime_1\dots d\theta^\prime_N}{N!(2\pi)^N}\frac{d\theta_1\dots d\theta_N}{N!(2\pi)^N}\frac{d\phi_1^\prime\dots d\phi_J^\prime}{J!(2\pi)^J}\frac{d\phi_1\dots d\phi_J}{J!(2\pi)^J}
\nonumber\\*
&&\qquad\times \prod_{n=1}^N\left(K^{a_nb_n}(\theta_n^\prime)\right)^*K^{c_nd_n}(\theta_n)\prod_{j=1}^{J}\left(K_B(\phi_j^\prime)\right)^*K_B(\phi_j)\nonumber\\*
&&\qquad\times \,_{b_1a_1\dots b_Na_N}\langle \phi^\prime_1,-\phi_1^\prime,\dots,\phi_J^\prime,-\phi_J^\prime,\theta_1^\prime,-\theta_1^\prime,\dots,\theta_N^\prime,-\theta_N^\prime\vert\nonumber\\*
&&\qquad\qquad\qquad\qquad\times\vert -\theta_N,\theta_N,\dots,-\theta_1,\theta_1,-\phi_J,\phi_J,\dots,-\phi_1,\phi_1\rangle_{c_N d_N\dots c_1d_1}\nonumber\\*
&+&\frac{\vert g\vert^2}{4}\sum_{N,J=0}^{\infty}\int_0^\infty\frac{d\theta^\prime_1\dots d\theta^\prime_N}{N!(2\pi)^N}\frac{d\theta_1\dots d\theta_N}{N!(2\pi)^N}\frac{d\phi_1^\prime\dots d\phi_J^\prime}{J!(2\pi)^J}\frac{d\phi_1\dots d\phi_J}{J!(2\pi)^J}\nonumber\\*
&&\qquad\times \prod_{n=1}^N\left(K^{a_nb_n}(\theta_n^\prime)\right)^*K^{c_nd_n}(\theta_n)\prod_{j=1}^{J}\left(K_B(\phi_j^\prime)\right)^*K_B(\phi_j)\nonumber\\*
&&\qquad\times \,_{b_1a_1\dots b_Na_N}\langle \phi^\prime_1,-\phi_1^\prime,\dots,\phi_J^\prime,-\phi_J^\prime,\theta_1^\prime,-\theta_1^\prime,\dots,\theta_N^\prime,-\theta_N^\prime,0_B\vert\nonumber\\*
&&\qquad\qquad\qquad\times\vert 0_B, -\theta_N,\theta_N,\dots,-\theta_1,\theta_1,-\phi_J,\phi_J,\dots,-\phi_1,\phi_1\rangle_{c_N d_N\dots c_1d_1},\label{expandeddem}
\eeq
where we have introduced the notation $\vert 0_B\rangle=B^\dagger(0)\vert 0\rangle$ to represent a state with one zero-momentum breather (this notation is introduced to distinguish such a state from the vacuum state, $\vert 0\rangle$). We note that the sums start at $N=J=0$. The norm of the initial state can be written more compactly by introducing the notation
\beq
\langle\Psi_0\vert\Psi_0\rangle\equiv\sum_{N=0}^\infty\sum_{J=0}^\infty Z_{2N,2J}+\sum_{N=0}^\infty\sum_{J=0}^\infty Z_{2N,2J+1},
\eeq
corresponding to the terms in (\ref{expandeddem}). For example, $Z_{0,0}=1$ or $Z_{0,0+1}=|g|^2/4$. In general we have $Z_{N,J}=\mathcal{O}(g^0)$ and $Z_{N,J+1}=\mathcal{O}(g^2)$. In the small quench limit we can write
\beq
\frac{1}{\langle \Psi_0\vert\Psi_0\rangle}&=&1-Z_{2,0}-Z_{0,2}-Z_{0,0+1}+\left(Z_{2,0}+Z_{0,2}+Z_{0,0+1}\right)^2\nonumber\\*
&&-Z_{4,0}-Z_{2,2}-Z_{2,0+1}-Z_{0,4}-Z_{0,2+1}+\mathcal{O}(K^6)+\mathcal{O}(g^2 K^4)\label{zexpand}
\eeq
where $K^{ab}(\theta)\sim K_B(\theta)\sim g\sim K$.

The numerator in (\ref{expectation}) can be expanded similarly,
\beq
\langle \Psi_t\vert \mathcal{O}\vert \Psi_t\rangle&=&\sum_{M,N,I,J=0}^{\infty}\int_0^\infty\frac{d\theta^\prime_1\dots d\theta^\prime_M}{M!(2\pi)^M}\frac{d\theta_1\dots d\theta_N}{N!(2\pi)^N}\frac{d\phi_1^\prime\dots d\phi_I^\prime}{I!(2\pi)^I}\frac{d\phi_1\dots d\phi_J}{J!(2\pi)^J}\nonumber\\*
&&\qquad\times \prod_{m=1}^M\left(K^{a_mb_m}(\theta_m^\prime)\right)^*\prod_{n=1}^NK^{c_nd_n}(\theta_n)\prod_{i=1}^{I}\left(K_B(\phi_i^\prime)\right)^*\prod_{j=1}^JK_B(\phi_j)\nonumber\\*
&&\qquad\times e^{2\Delta {\rm i} t\sum_m \cosh\theta_m^\prime}e^{-2\Delta {\rm i}t\sum_n\cosh\theta_n}e^{2\Delta_B{\rm i}t\sum_i\cosh\phi_i^\prime}e^{-2\Delta_B{\rm i}t\sum_j\cosh\phi_j}\nonumber\\*
&&\qquad\times \,_{b_1a_1\dots b_Ma_M}\langle \phi^\prime_1,-\phi_1^\prime,\dots,\phi_I^\prime,-\phi_I^\prime,\theta_1^\prime,-\theta_1^\prime,\dots,\theta_M^\prime,-\theta_M^\prime\vert\nonumber\\*
&&\qquad\qquad\qquad\times\mathcal{O}\vert -\theta_N,\theta_N,\dots,-\theta_1,\theta_1,-\phi_J,\phi_J,\dots,-\phi_1,\phi_1\rangle_{c_N d_N\dots c_1d_1}\nonumber\\
&+&\frac{g}{2}e^{-\Delta_B{\rm i}t}\sum_{M,N,I,J=0}^{\infty}\int_0^\infty\frac{d\theta^\prime_1\dots d\theta^\prime_M}{M!(2\pi)^M}\frac{d\theta_1\dots d\theta_N}{N!(2\pi)^N}\frac{d\phi_1^\prime\dots d\phi_I^\prime}{I!(2\pi)^I}\frac{d\phi_1\dots d\phi_J}{J!(2\pi)^J}\nonumber\\*
&&\qquad\times \prod_{m=1}^M\left(K^{a_mb_m}(\theta_m^\prime)\right)^*\prod_{n=1}^NK^{c_nd_n}(\theta_n)\prod_{i=1}^{I}\left(K_B(\phi_i^\prime)\right)^*\prod_{j=1}^JK_B(\phi_j)\nonumber\\*
&&\qquad\times e^{2\Delta {\rm i} t\sum_m \cosh\theta_m^\prime}e^{-2\Delta{\rm i}t\sum_n\cosh\theta_n}e^{2\Delta_B{\rm i}t\sum_i\cosh\phi_i^\prime}e^{-2\Delta_B{\rm i}t\sum_j\cosh\phi_j}\nonumber\\*
&&\qquad\times \,_{b_1a_1\dots b_Ma_M}\langle \phi^\prime_1,-\phi_1^\prime,\dots,\phi_I^\prime,-\phi_I^\prime,\theta_1^\prime,-\theta_1^\prime,\dots,\theta_M^\prime,-\theta_M^\prime\vert\nonumber\\*
&&\qquad\qquad\qquad\times\mathcal{O}\vert 0_B,-\theta_N,\theta_N,\dots,-\theta_1,\theta_1,-\phi_J,\phi_J,\dots,-\phi_1,\phi_1\rangle_{c_N d_N\dots c_1d_1}\nonumber\\
&+&\frac{g^*}{2}e^{\Delta_B{\rm i}t}\sum_{M,N,I,J=0}^{\infty}\int_0^\infty\frac{d\theta^\prime_1\dots d\theta^\prime_M}{M!(2\pi)^M}\frac{d\theta_1\dots d\theta_N}{N!(2\pi)^N}\frac{d\phi_1^\prime\dots d\phi_I^\prime}{I!(2\pi)^I}\frac{d\phi_1\dots d\phi_J}{J!(2\pi)^J}\nonumber\\*
&&\qquad\times \prod_{m=1}^M\left(K^{a_mb_m}(\theta_m^\prime)\right)^*\prod_{n=1}^NK^{c_nd_n}(\theta_n)\prod_{i=1}^{I}\left(K_B(\phi_i^\prime)\right)^*\prod_{j=1}^JK_B(\phi_j)\nonumber\\*
&&\qquad\times e^{2\Delta{\rm i} t\sum_m \cosh\theta_m^\prime}e^{-2\Delta {\rm i}t\sum_n\cosh\theta_n}e^{2\Delta_B{\rm i}t\sum_i\cosh\phi_i^\prime}e^{-2\Delta_B{\rm i}t\sum_j\cosh\phi_j}\nonumber\\*
&&\qquad\times \,_{b_1a_1\dots b_Ma_M}\langle \phi^\prime_1,-\phi_1^\prime,\dots,\phi_I^\prime,-\phi_I^\prime,\theta_1^\prime,-\theta_1^\prime,\dots,\theta_M^\prime,-\theta_M^\prime,0_B\vert\nonumber\\*
&&\qquad\qquad\qquad\times\mathcal{O}\vert -\theta_N,\theta_N,\dots,-\theta_1,\theta_1,-\phi_J,\phi_J,\dots,-\phi_1,\phi_1\rangle_{c_N d_N\dots c_1d_1}\nonumber\\
&+&\frac{\vert g\vert^2}{4}\sum_{M,N,I,J=0}^{\infty}\int_0^\infty\frac{d\theta^\prime_1\dots d\theta^\prime_M}{M!(2\pi)^M}\frac{d\theta_1\dots d\theta_N}{N!(2\pi)^N}\frac{d\phi_1^\prime\dots d\phi_I^\prime}{I!(2\pi)^I}\frac{d\phi_1\dots d\phi_J}{J!(2\pi)^J}\nonumber\\*
&&\qquad\times \prod_{m=1}^M\left(K^{a_mb_m}(\theta_m^\prime)\right)^*\prod_{n=1}^NK^{c_nd_n}(\theta_n)\prod_{i=1}^{I}\left(K_B(\phi_i^\prime)\right)^*\prod_{j=1}^JK_B(\phi_j)\nonumber\\*
&&\qquad\times e^{2\Delta{\rm i} t\sum_m \cosh\theta_m^\prime}e^{-2\Delta{\rm i}t\sum_n\cosh\theta_n}e^{2\Delta_B{\rm i}t\sum_i\cosh\phi_i^\prime}e^{-2\Delta_B{\rm i}t\sum_j\cosh\phi_j}\nonumber\\*
&&\qquad\times \,_{b_1a_1\dots b_Ma_M}\langle \phi^\prime_1,-\phi_1^\prime,\dots,\phi_I^\prime,-\phi_I^\prime,\theta_1^\prime,-\theta_1^\prime,\dots,\theta_M^\prime,-\theta_M^\prime,0_B\vert\nonumber\\*
&&\qquad\qquad\qquad\times\mathcal{O}\vert 0_B,-\theta_N,\theta_N,\dots,-\theta_1,\theta_1,-\phi_J,\phi_J,\dots,-\phi_1,\phi_1\rangle_{c_N d_N\dots c_1d_1}\nonumber\\
&\equiv& \sum_{M,N,I,J=0}^\infty C_{2M,2I;2N,2J}+\sum_{M,N,I,J=0}^\infty C_{2M,2I;2N,2J+1}\nonumber\\*
&&+\sum_{M,N,I,J=0}^\infty C_{2M,2I+1;2N,2J}+\sum_{M,N,I,J=0}^\infty C_{2M,2I+1;2N,2J+1}\label{numerator}.
\eeq

The linked-cluster expansion consists in combining the terms in the expansions (\ref{zexpand}) and (\ref{numerator}), term by term in orders of $K$ as
\beq
\frac{\langle\Psi_t\vert \mathcal{O}\vert \Psi_t\rangle}{\langle\Psi_t\vert\Psi_t\rangle}&=&\left( \sum_{M,N,I,J=0}^\infty C_{2M,2I;2N,2J}+\sum_{M,N,I,J=0}^\infty C_{2M,2I;2N,2J+1}\right.\nonumber\\*
&&\left.+\sum_{M,N,I,J=0}^\infty C_{2M,2I+1;2N,2J}+\sum_{M,N,I,J=0}^\infty C_{2M,2I+1;2N,2J+1}\right)\nonumber\\*
&&\times\left(\sum_{N=0}^\infty\sum_{J=0}^\infty Z_{2N,2J}+\sum_{N=0}^\infty\sum_{J=0}^\infty Z_{2N,2J+1}\right)^{-1}\nonumber\\
&\equiv& \sum_{M,N,I,J=0}^\infty D_{2M,2I;2N,2J}+\sum_{M,N,I,J=0}^\infty D_{2M,2I;2N,2J+1}\nonumber\\*
&&+\sum_{M,N,I,J=0}^\infty D_{2M,2I+1;2N,2J}+\sum_{M,N,I,J=0}^\infty D_{2M,2I+1;2N,2J+1},\label{linked}
\eeq
where the terms in the final expansion are regularised and finite in the infinite-volume limit. 

We have computed all the leading contributions to the linked-cluster expansion for large times after the quench, up to (and including) orders $K^4$, $gK^4$ and $g^2K^4$, for the operator $\mathcal{O}=\exp({\rm i}\beta\Phi/2)$. We show the explicit computation in Appendix~\ref{app:LCE}. In the next section we present this final result, and also argue that these leading, large-time contributions can be resummed for all orders of $K$. After resummation, it is evident that this observable exponentially decays at long times, with a set of different decay rates, related to soliton and breather contributions.

\section{The main result} \label{themainresult}
We have analysed the terms up to order $K^4$, $gK^4$ and $g^2K^4$ in the expansion (\ref{linked}), ie, 
\beq
\frac{\langle \Psi_t\vert e^{{\rm i}\beta\Phi/2}\vert\Psi_t\rangle}{\langle\Psi_0\vert\Psi_0\rangle}&=&D_{0,0;0,0}+D_{2,0;2,0}+D_{4,0;4,0}\nonumber\\*
&&+D_{0,0+1;0,0}+D_{2,0+1;2,0}+D_{4,0+1;4,0}+D_{0,2+1;0,2}+D_{0,4+1;0,4}+D_{2,2+1;2,2}\nonumber\\*
&&+D_{0,0;0,0+1}+D_{2,0;2,0+1}+D_{4,0;4,0+1}+D_{0,2;0,2+1}+D_{0,4;0,4+1}+D_{2,2;2,2+1}\nonumber\\*
&&+D_{0,0+1;0,0+1}+D_{2,0+1;2,0+1}+D_{4,0+1;4,0+1}+\dots,
\label{expansion}
\eeq
where the dots represent higher-order terms that have not been evaluated. The leading long-time behaviour of these terms can be extracted from the pole contributions, as explicitly done in Appendix~\ref{app:LCE}, with the result
\beq
\frac{\langle \Psi_t\vert e^{{\rm i}\beta\Phi/2}\vert\Psi_t\rangle}{\langle\Psi_0\vert\Psi_0\rangle}
&=&\mathcal{G}_{\beta/2}\left[1-\frac{t}{\tau}+\frac{1}{2}\left(\frac{t}{\tau}\right)^2+\dots\right]\nonumber\\*
&&+f_B^{\beta/2}\,{\rm Re}\left\{g\,e^{-\Delta_B{\rm i}t}\left[1-\frac{t}{\tau_B}-\frac{t}{\tau_{BB}}+\frac{t^2}{\tau_B\tau_{BB}}+\frac{1}{2}\left(\frac{t}{\tau_B}\right)^2+\frac{1}{2}\left(\frac{t}{\tau_{BB}}\right)^2+\dots\right]\right\}\nonumber\\*
&&+\frac{\vert g\vert^2}{4}\,f_{BB}^{\beta/2}({\rm i}\pi,0)\left[1-\frac{t}{\tau}+\frac{1}{2}\left(\frac{t}{\tau}\right)^2+\dots\right].\label{almostfinal}
\eeq
Here the dots represent sub-leading contributions contained in (\ref{expansion}) as well as higher-order terms. Furthermore, the appearing constants originate from the form factors,
\beq
\mathcal{G}_{\beta/2}=\langle 0\vert e^{{\rm i}\beta\Phi/2}\vert 0\rangle,\quad
f_B^{\beta/2}=\langle 0\vert e^{{\rm i}\beta\Phi/2}B^\dagger(0)\vert 0\rangle,\quad
f_{BB}^{\beta/2}({\rm i}\pi,0)=\langle 0\vert e^{{\rm i}\beta\Phi/2}B^\dagger({\rm i}\pi)B^\dagger(0)\vert 0\rangle,
\label{FFconstants}
\eeq
explicit expressions are given in Appendix~\ref{aap:explicit}. Here we just note that $f^{\beta/2}_B$ is purely imaginary. Furthermore, the relaxation parameters read 
\beq
\tau^{-1}&=&\frac{2\Delta}{\pi}\int_0^\infty d\theta\,G(\theta)\sinh\theta +\mathcal{O}(K^4),\label{decayrates1}\\
\tau_B^{-1}&=&\frac{\Delta}{\pi}\int_0^\infty d\theta\,G(\theta)\left[1+S_B(\theta)\right]\sinh\theta+\mathcal{O}(K^4),\label{decayrates2}\\
\tau_{BB}^{-1}&=&\frac{\Delta_B}{\pi}\int_0^\infty d\phi\,G_B(\phi)\left[1-S_{BB}(\phi)\right]\sinh\phi+\mathcal{O}(K^4).\label{decayrates3}
\eeq
In Ref.~\cite{repulsive} the same calculation was performed in the repulsive regime, leading the result above with $g=0$. In this reference it was suggested that these leading contributions for all higher orders of $K$ can be resummed into an exponential function, given that the series matches the expansion
\beq
e^{-x}=1-x+\frac{x^2}{2}+\dots\,.
\eeq
We note that a similar resummation was shown to take place in the Ising field theory~\cite{CEF,SE} where the leading long-time behaviour of all higher-order terms can be extracted explicitly. Motivated by the result for the repulsive regime we thus conjecture that the terms in (\ref{almostfinal}) can also be resummed as exponentials, leading to our final result
\begin{eqnarray}
\frac{\langle \Psi_t\vert e^{{\rm i}\beta\Phi/2}\vert\Psi_t\rangle}{\langle\Psi_0\vert\Psi_0\rangle}
&=&\left[\mathcal{G}_{\beta/2}+\frac{\vert g\vert^2}{4}f_{BB}^{\beta/2}({\rm i}\pi,0)\right]\,e^{-t/\tau}
+f_B^{\beta/2}\,{\rm Re}\left[g\,e^{-\Delta_B{\rm i}t}\,e^{-\left(1/\tau_B+1/\tau_{BB}\right)t}\right]+\dots\label{finalresult}\\*
&=&\left[\mathcal{G}_{\beta/2}+\frac{\vert g\vert^2}{4}f_{BB}^{\beta/2}({\rm i}\pi,0)\right]\,e^{-\Gamma_1 t}
+|g|f_B^{\beta/2}\,e^{-\Gamma_2t}\,\cos(\Omega t-\delta),
\label{finalresult2}
\end{eqnarray}
where
\begin{equation}
\Gamma_1=\frac{1}{\tau},\quad\Gamma_2={\rm Re}\left[\frac{1}{\tau_B}+\frac{1}{\tau_{BB}}\right],\quad\Omega=\Delta_B+{\rm Im}\left[\frac{1}{\tau_B}+\frac{1}{\tau_{BB}}\right],\quad g=|g|e^{\mathrm{i}\delta}.
\label{eq:rates}
\end{equation}
We recall that the dots represent sub-leading contributions containing further oscillatory terms as well as power-law corrections. For example, the terms $D_{2,0;0,0}+D_{0,0;2,0}$ and their higher-order descendants can be resummed~\cite{repulsive} into $\cos(2\Delta t)\,e^{-t/\tau}/(\Delta t)^{3/2}$. We stress that the resummed expression (\ref{finalresult2}) contains the leading long-time behaviour at all orders in $K$, the small-quench assumption reflects itself only in the fact that the relaxation parameters (\ref{decayrates1})--(\ref{decayrates3}) have been determined only in $\mathcal{O}(K^2)$.

\begin{figure}[t]
\begin{center}
\includegraphics[width=85mm]{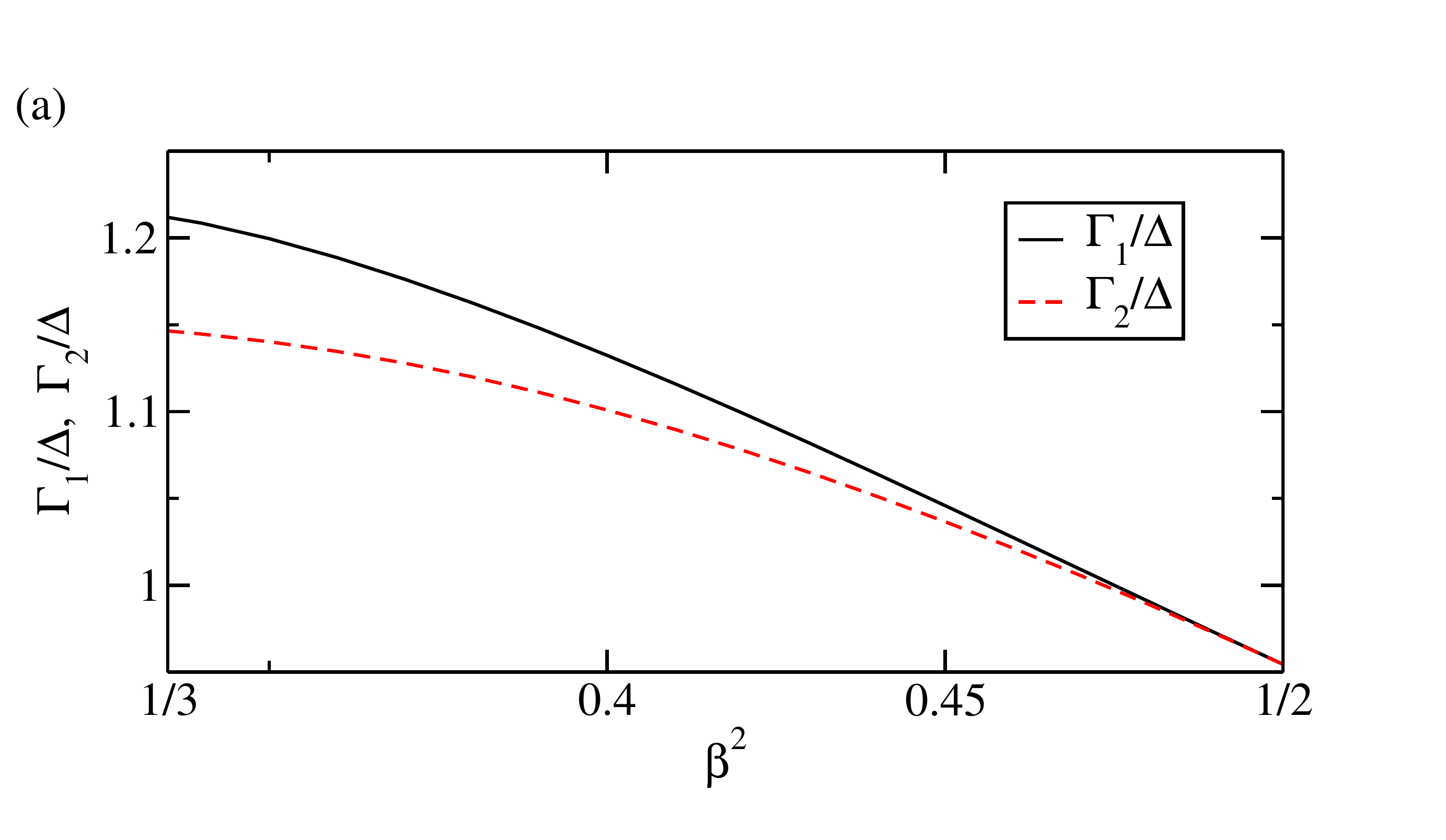}
\includegraphics[width=85mm]{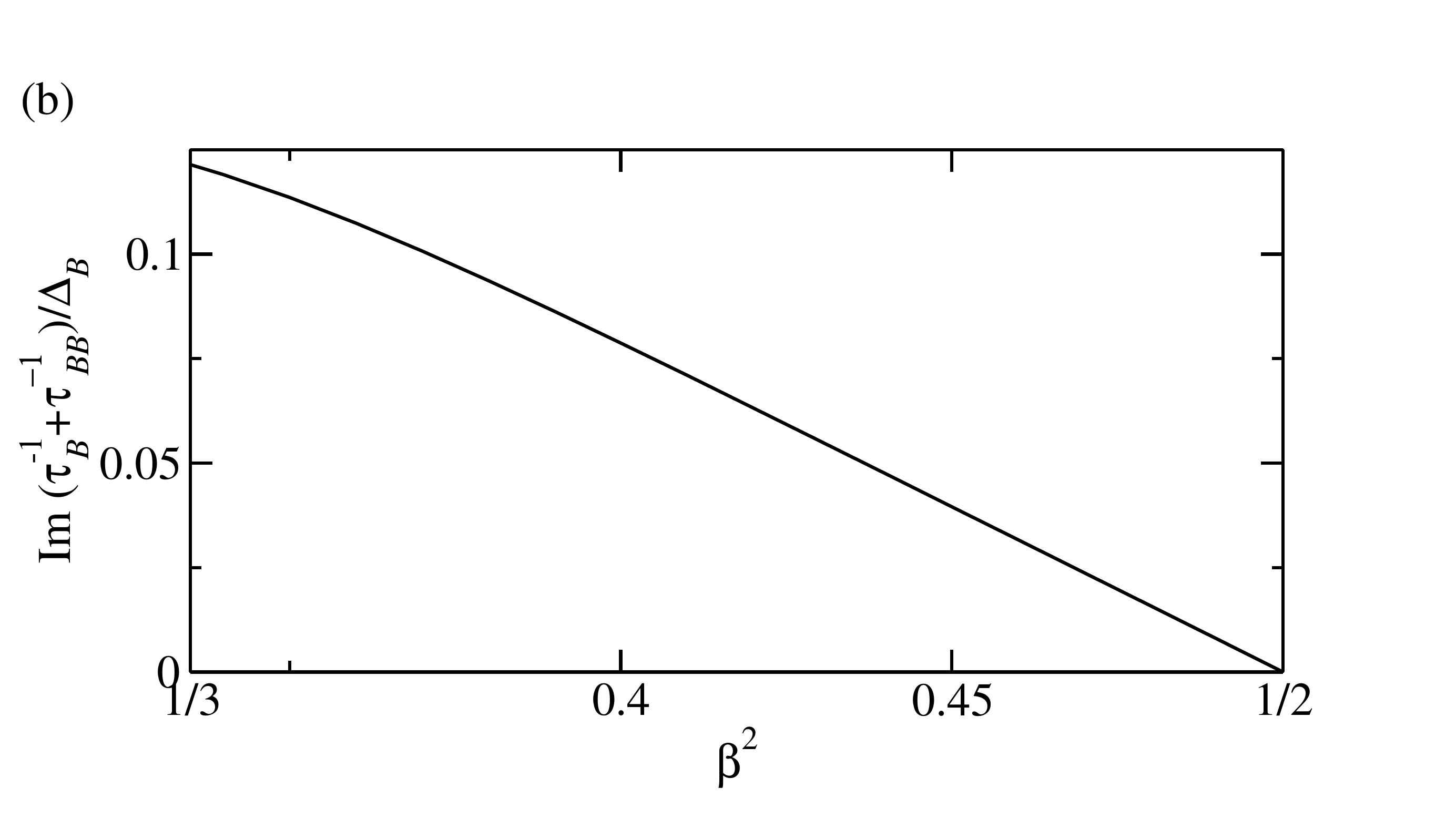}
\caption{(a) Decay rates of the two leading terms in the time evolution (\ref{finalresult}) as a function of the interaction strength $\beta^2$. We chose a Dirichlet-like initial state with $\Phi(t=0,x)=0$ and extrapolation time $\tau_0=0.1/\Delta$. We observe that the decay rate of the second, oscillating, term is smaller for $\beta^2<1/2$, ie, the oscillating behaviour is dominant at late times. (b) Relative $\mathcal{O}(K^2)$-corrections contained in the oscillation frequency $\Omega$ as compared to the bare value $\Delta_B$.}
\label{fig:rates}
\end{center}
\end{figure}

We stress that the second term in (\ref{finalresult2}) is explicitly oscillating with the frequency $\Omega$, ie, the leading terms at late times exhibit both, oscillations in time as well as exponential decay,\footnote{We note that undamped oscillations originating in one-particle contributions were observed in the form-factor calculation of Ref.~\cite{gritsev} and the perturbative treatment of Refs.~\cite{delfino,delfinotwo}. The new result we identify here is the additional exponential decay of the oscillations.} in contrast to the repulsive regime where oscillatory behaviour only appears as sub-leading corrections. The two leading terms decay with rates $\Gamma_1$ and $\Gamma_2$. In Figure~\ref{fig:rates}.(a) we show these decay rates for Dirichlet-like initial states (\ref{extrapolationtime}), the explicit expressions for $K^{ab}_{\rm Dirichlet}(\theta)$ and $K_{B\,{\rm Dirichlet}}(\phi)$ can be found in Appendix.~\ref{aap:explicit}. For the parameters chosen there the decay of the second term is slower. Furthermore, considering the explicit expressions for the soliton-breather and breather-breather scattering matrices (\ref{Smatrix}), respectively, one notes that the relaxation rates $\tau_B$ and $\tau_{BB}$ will in general be complex valued. This results in the $\mathcal{O}(K^2)$-corrections to the bare oscillation frequency given by the breather mass, $\Delta_B$, ie, $\Omega=\Delta_B+\mathcal{O}(K^2)$. Figure~\ref{fig:rates}.(b) shows this correction for Dirichlet-like initial states. Finally we note that when considering the time evolution of the observable $\cos(\beta\Phi/2)$ the oscillatory term in (\ref{finalresult2}) will cancel out because of $f^{\beta/2}_B+f^{-\beta/2}_B=0$. Thus in that case the leading time evolution will decay with the rate $\Gamma_1$, with oscillations only appearing in the sub-leading corrections.

\section{Power Spectrum} \label{thepowerspectrum}
Having determined the time evolution (\ref{finalresult2}) we can obtain the power spectrum,
 \beq
 P_\alpha(\omega)=\lim_{T\to\infty}\left\vert\int_0^T dt \,e^{{\rm i}\omega t}\frac{\langle \Psi_t\vert e^{{\rm i}\alpha\Phi}\vert\Psi_t\rangle}{\langle\Psi_0\vert\Psi_0\rangle}\right\vert^2.\label{powerformula}
 \eeq
This function was analysed by Gritsev et al.~\cite{gritsev} for $\alpha=\beta$, for a quench in the attractive regime of the sine-Gordon model, however, only terms up to order $gK^0$ in the linked-cluster expansion were considered. Since in this order there exist terms that oscillate permanently in time, the power spectrum was found to contain sharp delta peaks located at the excitation energies of the system. In turn it was suggested that these peaks in the power spectrum could be used to measure experimentally the breather spectrum of the sine-Gordon model. 

Our main result (\ref{finalresult2}) reveals that once we resum the leading terms to all orders in $K$,  the power spectrum may be qualitatively different. At least for the value $\alpha=\beta/2$ which we consider here, we obtain a sum of Lorentz peaks located at $\omega=0$ and $\omega=\pm\Omega$ with broadenings $\Gamma_1$ and $\Gamma_2$ respectively. Thus we conclude that while in principle it is still possible to approximately determine the breather spectrum by looking at the peaks in $P_{\beta/2}(\omega)$, these peaks will now be less sharp than the delta-functions predicted in Ref.~\cite{gritsev}, and therefore more difficult to detect experimentally. In particular, from Figure~\ref{fig:rates}.(a) we see that $\Gamma_i\sim\Delta\sim\Delta_B$, thus the position of the peaks and their broadening are of the same order. Furthermore, the location of the poles is shifted by the $\mathcal{O}(K^2)$-corrections away from the breather mass $\Delta_B$, which means one can experimentally determine the breather spectrum only up to  $\mathcal{O}(K^2)$ accuracy.

\section{Comparison with semiclassical methods} \label{semiclassicalmethods}
A semiclassical calculation of expectation values of vertex operators after a quantum quench in the sine-Gordon model has been performed by Kormos and Zar\'{a}nd~\cite{semiclassical}. This approach was used to study the repulsive regime, where there are only solitons and antisolitons. The considered semiclassical limit is motivated in the small quench (ie, small $K$) limit, and with the assumption that the solitons and antisolitons possess only small momenta. The latter assumption implies that the soliton-antisoliton scattering matrix can be approximated as
\beq
S_{a_1a_2}^{b_1b_2}(\theta)\approx S_{a_1a_2}^{b_1b_2}(0)=-\delta_{a_1}^{b_2}\delta_{a_2}^{b_1}.
\eeq
Thus the scattering matrix in this limit is purely reflective, which implies that for an initial state of the form (\ref{initial}) the spatial order of solitons and antisolitons is preserved during time evolution. 

Classically solitons and antisolitons are kinks that interpolate between two adjacent minima of the cosine potential. Expectation values of operators (such as the vertex operator we consider here) can be computed by studying the configuration of the field, $\Phi$, for a given state, $\vert\Psi_t\rangle$. One only needs to consider the statistics of the classical configurations interpolating between the different vacua of any given initial configuration, $\vert \Psi_0\rangle$, which evolves deterministically after the quench. This method was used to compute the expectation value of a general vertex operator, giving the result~\cite{semiclassical}
\beq
\frac{\langle\Psi_t\vert e^{{\rm i}\alpha\Phi}\vert \Psi_t\rangle}{\langle \Psi_0\vert\Psi_0\rangle}=\mathcal{G}_\alpha\Bigl[\cos^2(\pi\alpha/\beta)+\sin^2(\pi\alpha/\beta)e^{-t/\tau}\Bigr].
\label{eq:semiclassical}
\eeq
For the special case $\alpha=\beta/2$ this agrees with the result from the quantum treatment of Ref.~\cite{repulsive}.

In this semiclassical approach, only the configuration of the kinks and the arrangement of  different classical vacua are relevant to the computation of expectation values. This immediately implies that breathers can play no role in this approach, since these excitations have zero topological charge, and do not affect the classical vacuum configuration. Thus it was concluded~\cite{semiclassical} that breathers decouple from the soliton-antisoliton dynamics and should not affect correlation functions after the quench. 

However, this observation seems to be in conflict with the oscillating terms in our main result (\ref{finalresult2}). We recall that these originate from the presence of zero-momentum breathers in the initial state (\ref{initial}), a situation that was not considered in the semiclassical analysis~\cite{semiclassical}. However, even if zero-momentum breathers were considered in the initial state, because of their charge neutrality they are expected~\cite{semiclassical} to completely decouple from the soliton-antisoliton dynamics in the semiclassical analysis. Furthermore, since they do not change the classical value of the field $\Phi$, the expectation value (\ref{eq:semiclassical}) is unaffected. Thus we conclude that the appearance of oscillating terms is beyond the semiclassical approximation. 

\section{More than one species of breather} \label{morethanone}
In this section we briefly discuss the quench dynamics in the sine-Gordon model with $\beta^2\le 1/3$. In this case more than one breather state is present in the spectrum. If we assume $N$ breathers to exist, which is the case for interaction strengths $1/(N+2)<\beta^2\le 1/(N+1)$ corresponding to $1/\xi-1<N\le 1/\xi$. The breather masses are given by
\begin{equation}
\Delta_n=2\Delta\,\sin\frac{n\pi\xi}{2},\quad n=1,\ldots,N.
\end{equation}
We denote the corresponding creation and annihilation operators for the breathers by $B_n^\dagger(\theta)$ and $B_n(\theta)$ respectively. The direct generalisation of the squeezed initial state (\ref{initial}) is given by 
\begin{equation}
\vert\Psi_0\rangle=\left(1+\sum_{n=1}^N\frac{g_n}{2}B^\dag_n(0)\right)
\exp\left[\int_0^\infty\frac{d\theta}{2\pi}K^{ab}(\theta)Z_a^\dag(-\theta)Z_b^\dag(\theta)
+\sum_{n=1}^N\int_0^\infty \frac{d\phi}{2\pi}K_{B_n}(\phi)B^\dag_n(-\phi)B^\dag_n(\phi)\right]\vert 0\rangle.
\end{equation}
Again we assume Dirichlet-like initial states satisfying $K^{++}(\theta)=K^{--}(\theta)=0$, an appropriate regularisation at large rapidities and define the functions $G(\theta)=|K^{ab}(\theta)|^2$ and $G_{B_n}(\phi)=|K_{B_n}(\phi)|^2$. 

Now the result (\ref{finalresult}) is easy to generalise to the case of many breather species by performing $\mathcal{O}(K^4,\,gK^4,\, g^2K^4)$ calculations, similar to those we present in Appendix~\ref{app:LCE}.  We find 
\beq
&&\frac{\langle \Psi_t\vert e^{{\rm i}\beta\Phi/2}\vert\Psi_t\rangle}{\langle\Psi_0\vert\Psi_0\rangle}=\left(\mathcal{G}_{\beta/2}+\sum_n\frac{\vert g_n\vert^2}{4}f_{B_nB_n}^{\beta/2}({\rm i}\pi,0)\right)\left[1-\frac{t}{\tau}+\frac{1}{2}\left(\frac{t}{\tau}\right)^2+\dots\right]\nonumber\\
&&\quad+\sum_nf_{B_n}^{\beta/2}{\rm Re}\left\{g_ne^{-\Delta_n{\rm i}t}\left[1-\frac{t}{\tau_{n}}-\sum_m\frac{t}{\tau_{nm}}+\sum_{m}\frac{t^2}{\tau_{n}\tau_{nm}}+\sum_{m\neq k}\frac{t^2}{\tau_{nm}\tau_{nk}}+\frac{1}{2}\left(\frac{t}{\tau_{n}}\right)^2+\frac{1}{2}\sum_{m}\left(\frac{t}{\tau_{nm}}\right)^2+\dots\right]\right\}\nonumber\\
&&\quad+\sum_{n\neq m}{\rm Re}\left\{\frac{g_n^*g_m}{2}f_{B_nB_m}^{\beta/2}({\rm i}\pi,0)e^{\Delta_n{\rm i}t-\Delta_m{\rm i}t}\left[1-\frac{t}{\tau_{nms}}-\sum_k\frac{t}{\tau_{nmk}}\right.\right.\nonumber\\
&&\qquad\qquad\qquad\qquad\qquad\qquad\qquad
\left.\left.+\sum_{k}\frac{t^2}{\tau_{nms}\tau_{nmk}}+\sum_{k\neq l}\frac{t^2}{\tau_{nmk}\tau_{nml}}+\frac{1}{2}\left(\frac{t}{\tau_{nms}}\right)^2+\frac{1}{2}\sum_{k}\left(\frac{t}{\tau_{nmk}}\right)^2+\dots\right]\right\},\label{manybreathers}
\eeq 
where 
\beq
\tau^{-1}&=&\frac{2\Delta}{\pi}\int_0^\infty d\theta\,G(\theta)\sinh\theta+\mathcal{O}(K^4),\\
\tau_{n}^{-1}&=&\frac{\Delta}{\pi}\int_0^\infty d\theta\,G(\theta)\left[1+S_{B_n}(\theta)\right]\sinh\theta+\mathcal{O}(K^4),\\
\tau_{nm}^{-1}&=&\frac{\Delta_n}{\pi}\int_0^\infty d\phi\,G_{B_n}(\phi)\left[1-S_{B_nB_m}(\phi)\right]\sinh\phi+\mathcal{O}(K^4),\\
\tau_{nms}^{-1}&=&\frac{\Delta}{\pi}\int_0^\infty d\theta\,G(\theta)\left[1+S_{B_n}(\theta)S_{B_m}(-\theta)\right]\sinh\theta+\mathcal{O}(K^4),\\
\tau_{nmk}^{-1}&=&\frac{\Delta_k}{\pi}\int_0^\infty d\phi\,G_{B_k}(\phi)\left[1-S_{B_nB_k}(\phi)S_{B_mB_k}(-\phi)\right]\sinh\phi+\mathcal{O}(K^4),\label{eq:tau3breather}
\eeq
with $S_{B_n}(\theta)$ and $S_{B_nB_m}(\theta)$ denoting the corresponding breather-soliton and breather-breather scattering matrices~\cite{mussardo}. We note that the three-breather rate (\ref{eq:tau3breather}) vanishes when two of them are equal, $\tau_{nnk}^{-1}=0$. Assuming, as we did in the previous sections, that the leading contributions from the higher orders in $K$ at long times can be resummed as an exponential, we propose the final expression 
\beq
\frac{\langle \Psi_t\vert e^{{\rm i}\beta\Phi/2}\vert\Psi_t\rangle}{\langle\Psi_0\vert\Psi_0\rangle}&=&\left[\mathcal{G}_{\beta/2}+\sum_n\frac{\vert g_n\vert^2}{4}f_{B_nB_n}^{\beta/2}({\rm i}\pi,0)\right]e^{-t/\tau}+\sum_nf_{B_n}^{\beta/2}{\rm Re}\left[g_ne^{-\Delta_n{\rm i}t}\,e^{-(1/\tau_{n}+\sum_m 1/\tau_{nm})t}\right]\nonumber\\
&&\qquad+\sum_{n\neq m}{\rm Re}\left[\frac{g_n^*g_n}{2}f_{B_nB_m}^{\beta/2}({\rm i}\pi,0)e^{(\Delta_n{\rm i}-\Delta_m{\rm i})t}\,e^{-(1/\tau_{nms}+\sum_k 1/\tau_{nmk})t}\right].\label{manybreatherstwo}
\eeq 
Thus we conclude that the existence of more breather states results in the appearance of several relaxation rates as well as oscillation frequencies. Furthermore, since the parameters $\tau_n$, $\tau_{nm}$, $\tau_{nms}$ and $\tau_{nmk}$ are in general complex, we also expect several $\mathcal{O}(K^2)$-corrections to the oscillation frequencies. Finally we note that while the sine-Gordon model at $\beta^2=1/4$ possesses an enlarged SU(2) symmetry, this symmetry is not reflected in our result for the time evolution since the initial state explicitly breaks this symmetry.

\section{Conclusions} \label{conclusions}
We studied the time evolution of the expectation value of the vertex operator $\exp({\rm i}\beta\Phi/2)$ after a quantum quench into the attractive regime of the sine-Gordon model, where the particle spectrum consists of solitons, antisolitons and breathers. We assumed an initial state $\vert\Psi_0\rangle$ of the squeezed state form (\ref{initial}) that corresponds to integrable boundary conditions. The subsequent time evolution was computed by assuming a ``small quench", where the initial state (\ref{initial}) can be expanded in powers of the functions $K^{ab}(\theta)$ and $K_B(\theta)$ describing the amplitudes of soliton-antisoliton and breather pairs. The terms of this series can be computed with the knowledge of the exact form factors of the vertex operator. This expansion, however, exhibits several infrared divergences that need to be regularised. We then extracted and resum the leading contributions at late times after the quench.

Our result shows that form factors containing breathers only contribute to the leading late-time dynamics, if the initial state contains zero-momentum breather states. The difference in the qualitative behaviour of solitons (antisolitons) and breathers is due to their different semi-locality properties with respect to the vertex operator considered, which implies a different structure of the annihilation poles of the form factors. In the range of sine-Gordon couplings $1/3<\beta^2<1/2$, where there is only one species of breather, our main result for the behaviour at late times is given by
\begin{equation}
\frac{\langle \Psi_t\vert e^{{\rm i}\beta\Phi/2}\vert\Psi_t\rangle}{\langle\Psi_0\vert\Psi_0\rangle}
=\left[\mathcal{G}_{\beta/2}+\frac{\vert g\vert^2}{4}f_{BB}^{\beta/2}({\rm i}\pi,0)\right]\,e^{-\Gamma_1 t}
+|g|f_B^{\beta/2}\,e^{-\Gamma_2t}\,\cos(\Omega t-\delta),
\label{finalconclusion}
\end{equation}
where $\Gamma_1=\tau^{-1}$ and $\Gamma_2={\rm Re}\left[\tau_B^{-1}+\tau_{BB}^{-1}\right]$ are the relaxation rates given in terms of the parameters (\ref{decayrates1})--(\ref{decayrates3}), $\Omega=\Delta_B+{\rm Im}\left[\tau_B^{-1}+\tau_{BB}^{-1}\right]$ is the oscillation frequency, $g$ denotes the amplitude of the zero-momentum breathers in the initial state, and $\mathcal{G}_{\beta/2}$, $f_B^{\beta/2}$ and $f_{BB}^{\beta/2}({\rm i}\pi,0)$ are the respective form factors. Besides the exponential decay of both terms in (\ref{finalconclusion}) we see that the second shows oscillations with the renormalised frequency $\Omega$ which deviates from the bare breather mass $\Delta_B$. 

Our computations can be easily generalised to other values of the sine-Gordon coupling within the attractive regime, namely for values $\beta^2<1/3$, where there are more than one species of breathers. This result is given in Eq.~(\ref{manybreatherstwo}), which shows the same qualitative behaviour of exponential decay and oscillations, with decay rates and oscillation frequencies that depend on the given species of breather.

From the time evolution we have determined the power spectrum $P_{\beta/2}(\omega)$ defined in (\ref{powerformula}), which is given by a sum of Lorentzian peaks at $\omega=0$ and $\omega=\Omega$ with broadening $\Gamma_1$ and $\Gamma_2$ respectively. This is in contrast to previous results~\cite{gritsev} for the power spectrum $P_\beta(\omega)$ of the vertex operator $\exp({\rm i}\beta\Phi)$, where no broadenings were obtained. We attribute this to either to the locality of the operator $\exp({\rm i}\beta\Phi)$ (in contrast to the semi-locality of $\exp({\rm i}\beta\Phi/2)$) or the fact that also for $P_\beta(\omega)$ a resummation of the leading long-time behaviour should be performed. 

Furthermore, we discussed our results in light of a semiclassical approach to quantum quenches developed in Ref.~\cite{semiclassical}. We saw that in particular the decay rate $\tau_B^{-1}$ cannot be obtained within this approach, since the non-trivial scattering of solitons and antisolitons off breathers is essential for its derivation in our quantum treatment. Thus the further investigation of the applicability and limitations of the semiclassical approach in the attractive sine-Gordon model seems desirable.

It should be possible, in principle, to compute the expectation values of the general vertex operator, $\exp\left({\rm i}\alpha\Phi\right)$, with the same techniques discussed in this paper. This computation would only be more tedious, as one needs to keep track of non-trivial semi-locality factors, $l_a^\alpha$, but it is not an impossible task. The $\mathcal{O}(K^2)$ contributions to this expectation value were computed in the repulsive regime in Ref.~\cite{repulsive}. The computation of higher-order terms at long times seems much more difficult at this point, and the result may not simply exponentiate, as is indicated by the semiclassical result (\ref{eq:semiclassical}) derived in Ref.~\cite{semiclassical}. 

\section*{Acknowledgement}
We would like to thank Bruno Bertini, Fabian Essler, Vladimir Gritsev and Marton Kormos for useful comments. This work is part of the D-ITP consortium, a program of the Netherlands Organisation for Scientific Research (NWO) that is funded by the Dutch Ministry of Education, Culture and Science (OCW). This work was supported by the Foundation for Fundamental Research on Matter (FOM), which is part of the Netherlands Organisation for Scientific Research (NWO), under 14PR3168. ACC acknowledges support from the European UnionÕs Horizon 2020 under the Marie Sklodowska-Curie grant agreement 750092.

\appendix
\section{Various explicit expressions}\label{aap:explicit}
In this appendix we list for completeness the explicit expressions of some of the relevant objects appearing in the main text. First, the soliton-antisoliton scattering matrix is given by~\cite{mussardo}
\begin{eqnarray}
S^{++}_{++}(\theta)&=&S^{--}_{--}(\theta)=S_0(\theta)=
-\exp\left[\mathrm{i}\int_0^\infty\frac{dt}{t}
\sin\left(\frac{t\theta}{\pi\xi}\right)\frac{\sinh\big(\frac{\xi-1}{2\xi}t\big)}
{\sinh\big(\frac{t}{2}\big)\cosh\big(\frac{t}{2\xi}\big)}\right],\nonumber\\
S^{+-}_{+-}(\theta)&=&S^{-+}_{-+}(\theta)
= S_T(\theta)S_0(\theta),\quad S_T(\theta)=-\frac{\sinh\big(\frac{\theta}{\xi}\big)}
{\sinh\big(\frac{\theta-\mathrm{i}\pi}{\xi}\big)},\nonumber\\
S^{+-}_{-+}(\theta)&=&S^{-+}_{+-}(\theta)= S_R(\theta)S_0(\theta),\quad
S_R(\theta)=-\frac{\mathrm{i}\sin\big(\frac{\pi}{\xi}\big)}
{\sinh\big(\frac{\theta-\mathrm{i}\pi}{\xi}\big)},
\end{eqnarray}
while the soliton-breather and breather-breather scattering matrices read
\begin{equation}
S_B(\theta)=\frac{\sinh\theta+{\rm i}\cos\frac{\pi\xi}{2}}{\sinh\theta-{\rm i}\cos\frac{\pi\xi}{2}},\quad
S_{BB}(\theta)=\frac{\sinh\theta+{\rm i}\sin(\pi\xi)}{\sinh\theta-{\rm i}\sin(\pi\xi)}.
\label{Smatrix}
\end{equation}

Next we state explicit expressions for the K-matrices provided we assume the initial state to correspond to an integrable boundary state with Dirichlet boundary conditions $\Phi(t=0,x)=\Phi_0=0$. We find
\begin{equation}
K^{ab}_\mathrm{Dirichlet}(\theta)=R^b_{\bar{a}}\left(\frac{\mathrm{i}\pi}{2}-\theta\right),\quad
K_{B\,\mathrm{Dirichlet}}(\theta)=R_B\left(\frac{\mathrm{i}\pi}{2}-\theta\right)
\end{equation}
for the soliton-antisoliton and breather matrices. Explicit expressions for the reflection matrices are given by~\cite{gloshal,gloshaltwo}
\begin{eqnarray}
R_\pm^\pm(\theta)&=&\cosh\left(\frac{\theta}{\xi}\right)\,R_0(\theta)\,\sigma(\alpha=0,\theta),\\
R_0(\theta)&=&\frac{\Gamma(1+\frac{\mathrm{i}2\theta}{\pi\xi})}{\Gamma(1-\frac{\mathrm{i}2\theta}{\pi\xi})}
\frac{\Gamma(\frac{1}{\xi}-\frac{\mathrm{i}2\theta}{\pi\xi})}{\Gamma(\frac{1}{\xi}+\frac{\mathrm{i}2\theta}{\pi\xi})}
\prod_{k=1}^\infty\frac{\Gamma(\frac{4k}{\xi}+\frac{\mathrm{i}2\theta}{\pi\xi})}{\Gamma(\frac{4k}{\xi}-\frac{\mathrm{i}2\theta}{\pi\xi})}\frac{\Gamma(\frac{4k+1}{\xi}-\frac{\mathrm{i}2\theta}{\pi\xi})}{\Gamma(\frac{4k+1}{\xi}+\frac{\mathrm{i}2\theta}{\pi\xi})}\frac{\Gamma(1+\frac{4k}{\xi}+\frac{\mathrm{i}2\theta}{\pi\xi})}{\Gamma(1+\frac{4k}{\xi}-\frac{\mathrm{i}2\theta}{\pi\xi})}\frac{\Gamma(1+\frac{4k-1}{\xi}-\frac{\mathrm{i}2\theta}{\pi\xi})}{\Gamma(1+\frac{4k-1}{\xi}+\frac{\mathrm{i}2\theta}{\pi\xi})},\\
\sigma(\alpha=0,\theta)&=&\left[\prod_{k=0}^\infty\frac{\Gamma(\frac{1}{2}+\frac{2k}{\xi}-\frac{\mathrm{i}\theta}{\pi\xi})}{\Gamma(\frac{1}{2}+\frac{2k+1}{\xi}-\frac{\mathrm{i}\theta}{\pi\xi})}\frac{\Gamma(\frac{1}{2}+\frac{2k+1}{\xi}+\frac{\mathrm{i}\theta}{\pi\xi})}{\Gamma(\frac{1}{2}+\frac{2k+2}{\xi}+\frac{\mathrm{i}\theta}{\pi\xi})}\frac{\Gamma(\frac{1}{2}+\frac{2k+2}{\xi})}{\Gamma(\frac{1}{2}+\frac{2k}{\xi})}\right]^2,\\
R_B(\theta)&=&-\frac{1+\mathrm{i}\sinh\theta}{1-\mathrm{i}\sinh\theta}\frac{\cos\left(\frac{\pi\xi}{4}-\frac{\mathrm{i}\theta}{2}\right)\cos\left(\frac{\pi\xi}{4}+\frac{\pi}{4}+\frac{\mathrm{i}\theta}{2}\right)\sin\left(\frac{\pi\xi}{4}-\frac{\mathrm{i}\theta}{2}\right)}{\cos\left(\frac{\pi\xi}{4}+\frac{\mathrm{i}\theta}{2}\right)
\cos\left(\frac{\pi\xi}{4}+\frac{\pi}{4}-\frac{\mathrm{i}\theta}{2}\right)\sin\left(\frac{\pi\xi}{4}+\frac{\mathrm{i}\theta}{2}\right)}.
\end{eqnarray}
Integral representations for the above expressions can be found in Ref.~\cite{Caux-03}.

The constants (\ref{FFconstants}) originating from the form factors are explicitly given by~\cite{Lukyanov97}
\begin{eqnarray}
\mathcal{G}_{\beta/2}&=&\langle 0\vert e^{\pm{\rm i}\beta\Phi/2}\vert 0\rangle\nonumber\\
&=&\left(\frac{\Delta\sqrt{\pi}\Gamma\left(\frac{1}{2-2\beta^2}\right)}{2\Gamma\left(\frac{\beta^2}{2-2\beta^2}\right)}\right)^{\beta^2/2}\,\exp\left(\int_0^\infty\frac{dt}{t}\left[\frac{\sinh(\beta^2 t)}{2\sinh t\,\cosh[(1-\beta^2)t]}-\frac{\beta^2}{2}e^{-2t}\right]\right),\\
f_B^{\pm\beta/2}&=&\langle 0\vert e^{\pm{\rm i}\beta\Phi/2}\vert 0_B\rangle=\pm\mathrm{i}\mathcal{G}_{\beta/2}\lambda\frac{\sin\frac{\pi\xi}{2}}{\sin(\pi\xi)},\\
f_{BB}^{\pm\beta/2}({\rm i}\pi,0)&=&\langle 0\vert e^{\pm{\rm i}\beta\Phi/2}B^\dagger({\rm i}\pi)B^\dagger(0)\vert 0\rangle=-\mathcal{G}_{\beta/2}\mathcal{N}\lambda^2\frac{\sin^2\frac{\pi\xi}{2}}{\sin^2(\pi\xi)},\\
\end{eqnarray}
where 
\begin{eqnarray}
\lambda&=&2\cos\frac{\pi\xi}{2}\,\sqrt{2\sin\frac{\pi\xi}{2}}\,\exp\left(-\int_0^{\pi\xi}\frac{dt}{2\pi}\frac{t}{\sin t}\right),\\
\mathcal{N}&=&\exp\left(4\int_0^\infty\frac{dt}{t}\frac{\sinh t\,\sinh(t\xi)\,\sinh[t(1+\xi)]}{\sinh^2(2t)}\right).
\end{eqnarray}
In particular we find $f^{\beta/2}_B+f^{-\beta/2}_B=0$. 

\section{Form-factor axioms}\label{app:FFA}
In this appendix, we give a brief overview of the form-factor axioms, see Refs.~\cite{bootstrap,mussardo} for a more detailed discussion. For compactness, we introduce the particle creation operators
$A_{a}^\dag(\theta)$ with an index that can take the values $a=\pm, B$. For $a=\pm$, we define $A_{a}^\dag(\theta)=Z^\dag_{a}(\theta)$, and for $a=B$, we define $A_B^\dag(\theta)=B^\dag(\theta)$. We also define the generalised scattering matrix, $\mathcal{S}_{ab}^{cd}(\theta)$, such that 
\beq
A_{a_1}^\dag(\theta_1)A_{a_2}^\dag(\theta_2)=\mathcal{S}_{a_1a_2}^{b_1b_2}(\theta_1-\theta_2)A_{b_2}^\dag(\theta_2)A_{b_2}^\dag(\theta_1).
\eeq

We can now define the $n$-particle form factor of some operator, $\mathcal{O}$, as
\beq
f_{a_1\dots a_n}^{\mathcal{O}}(\theta_1,\dots,\theta_n)=\langle 0\vert \mathcal{O}\vert\theta_1\dots,\theta_n\rangle_{a_1\dots a_n}=\langle 0\vert \mathcal{O}A_{a_1}^\dag(\theta_1)\dots A_{a_n}^\dag(\theta_n)\vert 0\rangle.\label{ffdef}
\eeq
These form factors satisfy the following axioms:
\begin{enumerate}
\item  The functions $f_{a_1\dots a_n}^{\mathcal{O}}(\theta_1,\dots,\theta_n)$ are meromorphic functions in the interval $0<\mathrm{Im}\,\theta_i<2\pi$ for all $i=1,\ldots,n$. 
There exist only simple poles in this so-called ``physical strip", which correspond to annihilation and bound-state poles, as described below.
\item Scattering axiom:
\beq
&&f_{a_1\dots,a_i,a_{i+1},\dots,a_n}^{\mathcal{O}}(\theta_1,\dots,\theta_i,\theta_{i+1},\dots,\theta_n)\nonumber\\
&&\,\,\,\,\,\,\,\,\,\,=\mathcal{S}_{a_ia_{i+1}}^{b_ib_{i+1}}(\theta_i-\theta_{i+1})f_{a_1,\dots,b_{i+1},b_{i},\dots,a_n}^{\mathcal{O}}(\theta_1,\dots,\theta_{i+1},\theta_i,\dots,\theta_n).
\eeq
\item Periodicity axiom:
\beq
&&f_{a_1\dots a_n}(\theta_1+2\pi{\rm i},\theta_2,\dots,\theta_n)=l_{a_1}(\mathcal{O})f_{a_2\dots a_na_1}^{\mathcal{O}}(\theta_2,\dots,\theta_n,\theta_1),
\eeq
where $l_{a}(\mathcal{O})$ is the mutual semi-locality factor between the operator $\mathcal{O}$ and the fundamental fields associated with the particle created by $A_a^\dag(\theta)$. In our particular example, for the operator $\mathcal{O}=\exp\left({\rm i}\alpha\Phi\right)$, this factor is $l_{\pm}^\alpha=e^{\pm {\rm i}2\pi\alpha/\beta}$, $l_{B}^\alpha=1$. For $\alpha=\beta/2$, we obtain the particularly simple value $l_{\pm}^{\beta/2}=-1$, which is the technical reason we evaluate only this  vertex operator.
\item Lorentz transformations:
\beq
f_{a_1\dots a_n}^{\mathcal{O}}(\theta_1+\Lambda,\dots,\theta_n+\Lambda)=e^{s(\mathcal{O})\Lambda}f_{a_1\dots a_N}(\theta_1,\dots,\theta_n),
\eeq
where $s(\mathcal{O})$ is the Lorentz spin of the operator $\mathcal{O}$. For the scalar operator $\mathcal{O}=\exp({\rm i}\alpha\Phi)$, we have $s(\mathcal{O})=0$.
\item Annihilation pole axiom:
\beq
&&{\rm Res}\left[f_{aba_1\dots a_n}^{\mathcal{O}}(\theta^\prime,\theta,\theta_1,\dots,\theta_n),\theta^\prime=\theta+{\rm i}\pi\right]\nonumber\\
&&\,\,\,\,\,={\rm i}C_{ac}f_{b_1\dots b_n}^\mathcal{O}(\theta_1,\dots,\theta_n)\left[\delta_{a_1}^{b_1}\dots\delta_{a_n}^{b_n}\delta_{b}^{c}-l_a(\mathcal{O})\mathcal{S}_{ba_1}^{c_1b_1}(\theta-\theta_1)\mathcal{S}_{c_1a_2}^{c_2b_2}(\theta-\theta_2)\dots \mathcal{S}_{c_{n-1}a_n}^{cb_n}(\theta-\theta_n)\right],\label{annihilationpoleaxiom}
\eeq
where we introduced the charge conjugation matrix, given in the sine-Gordon model by $C_{ab}=\delta_{a\bar{b}}$ (note that breathers are their own antiparticles, so $\bar{B}=B$).
\item Bound state pole axiom: 

Suppose the particles created by $A_{a}^\dag(\theta)$ and $A_{b}^\dag(\theta^\prime)$ can form a physical bound state, $A_{c}^\dag(\theta^{\prime\prime})$, with mass given by
\beq
\Delta_c^2=\Delta_a^2+\Delta_b^2+2\Delta_a\Delta_b\cos u_{ab}^c.
\eeq
This means the scattering matrix must have a simple pole such that
\begin{equation}
\mathcal{S}_{ab}^{a'b'}(\theta)\sim\frac{\mathrm{i}\,\Gamma_{ab}^c\Gamma_c^{a'b'}}{\theta-\mathrm{i}u_{ab}^c},
\end{equation}
where the $\Gamma_{ab}^c$ and $\Gamma_c^{a'b'}$ are the corresponding couplings. For example, the first breather state formed as a bound state of one soliton and one antisoliton is obtained with $u_{+-}^B=u_{-+}^B=\pi(1-\xi)$. In the presence of bound states, the form factors also have simple poles, whose residues are given by
\beq
{\rm Res}\left[f_{aba_1\dots a_n}^{\mathcal{O}}(\theta',\theta,\theta_1,\dots,\theta_n),\theta'=\theta+\mathrm{i}u_{ab}^c\right]={\rm i}\,\Gamma_{ab}^cf_{ca_1\dots a_n}^{\mathcal{O}}(\theta^{\prime\prime},\theta_1,\dots,\theta_n),
\eeq
where $\theta^{\prime\prime}=(\bar{u}_{bc}^a\theta'+\bar{u}_{ca}^b\theta)/u_{ab}^c$ with $\bar{u}_{ab}^c=\pi-u_{ab}^c$.
\end{enumerate}

\section{Terms of the linked-cluster expansion}\label{app:LCE}
In this appendix we compute the terms up to order $K^4$, $gK^4$, and $g^2K^2$ of the linked-cluster expansion. We focus on the leading contributions at late times. We disregard terms which decay faster and thus lead to sub-leading corrections.

\subsection{Order $\boldsymbol{K^0}$}
The only contribution to this order is
\beq
D_{0,0;0,0}=C_{0,0;0,0}=\mathcal{G}_{\beta/2},
\eeq
ie, the vacuum expectation value of the vertex operator.

\subsection{Order $\boldsymbol{K^1}$ and $\boldsymbol{gK^0}$}
The contributions to this order are $D_{2,0;0,0}$, $D_{0,2;0,0}$, $D_{0,0+1;0,0}$, $D_{0,0;2,0}$, $D_{0,0;0,2}$ and $D_{0,0;0,0+1}$, which are given explicitly by
\beq
D_{2,0;0,0}&=&C_{2,0;0,0}=\int_0^\infty\frac{d\theta}{2\pi}\left(K^{ab}(\theta)\right)^* f_{ab}^{-\beta/2}(-\theta,\theta)^*\,e^{2\Delta{\rm  i}t \cosh\theta},\\
&&\nonumber\\
D_{0,2;0,0}&=&C_{0,2;0,0}=\int_0^\infty\frac{d\phi}{2\pi}\left(K_B(\phi)\right)^*f_{BB}^{-\beta/2}(-\phi,\phi)^*\,e^{2\Delta_B{\rm i}t\cosh \theta},\\
&&\nonumber\\
D_{0,0+1;0,0}&=&C_{0,0+1;0,0}=\frac{g^*}{2}f^{\beta/2}_B\,e^{\Delta_B {\rm i}t},\label{kone}
\eeq
where we introduced the notation for the form factors
\begin{equation}
f_{ab}^{\alpha}(\theta_1,\theta_2)=\langle 0\vert e^{{\rm i}\alpha\Phi}\vert \theta_1,\theta_2\rangle_{ab},\quad
f_{BB}^{\alpha}(\phi_1,\phi_2)=\langle 0\vert e^{{\rm i}\alpha\Phi}\vert\phi_1,\phi_2\rangle,\quad
f^\alpha_B=\langle 0\vert e^{{\rm i}\alpha\Phi}\vert \phi\rangle,
\end{equation}
The remaining terms are $D_{0,0;2,0}=D_{2,0;0,0}^*$ and $D_{0,0;0,2}=D_{0,2;0,0}^*$ upon replacing $\beta\to-\beta$ as well as $D_{0,0;0,0+1}=gf^{\beta/2}_B\,e^{-\Delta_B\mathrm{i}t}/2$.

As can easily be seen from a stationary-phase approximation, the contributions which contain two particles (either a soliton-antisoliton pair or two breathers), are suppressed at long times. On the other hand, the one-breather contributions do not decay with time, but instead continues to oscillate. This is a new phenomenon which does not occur in the repulsive regime. Hence the leading contribution in this order at late times is 
\beq
D_{0,0+1;0,0}+D_{0,0;0,0+1}=f^{\beta/2}_B \,{\rm Re}[g e^{-\Delta_B{\rm i}t}].
\eeq

\subsection{Order $\boldsymbol{K^2}$, $\boldsymbol{gK^1}$ and $\boldsymbol{g^2 K^0}$}
In this order there are several contributions. We focus on the ones leading at late times, which are given by
\beq
D_{2,0;2,0}&=&C_{2,0;2,0}-Z_{2,0}C_{0,0;0,0},\\
D_{0,2;0,2}&=&C_{0,2;0,2}-Z_{0,2}C_{0,0;0,0},\\
D_{0,0+1;0,0+1}&=&C_{0,0+1;0,0+1}-Z_{0,0+1}C_{0,0;0,0}.\label{linkedclustersktwo}
\eeq
We first examine the contributions to the denominator of (\ref{expectation}) at this order, namely $Z_{2,0}$, $Z_{0,2}$ and $Z_{0,0+1}$. These terms contain infinite-volume divergences that have to be regularised as discussed in Appendix~\ref{app:regularisation}. After regularisation, these terms read
\beq
Z_{2,0}&=&\int d\kappa P(\kappa)\int_0^\infty\frac{d\theta^\prime d\theta}{(2\pi)^2}\left(K^{ab}(\theta^\prime)\right)^* K^{cd}(\theta)\, \,_{ba}\langle \theta^\prime,-\theta^\prime\vert -\theta+\kappa,\theta+\kappa\rangle_{cd}\nonumber\\*
&=&\int d\kappa P(\kappa)\delta(-2\kappa)\int_0^\infty d\theta\left(K^{ab}(\theta+\kappa)\right)^*K^{ab}(\theta)=\frac{L}{2}\int_0^\infty d\theta\, G(\theta),\\
Z_{0,2}&=&\int d\kappa P(\kappa) \int_0^\infty \frac{d\phi^\prime d \phi}{(2\pi)^2}\left(K_B(\phi^\prime)\right)^* K_B(\phi)\,\langle \phi^\prime,-\phi^\prime\vert -\phi+\kappa,\phi+\kappa\rangle,\nonumber\\*
&=&\int d\kappa P(\kappa)\delta(-2\kappa)\int_0^\infty \left(K_B(\phi+\kappa)\right)^*K_B(\phi)=\frac{L}{2}\int_0^\infty d\phi\, G_B(\phi),\\
Z_{0,0+1}&=&\int d\kappa P(\kappa) \frac{\vert g\vert^2}{4} \langle 0_B\vert 0_B+\kappa\rangle=L\,\frac{\pi\vert g\vert^2}{2},
\eeq
where in the last line we have used $\langle 0_B\vert 0_B+\kappa\rangle=\langle 0\vert B(0)B^\dagger(\kappa)\vert 0\rangle=2\pi\delta(\kappa)$. 

\subsubsection{The term $\boldsymbol{D_{2,0;2,0}}$}
We first consider the contribution $D_{2,0;2,0}$. This has been computed in Ref.~\cite{repulsive} for the repulsive regime; we repeat that calculation here and comment on the changes in the attractive regime. We start by considering
\beq
C_{2,0;2,0}=\int_0^\infty \frac{d\theta^\prime d\theta}{(2\pi)^2}\left(K^{ab}(\theta^\prime)\right)^*K^{cd}(\theta)\,_{ba}\langle \theta^\prime,-\theta^\prime\vert e^{{\rm i} \beta\Phi/2}\vert -\theta,\theta\rangle_{cd}\,e^{2\Delta {\rm i} t(\cosh\theta^\prime-\cosh\theta)}.\label{ctwosolitons}
\eeq
We can write explicitly the connected and disconnected pieces of the form factor in (\ref{ctwosolitons}), and regularise using the $\kappa$ parameter such that
\beq
\,_{ba}\langle \theta^\prime,-\theta^\prime\vert e^{{\rm i}\beta\Phi/2}\vert -\theta,\theta\rangle_{cd}&=&
\,_{ba}\langle \theta^\prime,-\theta^\prime\vert e^{{\rm i}\beta\Phi/2}\vert -\theta+\kappa,\theta+\kappa\rangle_{cd}\nonumber\\
&=&(2\pi)^2\mathcal{G}_{\beta/2} \delta_{a}^c\delta_b^d\delta(-2\kappa)\delta(\theta^\prime-\theta+\kappa)\nonumber\\
&&+2\pi\, S_{ba}^{ef}(2\theta-2\kappa)S_{cd}^{fh}(-2\theta)\delta(\theta^\prime-\theta+\kappa)\,_e\langle \theta^\prime+{\rm i} 0\vert e^{{\rm i} \beta\Phi/2}\vert \theta+\kappa\rangle_h\nonumber\\
&&+2\pi \,\delta_b^d\delta(\theta^\prime-\theta-\kappa)\,_a\langle-\theta^\prime+{\rm i}0\vert e^{{\rm i} \beta\Phi/2}\vert -\theta+\kappa\rangle_c\nonumber\\
&&+\,_{ba}\langle \theta^\prime+{\rm i}0,-\theta^\prime+{\rm i}0\vert e^{{\rm i}\beta\Phi/2}\vert-\theta+\kappa, \theta+\kappa\rangle_{cd}.\label{foursolitonform}
\eeq
We can now insert (\ref{foursolitonform}) into (\ref{ctwosolitons}) to obtain three contributions which can be described as disconnected, semi-connected, and fully connected, respectively,
\beq
C_{2,0;2,0}=C_{2,0;2,0}^0+C_{2,0;2,0}^1+C_{2,0;2,0}^2.
\eeq

The disconnected term can be easily computed with the result
\beq
C_{2,0;2,0}^0&=&\int_0^\infty \frac{d\theta^\prime d\theta}{(2\pi)^2}\left(K^{ab}(\theta^\prime)\right)^*K^{cd}(\theta)(2\pi)^2\mathcal{G}_{\beta/2} \delta_{a}^c\delta_b^d\delta(-2\kappa)\delta(\theta^\prime-\theta+\kappa)\,e^{2\Delta {\rm i} t(\cosh\theta^\prime-\cosh\theta)}\nonumber\\
&=&\mathcal{G}_{\beta/2}\delta(-2\kappa)\int_0^\infty d\theta\,\vert K^{ab}(\theta)\vert^2=\frac{L}{2}\mathcal{G}_{\beta/2}\int_0^\infty d\theta G(\theta)=Z_{2,0} \mathcal{G}_{\beta/2},
\eeq
which is identical to $Z_{2,0}C_{0,0;0,0}$, therefore the contributions from these two terms cancel.

We now consider the semi-connected contribution
\beq
C_{2,0;2,0}^1&=&\int_0^\infty\frac{d\theta}{2\pi}\left(K^{ab}(\theta)\right)^*K^{cd}(\theta+\kappa)S_{ba}^{ef}(2\theta)S_{cd}^{fh}(-2\theta-2\kappa)\,_e\langle\theta+{\rm i}0\vert e^{{\rm i}\beta\Phi/2}\vert \theta+2\kappa\rangle_h\,e^{2\Delta {\rm i}t\left[\cosh\theta-\cosh(\theta+\kappa)\right]}\nonumber\\
&&+\int_0^\infty\frac{d\theta}{2\pi}\left(K^{ab}(\theta)\right)^*K^{cb}(\theta-\kappa)\,_a\langle-\theta+{\rm i}0\vert e^{i\beta\Phi/2}\vert -\theta+2\kappa\rangle_c\,e^{2\Delta{\rm i}t\left[\cosh\theta-\cosh(\theta-\kappa)\right]}\nonumber\\
&=&-f_{\bar{e}h}^{\beta/2}({\rm i}\pi +{\rm i}0,2\kappa)\int_0^\infty\frac{d\theta}{2\pi}\left(K^{ab}(\theta)\right)^*K^{cd}(\theta+\kappa)S_{ba}^{ef}(2\theta)S_{cd}^{fh}(-2\theta-2\kappa)\,e^{2\Delta{\rm i}t\left[\cosh\theta-\cosh(\theta+\kappa)\right]}\nonumber\\
&&-f_{\bar{a}c}^{\beta/2}({\rm i}\pi+{\rm i}0,2\kappa)\int_0^{\infty}\frac{d\theta}{2\pi}\left(K^{ab}(\theta)\right)^*K^{cb}(\theta-\kappa)\,e^{2\Delta{\rm i}t\left[\cosh\theta-\cosh(\theta-\kappa)\right]}.
\eeq
From the annihilation pole axiom, we know
\beq
f_{ab}^{\beta/2}({\rm i}\pi+{\rm i}0,2\kappa)=-2{\rm i}\frac{C_{ab}\,\mathcal{G}_{\beta/2}}{2\kappa-{\rm i}0}-F_{ba}^{\beta/2}(\kappa),
\eeq
where $F_{ba}^{\beta/2}(\kappa)$ is analytic for $\kappa\to0$. We can now expand $C_{2,0;2,0}^1$ for small $\kappa$, and discard any terms that go to zero as $\kappa\to0$,
\beq
C_{2,0;2,0}^1&=&\frac{2{\rm i}\mathcal{G}_{\beta/2}}{2\kappa-{\rm i}0}\int_0^\infty \frac{d\theta}{2\pi}\left(K^{ab}(\theta)\right)^*K^{cd}(\theta)S_{ab}^{ef}(2\theta)S_{ef}^{cd}(-2\theta)\left[1-2\Delta{\rm i}t\kappa\,\sinh\theta\right]\nonumber\\
&&+\frac{2{\rm i}\mathcal{G}_{\beta/2}}{2\kappa-{\rm i}0}\int_0^\infty \frac{d\theta}{2\pi}\vert K^{ab}(\theta)\vert^2\left[1+2\Delta{\rm i}t\kappa\sinh\theta\right]+\dots\label{ctwotwoone}
\eeq
This term is divergent, as it contains $1/\kappa$ contributions. These divergences will completely cancel with similar terms from the fully connected term.

So far everything has been identical to the analog calculation in the repulsive regime. We now consider the fully connected term, 
\beq
C_{2,0;2,0}^2=\int_0^\infty\frac{d\theta^\prime d\theta}{(2\pi)^2}\left(K^{ab}(\theta^\prime)\right)^*K^{cd}(\theta) f_{\bar{b }\bar{a}cd}^{\beta/2}(\theta^\prime+{\rm i}\pi +{\rm i}0,-\theta^\prime+{\rm i}\pi+{\rm i}0,-\theta+\kappa,\theta+\kappa)e^{2\Delta {\rm i}t(\cosh\theta^\prime-\cosh\theta)}.
\eeq
The form factor has annihilation poles at $\theta=\pm\theta^\prime-\kappa+{\rm i}0$ and $\theta=\mp\theta^\prime+\kappa-{\rm i}0$ as well as bound-state poles at $\theta=-\mathrm{i}\pi(1-\xi)/2$, while the matrix $K^{cd}(\theta)$ has poles at $\theta=\pm\mathrm{i}\pi(1-\xi)$. The important difference fact here is that the bound-state poles have a finite imaginary part, while the annihilation poles lie close to the real axis. Thus when shifting the contour of integration over $\theta$ to the lower half plane to, say, $\theta\to\theta-\mathrm{i}\pi(1-\xi)/4$ we will pick up contributions from the annihilation poles only. This results in the decomposition of the term $C_{2,0;2,0}$ into a contribution coming from the annihilation poles of the form factor denoted by $C_{2,0;2,0}^p$ and a finite remainder $C_{2,0;2,0}^\prime$, ie, 
\beq
C_{2,0;2,0}^2=C_{2,0;2,0}^\prime+C_{2,0;2,0}^p.
\eeq
First, the finite contribution is given by the integral over the shifted contour,
\beq
C_{2,0;2,0}^\prime=\int_0^\infty \frac{d\theta^\prime}{2\pi}\int_{\gamma_-}\frac{d\theta}{2\pi}\left(K^{ab}(\theta')\right)^*K^{cd}(\theta)f_{\bar{b}\bar{a}cd}^{\beta/2}(\theta^\prime+{\rm i}\pi+{\rm i}0,-\theta^\prime+{\rm i}\pi+{\rm i}0,-\theta+\kappa,\theta+\kappa)e^{2\Delta{\rm i}t(\cosh\theta^\prime-\cosh\theta)},
\eeq
where the shifted contour of integration $\gamma_-$ can be written as
\beq
\gamma_-(s)=\left\{\begin{array}{cc}
-{\rm i}s,&0\leq s\leq \phi_0,\\
(s-\phi_0)-{\rm i}\phi_0,&\phi_0\leq s< \infty,\end{array}\right.
\eeq
for some fixed $\phi_0$ in the interval $0<\phi_0<\pi(1-\xi)/2$. This contribution is sub-leading at long times; we will thus not analyse it further. Second, the contribution originating from the annihilation poles of the form factor reads
\beq
C_{2,0;2,0}^p&=&-{\rm i}\int_0^\infty\frac{d\theta}{2\pi}\left(K^{ab}(\theta^\prime)\right)^*K^{cd}(\theta^\prime+\kappa)e^{2\Delta{\rm i}t\left[\cosh\theta^\prime-\cosh(\theta^\prime+\kappa)\right]}\nonumber\\
&&\times {\rm Res} \left[ f_{\bar{b} \bar{a}cd}^{\beta/2}(\theta^\prime+{\rm i}\pi+{\rm i}0,-\theta^\prime+{\rm i}\pi+{\rm i}0,-\theta+\kappa,\theta+\kappa),\,\theta=\theta^\prime+\kappa-{\rm i}0\right],
\eeq
which by evaluating the residue using $\mathrm{Res}[f(z),z=z_0]=-\textrm{Res}[f(-z),z=-z_0]$ gives
\beq
C_{2,0;2,0}^p&=&C_{gk}f_{ij}^{\beta/2}({\rm i}\pi+{\rm i}0,2\kappa)\int_0^\infty\frac{d\theta}{2\pi}\left(K^{ab}(\theta)\right)^*K^{cd}(\theta+\kappa)S_{\bar{b}\bar{a}}^{ef}(2\theta)S_{cd}^{gh}(-2\theta-2\kappa)\nonumber\\
&&\times\left[\delta_e^i\delta_h^j\delta_f^k+S_{fe}^{li}(-2\theta)S_{lh}^{kj}(-2\theta-2\kappa+{\rm i}\pi)\right]e^{2\Delta{\rm i}t\left[\cosh\theta-\cosh(\theta+\kappa)\right]}.
\eeq
Finally expanding this for small $\kappa$ we find
\beq
C_{2,0;2,0}^p&=&-{\rm i}\frac{2\mathcal{G}_{\beta/2}}{2\kappa-{\rm i}0}\int_0^\infty\frac{d\theta}{2\pi}\left(K^{ab}(\theta))\right)^*K^{cd}(\theta)S_{ab}^{ef}(2\theta)S_{ef}^{cd}(-2\theta)\left[1-2\Delta{\rm i}t\kappa\sinh\theta\right]\nonumber\\
&&-{\rm i}\frac{2\mathcal{G}_{\beta/2}}{2\kappa-{\rm i}0}\int_0^\infty\frac{d\theta}{2\pi}\vert K^{ab}(\theta)\vert^2\left[1-2\Delta {\rm i}t\kappa\sinh\theta\right]+\dots\,.\label{ctwotwosolitonp}
\eeq

We can now add together all the contributions to $D_{2,0;2,0}$. It is easy to see that the divergent terms from $C_{2,0;2,0}^1$ and $C_{2,0;2,0}^p$ exactly cancel each other. Thus, at late times the leading contribution is
\beq
D_{2,0;2,0}=-\frac{2\kappa}{2\kappa-{\rm i}0}4\mathcal{G}_{\beta/2}\Delta t\int_0^\infty\frac{d\theta}{2\pi}G(\theta)\sinh\theta.
\eeq
Multiplying by $P(\kappa)$ and integrating over $\kappa$, we find the result for large $t$,
\beq
D_{2,0;2,0}=-\mathcal{G}_{\beta/2}\frac{t}{\tau},\quad
\tau^{-1}=\frac{2\Delta}{\pi}\int_0^\infty d\theta\,G(\theta)\sinh\theta+\mathcal{O}(K^4).
\eeq

\subsubsection{The term $\boldsymbol{D_{0,2;0,2}}$}
The contribution $D_{0,2;0,2}$ looks very similar to the contribution $D_{2,0;2,0}$ we have just discussed. We will see, however, that this term does not give a leading contribution to the expectation values at long times. This is due to the fact that, unlike solitons, the breathers are local particles with respect to the vertex operator. This implies that the two-breather form factor does not have an annihilation pole. Using the annihilation pole axiom, we see that the contribution from a four-breather form factor is therefore less divergent than the contribution from a four soliton/antisoliton form factor, and thus gives a sub-leading contribution to the expectation value at long times.

To make this line of argument more explicit we consider 
\beq
C_{0,2;0,2}=\int_0^\infty\frac{d\phi^\prime d\phi}{(2\pi)^2}\left(K_B(\phi^\prime)\right)^*K_B(\phi)\,\langle \phi^\prime,-\phi^\prime\vert e^{{\rm i}\beta\Phi/2}\vert -\phi,\phi\rangle\,e^{2\Delta_B {\rm i}t(\cosh\phi^\prime-\cosh\phi)}.
\eeq
We can write the four-breather form factor as
\beq
\langle \phi^\prime,-\phi^\prime\vert e^{{\rm i}\beta\Phi/2}\vert -\phi,\phi\rangle&=&\langle \phi^\prime,-\phi^\prime\vert e^{{\rm i}\beta\Phi/2}\vert -\phi+\kappa,\phi+\kappa\rangle\nonumber\\
&=&(2\pi)^2\mathcal{G}_{\beta/2}\delta(-2\kappa)\delta(\phi^\prime-\phi+\kappa)\nonumber\\
&&+2\pi S_{BB}(2\phi-2\kappa)S_{BB}(-2\phi)\delta(\phi^\prime-\phi+\kappa)\,\langle \phi^\prime+{\rm i}0\vert e^{{\rm i}\beta\Phi/2}\vert \phi+\kappa\rangle\nonumber\\
&&+2\pi\delta(\phi^\prime-\phi-\kappa)\,\langle -\phi^\prime+{\rm i}0\vert e^{{\rm i}\beta\Phi/2}\vert -\phi+\kappa\rangle\nonumber\\
&&+\langle \phi^\prime+{\rm i}0,-\phi^\prime+{\rm i}0\vert e^{{\rm i}\beta\Phi/2}\vert -\phi+\kappa,\phi+\kappa\rangle.
\eeq
Using this, we can again separate $C_{0,2;0,2}$ into disconnected, semi-connected and fully connected pieces
\beq
C_{0,2;0,2}=C_{0,2;0,2}^0+C_{0,2;0,2}^1+C_{0,2;0,2}^2.
\eeq

The disconnected contribution is
\beq
C_{0,2;0,2}^0=\mathcal{G}_{\beta/2}\delta(-2\kappa)\int_0^\infty d\phi \vert K_B(\phi)\vert^2=\frac{L}{2}\mathcal{G}_{\beta/2}\int_0^\infty d\phi \,G_B(\phi),
\eeq
after multiplying by $P(\kappa)$ and integrating over $\kappa$. This contribution exactly cancels out with $Z_{0,2}\,C_{0,0;0,0}$.

We consider now the semi-connected contribution
\beq
C_{0,2;0,2}^1&=&\int_0^\infty\frac{d\phi}{2\pi}\left(K_B(\phi)\right)^*K_B(\phi+\kappa)S_B(2\phi)S_B(-2\phi-2\kappa)\,\langle \phi+{\rm i}0\vert e^{{\rm i}\beta\Phi/2}\vert \phi+2\kappa\rangle e^{2\Delta_B{\rm i}t\left[\cosh\phi-\cosh(\phi+\kappa)\right]}\nonumber\\
&&+\int_0^\infty \frac{d\phi}{2\pi}\left(K_B(\phi)\right)^*K_B(\phi-\kappa)\,\langle -\phi+{\rm i}0\vert e^{{\rm i}\beta\Phi/2}\vert -\phi+2\kappa\rangle e^{2\Delta_B {\rm i}t\left[\cosh\phi-\cosh(\theta-\kappa)\right]}\nonumber\\
&=&f_{BB}^{\beta/2}({\rm i}\pi+{\rm i}0,2\kappa)\int_0^\infty \frac{d\theta}{2\pi}\left(K_B(\phi)\right)^*K_B(\phi+\kappa)S_{BB}(2\phi)S_{BB}(-2\phi-2\kappa)\,e^{2\Delta_B{\rm i}t\left[\cosh\theta-\cosh(\theta+\kappa)\right]}\nonumber\\
&&+f_{BB}^{\beta/2}({\rm i}\pi+{\rm i}0,2\kappa)\int_0^\infty\frac{d\theta}{2\pi}\left(K_B(\phi)\right)^*K(\phi-\kappa)\,e^{2\Delta_B{\rm i}t\left[\cosh\phi-\cosh(\phi-\kappa)\right]}.
\eeq
The key difference we now encounter from the previous subsection is that the two-particle form factors
$f^{\beta/2}({\rm i}\pi+{\rm i}0,2\kappa)$ are finite as $\kappa\to0$, since the right-hand side of the annihilation-pole axiom vanishes due to the locality of the operator with respect to the breather, $l_B^{\beta/2}=1$. Thus the only non-vanishing (as $\kappa\to0$) contribution to the semi-connected piece is
\beq
C_{0,2;0,2}^1=2f_{BB}^{\beta/2}({\rm i}\pi,0)\int_0^\infty \frac{ d\phi}{2\pi}\vert K_B(\phi)\vert^2,
\label{semiC0202}
\eeq
where we have used $S_{BB}(2\phi)S_{BB}(-2\phi)=1$.

The fully connected term can again be separated by shifting the integration contour to the lower half plane and picking up an annihilation pole. When shifting the contour one has to make sure not to pick up addititional contributions from the bound-state poles. Doing this, the contribution from the shifted contour,
\beq
C_{0,2;0,2}^\prime=\int_0^\infty\frac{d\phi^\prime}{2\pi}\int_{\gamma_-}\frac{d\phi}{2\pi}\vert K_B(\phi)\vert^2 f_{BBBB}^{\beta/2}(\phi^\prime+{\rm i}\pi+{\rm i}0,-\phi^\prime+{\rm i}\pi+{\rm i}0,-\phi+\kappa,\phi+\kappa)\,e^{2\Delta_B{\rm i}t(\cosh\theta^\prime-\cosh\theta)}
\eeq
is finite, and thus sub-leading at long times, so we will ignore it. The contribution from the annihilation pole for $\kappa\to0$ can be written using $S_{BB}(\phi)S_{BB}(\phi+\mathrm{i}\pi)=1$ as
\beq
C_{0,2;0,2}^p=-2f_{BB}^{\beta/2}({\rm i}\pi,0)\int_0^\infty\frac{d\phi}{2\pi}\vert K_B(\phi)\vert^2,
\eeq
and thus completely cancels with the semi-connected contribution (\ref{semiC0202}).

There is therefore no contribution linear in time to the expectation value (\ref{expectation}) at long times from $D_{0,2;0,2}$.

\subsubsection{The term $\boldsymbol{D_{0,0+1;0,0+1}}$}
This term gives a contribution that does not depend in time, and cannot be ignored in the long-time limit. We  find
\beq
C_{0,0+1;0,0+1}=\frac{\vert g\vert^2}{4}\langle 0_B\vert e^{{\rm i}\beta\Phi/2}\vert 0_B\rangle,
\eeq
where the form factor can be written using the $\kappa$-regularisation as
\begin{equation}
\langle 0_B\vert e^{{\rm i}\beta\Phi/2}\vert 0_B\rangle=\langle 0\vert B(0)e^{{\rm i}\beta\Phi/2}B^\dagger(\kappa)\vert 0\rangle=\langle 0\vert B({\rm i}0) e^{{\rm i}\beta\Phi/2}B^\dagger(\kappa)\vert 0\rangle+2\pi\delta(-\kappa)\mathcal{G}_{\beta/2}.\label{twobreather}
\end{equation}
The first term is finite as $\kappa\to0$, because the two-breather form factor does not have an annihilation pole. The second term gives a contribution that exactly cancels with $Z_{0,0+1}C_{0,0;0,0}$. Thus we find in total
\beq
D_{0,0+1;0,0+1}=\frac{\vert g\vert^2}{4}f_{BB}^{\beta/2}({\rm i}\pi,0).\label{done}
\eeq

\subsection{Order $\boldsymbol{K^3}$, $\boldsymbol{gK^2}$ and $\boldsymbol{g^2 K^1}$}
The only contributions to the expectation value (\ref{expectation}) which are not sub-leading at long times are 
\beq
D_{2,0+1;2,0}&=&C_{2,0+1;2,0}-Z_{2,0}C_{0,0+1;0,0},\\
D_{0,2+1;0,2} &=&C_{0,2+1;0,2}-Z_{0,2}C_{0,0+1;0,0},
\eeq
as well as $D_{2,0;2,0+1}$ and $D_{0,2;0,2+1}$. The computation of these two contributions is very similar to that of the terms $D_{2,0;2,0}$ and $D_{0,2;0,2}$ from last section.

\subsubsection{The term $\boldsymbol{D_{2,0+1;2,0}}$}
We consider
\begin{equation}
C_{2,0+1;2,0}=\frac{g^*}{2}\int_0^\infty \frac{d\theta^\prime d\theta}{(2\pi)^2}\left(K^{ab}(\theta^\prime)\right)^*K^{cd}(\theta)\,_{ba}\langle \theta^\prime,-\theta^\prime,0_B\vert e^{{\rm i} \beta\Phi/2}\vert -\theta,\theta\rangle_{cd}\,e^{2\Delta {\rm i} t(\cosh\theta^\prime-\cosh\theta)+\Delta_B{\rm i}t}.
\end{equation}
The form factor can be expressed as
\beq
\,_{ba}\langle \theta^\prime,-\theta^\prime,0_B\vert e^{{\rm i} \beta\Phi/2}\vert -\theta,\theta\rangle_{cd}&=&(2\pi)^2\delta_a^c\delta_b^d\delta(-2\kappa)\delta(\theta^\prime-\theta+\kappa)f_B^{\beta/2}\nonumber\\
&&+2\pi S_{ba}^{ef}(2\theta-2\kappa)S_{cd}^{fh}(-2\theta)\delta(\theta^\prime-\theta+\kappa)\,_e\langle\theta^\prime+{\rm i}0, 0_B\vert e^{{\rm i}\beta\Phi/2}\vert \theta+\kappa\rangle_h\nonumber\\
&&+2\pi\delta_b^d\delta(\theta^\prime-\theta-\kappa)\,_a\langle-\theta^\prime+{\rm i}0,0_B\vert e^{{\rm i}\beta\Phi/2}\vert -\theta+\kappa\rangle_c\nonumber\\
&&+\,_{ba}\langle \theta^\prime+{\rm i}0,-\theta^\prime+{\rm i}0,0_B\vert e^{{\rm i}\beta\Phi/2}\vert -\theta+\kappa,\theta+\kappa\rangle_{cd}.
\eeq
These terms give disconnected, semi-connected and fully connected contributions to $C_{2,0+1;2,0}$, but now all the terms have a zero-momentum breather in the form factor.

The disconnected term is
\beq
C_{2,0+1;2,0}^0=\frac{g^*}{2}\frac{L}{2}f_{B}^{\beta/2}e^{\Delta_B{\rm i}t}\int_0^\infty d\theta\,G(\theta),
\eeq
which cancels exactly with the term $Z_{2,0}\,C_{0,0+1;0,0}$.

The semi-connected piece is
\beq
C_{2,0+1;2,0}^1&=&\frac{g^*}{2}\int_0^\infty\frac{d\theta}{2\pi}\left(K^{ab}(\theta)\right)^*K^{cd}(\theta+\kappa)\,f_{\bar{e}Bh}^{\beta/2}({\rm i}\pi+{\rm i}0,{\rm i}\pi-\theta,2\kappa)\nonumber\\
&&\,\,\,\,\,\,\,\,\,\,\,\,\,\,\,\,\,\,\,\,\,\,\,\,\,\,\,\,\,\,\,\,\,\,\,\,\,\,\,\,\,\,\,\,\,\,\,\,\,\,\,\,\,\,\,\,\,\,\,\,\,\,\,\,\,\,\,\,\,\,\,\,\,\times S_{ba}^{ef}(2\theta)S_{cd}^{fh}(-2\theta-2\kappa)e^{2\Delta{\rm i}t\left[\cosh\theta-\cosh(\theta+\kappa)\right]+\Delta_B{\rm i}t}\nonumber\\
&&+\frac{g^*}{2}\int_0^\infty\frac{d\theta}{2\pi}\left(K^{ab}(\theta)\right)^*K^{cb}(\theta-\kappa)\, f_{\bar{a}Bc}^{\beta/2}({\rm i}\pi+{\rm i}0,{\rm i}\pi+\theta,2\kappa)\nonumber\\
&&\,\,\,\,\,\,\,\,\,\,\,\,\,\,\,\,\,\,\,\,\,\,\,\,\,\,\,\,\,\,\,\,\,\,\,\,\,\,\,\,\,\,\,\,\,\,\,\,\,\,\,\,\,\,\,\,\,\,\,\,\,\,\,\,\,\,\,\,\,\,\,\,\,\times e^{2\Delta{\rm i}t\left[\cosh\theta-\cosh(\theta-\kappa)\right]+\Delta_B{\rm i}t}
\eeq
Using the annihilation-pole axiom, we can write the three-particle form factor in the previous expression as
\beq
f_{\bar{a}Bc}^{\beta/2}({\rm i}\pi+{\rm i}0,{\rm i}\pi\pm\theta,2\kappa)&=&S_B({\rm i}\pi\pm\theta-2\kappa)f_{\bar{a}cB}^{\beta/2}({\rm i}\pi+{\rm i}0,2\kappa,{\rm i}\pi\pm\theta)\nonumber\\
&=&S_B({\rm i}\pi\pm\theta-2\kappa)\frac{-{\rm i}C_{\bar{a}c}}{2\kappa-{\rm i}0}\,f_B^{\beta/2}\left[1+S_B(2\kappa-{\rm i}\pi\mp\theta)\right]+F(\kappa)\nonumber\\
&=&\frac{{\rm i}C_{\bar{a}c}}{2\kappa-{\rm i}0}\,f_B^{\beta/2}\left[1+S_B({\rm i}\pi -2\kappa\pm\theta)\right]+F(\kappa),\label{residue}
\eeq
where $f_B^{\beta/2}$ is the one-breather form factor (which is a constant, independent of the breather's rapidity, due to translation invariance), and $F(\kappa)$ are finite terms that will not contribute when $\kappa\to0$. Using this expression, we find for small $\kappa$,
\beq
C_{2,0+1;2,0}^1&=&\frac{\mathrm{i}}{2}\frac{g^*\,f_B^{\beta/2}}{2\kappa-{\rm i}0}e^{\Delta_B{\rm i}t}\int_0^\infty\frac{d\theta}{2\pi}\left(K^{ab}(\theta)\right)^*K^{cd}(\theta)S_{ab}^{ef}(2\theta)S_{ef}^{cd}(-2\theta)
\left[1+S_B({\rm i}\pi -\theta)\right]\left[1-2\Delta{\rm i}t\kappa\sinh\theta\right]\nonumber\\
&&+\frac{\mathrm{i}}{2}\frac{g^*\,f_B^{\beta/2}}{2\kappa-{\rm i}0}e^{\Delta_B{\rm i}t}\int_0^\infty \frac{d\theta}{2\pi}\vert K^{ab}(\theta)\vert^2\left[1+S_B({\rm i}\pi+\theta)\right]\left[1+2\Delta{\rm i}t\kappa\sinh\theta\right]
\eeq

We now consider the fully connected term $C_{2,0+1;2,0}^2$. Again we can deform the integration contour and separate this term into a finite part and a contribution from the region around the annihilation poles. The finite part is
\beq
C_{2,0+1;2,0}^\prime&=&\frac{g^*}{2}e^{\Delta_B{\rm i}t}\int_0^\infty\frac{d\theta^\prime}{2\pi}\int_{\gamma_-}\frac{d\theta}{2\pi}\left(K^{ab}(\theta^\prime)\right)^*K^{cd}(\theta)e^{2\Delta{\rm i}t(\cosh\theta^\prime-\cosh\theta)}\nonumber\\
&&\,\,\,\,\,\,\,\,\times 
f_{\bar{b}\bar{a}Bcd}^{\beta/2}(\theta^\prime+{\rm i}\pi+{\rm i}0,-\theta^\prime+{\rm i}\pi+{\rm i}0,{\rm i}\pi,-\theta+\kappa,\theta+\kappa).
\eeq
This term is sub-leading at long times, so we will not compute it. The contribution from the annihilation poles is
\beq
C_{2,0+1;2,0}^p&=&-{\rm i}\frac{g^*}{2}e^{\Delta_B{\rm i}t}\int_0^\infty\frac{d\theta^\prime}{2\pi}\left(K^{ab}(\theta^\prime)\right)^*K^{cd}(\theta^\prime+\kappa)e^{2\Delta{\rm i}t\left[\cosh\theta^\prime-\cosh(\theta^\prime+\kappa)\right]}\nonumber\\
&&\times{\rm Res}\left[f_{\bar{b}\bar{a}Bcd}^{\beta/2}(\theta^\prime+{\rm i}\pi+{\rm i}0,-\theta^\prime+{\rm i}\pi+{\rm i}0,{\rm i}\pi,-\theta+\kappa,\theta+\kappa),\theta=\theta^\prime+\kappa-{\rm i}0\right],
\eeq
where the residue is given by
\beq
&&{\rm Res}\left[f_{\bar{b}\bar{a}Bcd}^{\beta/2}(\theta^\prime+{\rm i}\pi+{\rm i}0,-\theta^\prime+{\rm i}\pi+{\rm i}0,{\rm i}\pi,-\theta+\kappa,\theta+\kappa),\theta=\theta^\prime+\kappa-{\rm i}0\right]\nonumber\\
&&\,\,\,\,\,\,\,\,\,\,={\rm i}C_{gk}S_{\bar{b}\bar{a}}^{ef}(2\theta^\prime)S_{cd}^{gh}(-2\theta^\prime-2\kappa)f_{iBj}^{\beta/2}({\rm i}\pi+{\rm i}0,{\rm i}\pi-\theta^\prime,2\kappa)\nonumber\\
&&\,\,\,\,\,\,\,\,\,\,\,\,\,\,\,\times\left[\delta_e^i\delta_h^j\delta_f^k+S_{fe}^{li}(-2\theta^\prime)S_B(-\theta^\prime)S_{lh}^{kj}({\rm i}\pi-2\theta^\prime-2\kappa)\right].
\eeq
We then find
\beq
C_{2,0+1;2,0}^p&=&C_{gk}\frac{g^*}{2}e^{\Delta_B{\rm i}t}\int_0^\infty\frac{d\theta}{2\pi}\left(K^{ab}(\theta)\right)^*K^{cd}(\theta+\kappa)S_{\bar{b}\bar{a}}^{ef}(2\theta)S_{cd}^{gh}(-2\theta-2\kappa)f_{iBj}^{\beta/2}({\rm i}\pi+{\rm i}0,{\rm i}\pi-\theta,2\kappa)\nonumber\\
&&\times\left[\delta_e^i\delta_h^j
\delta_f^k+S_B(-\theta)S_{fe}^{li}(-2\theta)S_{lh}^{kj}({\rm i}\pi-2\theta-2\kappa)\right]e^{2\Delta{\rm i}t\left[\cosh\theta-\cosh(\theta+\kappa)\right]}.
\eeq
Using the expression (\ref{residue}), crossing (\ref{Scrossing}) and expanding for small $\kappa$, we find after straightforward simplifications
\begin{equation}
C_{2,0+1;2,0}^p=-\frac{{\rm i}}{2}\frac{g^*f_B^{\beta/2}}{2\kappa-{\rm i}0}e^{\Delta_B{\rm i}t}\int_0^\infty\frac{d\theta}{2\pi}\vert K^{ab}(\theta)\vert^2\left[2+S_B(\theta)+S_B(-\theta)\right]\left[1-2\Delta{\rm i}t\kappa\sinh\theta\right].
\label{C2120almost}
\end{equation}

It is easy to see that the part of (\ref{C2120almost}) that diverges as $1/\kappa$ cancels with the divergent part of the semi-connected term. Combining all the terms, the leading contribution at long times to $D_{2,0+1;2,0}$ is 
\beq
D_{2,0+1;2,0}=-\frac{g^*}{2}f_B^{\beta/2}e^{\Delta_B{\rm i}t}\frac{t}{\tau_B^*}+\dots,\label{tautwo}
\eeq
where
\beq
\frac{1}{\tau_B^*}=2\Delta \int_0^\infty \frac{d\theta}{2\pi} G(\theta)\left[1+S_{B}({\rm i}\pi+\theta)\right]\sinh\theta,
\eeq
Taking this together with the term $D_{2,0;2,0+1}$ obtained in a similar manner we finally get 
\beq
D_{2,0+1;2,0}+D_{2,0;2,0+1}=-2\,f_B^{\beta/2}{\rm Re}\left[\frac{g}{2}e^{-\Delta_B{\rm i}t}\frac{t}{\tau_B}\right],
\eeq
where we have used $S_B(\mathrm{i}\pi+\theta)^*=S_B(-\theta)^*=S_B(\theta)$. 

\subsubsection{The term $\boldsymbol{D_{0,2+1;0,2}}$}
The computation of this term closely parallels that of the previous subsection. We start by considering
\begin{equation}
C_{0,2+1;0,2}=\frac{g^*}{2}\int_0^\infty \frac{d\phi^\prime d\phi}{(2\pi)^2}\left(K_B(\phi^\prime)\right)^*K_B(\phi)\,\langle \phi^\prime,-\phi^\prime,0_B\vert e^{{\rm i}\beta\Phi/2}\vert -\phi,\phi\rangle e^{2\Delta_B{\rm i}t(\cosh\phi^\prime-\cosh\phi)+\Delta_B{\rm i}t}.
\end{equation}
The form factor can be written as
\beq
\langle\phi^\prime,-\phi^\prime,0_B\vert e^{{\rm i}\beta\Phi/2}\vert -\phi,\phi\rangle&=&(2\pi)^2\delta(-2\kappa)\delta(\phi^\prime-\phi+\kappa)f_B^{\beta/2}\nonumber\\
&&+2\pi S_{BB}(2\phi-2\kappa)S_{BB}(-2\phi)\delta(\phi^\prime-\phi+\kappa)\,\langle\phi^\prime+{\rm i} 0,0_B\vert e^{{\rm i}\beta\Phi/2}\vert \phi+\kappa\rangle\nonumber\\
&&+2\pi\delta(\phi^\prime-\phi-\kappa)\,\langle-\phi^\prime+{\rm i}0,0_B\vert e^{{\rm i}\beta\Phi/2}\vert -\phi+\kappa\rangle\nonumber\\
&&+\langle \phi^\prime+{\rm i}0,-\phi^\prime+{\rm i}0,0_B\vert e^{{\rm i}\beta\Phi/2}\vert-\phi+\kappa,\phi+\kappa\rangle,
\eeq
which we use to split $C_{0,2+1;0,2}$ into disconnected, semi-connected, and fully connected pieces.

The disconnected term is
\beq
C_{0,2+1;0,2}^0=\frac{g^*}{2}\frac{L}{2}f_B^{\beta/2}e^{\Delta_B{\rm i}t}\int_0^\infty d\phi\,G_B(\phi),
\eeq
which cancels exactly with the term $Z_{0,2}C_{0,0+1;0,0}$.

The semi-connected term is
\beq
C_{0,2+1;0,2}^1&=&\frac{g^*}{2}\int_0^\infty\frac{d\phi}{2\pi}\left(K_B(\phi)\right)^*K_B(\phi+\kappa)f_{BBB}^{\beta/2}({\rm i}\pi+{\rm i}0,{\rm i}\pi-\phi,2\kappa)\nonumber\\
&&\qquad\times S_{BB}(2\phi)S_{BB}(-2\phi-2\kappa)\,e^{2\Delta_B{\rm i}t\left[\cosh\phi-\cosh(\phi+\kappa)\right]+\Delta_B{\rm i}t}\nonumber\\
&&+\frac{g^*}{2}\int_0^\infty\frac{d\phi}{2\pi}\left(K_B(\phi)\right)^*K_B(\phi-\kappa)f_{BBB}^{\beta/2}({\rm i}\pi+{\rm i}0,{\rm i}\pi+\phi,2\kappa) e^{2\Delta_B{\rm i}t\left[\cosh\phi-\cosh(\phi-\kappa)\right]+\Delta_B{\rm i}t}.
\eeq
Using the annihilation-pole axiom, the three-breather form factor can be written as
\beq
f_{BBB}^{\beta/2}({\rm i}\pi+{\rm i}0,{\rm i}\pi\pm\phi,2\kappa)&=&S_{BB}({\rm i}\pi\pm\phi-2\kappa)f_{BBB}^{\beta/2}({\rm i}\pi+{\rm i}0,2\kappa,{\rm i}\pi\pm\phi)\nonumber\\
&=&S_{BB}({\rm i}\pi\pm\phi-2\kappa)\frac{-{\rm i}}{2\kappa-{\rm i}0}f_{B}^{\beta/2}\left[1-S_{BB}(2\kappa-{\rm i}\pi\mp\phi)\right]+F(\kappa)\nonumber\\
&=&\frac{{\rm i}}{2\kappa-{\rm i}0}f_B^{\beta/2}\left[1-S_{BB}({\rm i}\pi-2\kappa\pm\phi)\right]+F(\kappa).\label{threebreatherpole}
\eeq
We can then use this to find for $\kappa\to0$,
\beq
C_{0,2+1;0,2}^1&=&\frac{\mathrm{i}}{2}\frac{g^*f_B^{\beta/2}}{2\kappa-{\rm i}0}e^{\Delta_B{\rm i}t}\int_0^\infty\frac{d\phi}{2\pi}\vert K_B(\phi)\vert^2\left[1-S_{BB}({\rm i}\pi-\phi)\right]\left[1-2\Delta_B{\rm i}t\kappa\sinh\phi\right]\nonumber\\
&&+\frac{{\rm i}}{2}\frac{g^*f_B^{\beta/2}}{2\kappa-{\rm i}0}e^{\Delta_B{\rm i}t}\int_0^\infty\frac{d\phi}{2\pi}\vert K_B(\phi)\vert^2\left[1-S_{BB}({\rm i}\pi+\phi)\right]\left[1+2\Delta_B{\rm i}t\kappa\sinh\phi\right],
\eeq
which still contains some divergent parts that need to cancel with the fully connected term.

The fully connected term $C_{0,2+1;0,2}$ can again be divided into a finite part, and contributions coming from the annihilation poles. The finite part is given by
\beq
C_{0,2+1;0,2}^\prime&=&\frac{g^*}{2}e^{\Delta_B{\rm i}t}\int_0^\infty\frac{d\phi^\prime}{2\pi}\int_{\gamma_-}\frac{d\phi}{2\pi}\left(K_B(\phi^\prime)\right)^*K_B(\phi)\,e^{2\Delta_B{\rm i}t(\cosh\phi^\prime-\cosh\phi)}\nonumber\\
&&\times f_{BBBBB}^{\beta/2}(\phi^\prime+{\rm i}\pi+{\rm i}0,-\phi^\prime+{\rm i}\pi+{\rm i}0,{\rm i}\pi,-\phi+\kappa,\phi+\kappa),
\eeq
which gives sub-leading corrections at long times and will be ignored. The contribution from the poles is
\beq
C_{0,2+1;0,2}^p&=&-{\rm i}\frac{g^*}{2}e^{\Delta_B{\rm i}t}\int_0^\infty}\frac{d\phi^\prime}{2\pi}\left(K_B(\phi^\prime)\right)^*K(\phi^\prime+\kappa)\,e^{2\Delta_B{\rm i}t\left[\cosh\phi^\prime-\cosh(\phi^\prime+\kappa)\right]\nonumber\\
&&\times {\rm Res}\left[f_{BBBBB}^{\beta/2}(\phi^\prime+{\rm i}\pi+{\rm i}0,-\phi^\prime+{\rm i}\pi+{\rm i}0,{\rm i}\pi,-\phi+\kappa,\phi+\kappa),\phi=\phi^\prime+\kappa-{\rm i}0\right],
\eeq
and the residue is
\beq
&&{\rm Res}\left[f_{BBBBB}^{\beta/2}(\phi^\prime+{\rm i}\pi+{\rm i}0,-\phi^\prime+{\rm i}\pi+{\rm i}0,{\rm i}\pi,-\phi+\kappa,\phi+\kappa),\phi=\phi^\prime+\kappa-{\rm i}0\right]\nonumber\\
&&\,\,\,\,\,\,\,\,\,=-{\rm i}S_{BB}(2\phi^\prime)S_{BB}(-2\phi^\prime-2\kappa)f_{BBB}^{\beta/2}({\rm i}\pi+{\rm i}0,\mathrm{i}\pi-\phi^\prime,2\kappa)\nonumber\\
&&\,\,\,\,\,\,\,\,\,\,\,\,\,\times\left[1-S_{BB}(-2\phi^\prime)S_{BB}(-\phi^\prime+{\rm i}0)S_{BB}(-2\phi^\prime-2\kappa+{\rm i}\pi)\right],
\eeq
such that
\beq
C_{0,2+1;0,2}^p&=&-\frac{g^*}{2}e^{\Delta_B{\rm i}t}\int_0^\infty\frac{d\phi}{2\pi}\left(K_B(\phi)\right)^*K_B(\phi+\kappa)S_{BB}(2\phi)S_{BB}(-2\phi-2\kappa)f_{BBB}^{\beta/2}({\rm i}\pi+{\rm i}0,{\rm i}\pi-\phi,2\kappa)\nonumber\\
&&\times\left[1-S_{BB}(-\phi)S_{BB}(-2\phi)S_{BB}(-2\phi-2\kappa+{\rm i}\pi)\right]e^{2\Delta_B{\rm i}t\left[\cosh\phi-\cosh(\phi+\kappa)\right]}.
\eeq
Using (\ref{threebreatherpole}) and expanding for small $\kappa$, we find
\begin{equation}
C_{0,2+1;0,2}^p=-\frac{{\rm i}}{2}\frac{g^*f_B^{\beta/2}}{2\kappa-{\rm i}0}\,e^{\Delta_B{\rm i}t}\int_0^\infty\frac{d\phi}{2\pi}\vert K_B(\phi)\vert^2\left[2-S_{BB}({\rm i}\pi-\phi)-S_{BB}({\rm i}\pi+\phi)\right]\left[1-2\Delta_B{\rm i}t\kappa\sinh\phi\right].
\end{equation}
Combining all the terms, the leading contribution at long times is
\beq
D_{0,2+1;0,2}=-\frac{g^*}{2}f_B^{\beta/2}\,e^{\Delta_B{\rm i}t}\frac{t}{\tau_{BB}^*}+\dots,
\eeq
where
\beq
\frac{1}{\tau_{BB}^*}=2\Delta_B\int_0^\infty\frac{d\phi}{2\pi}G_B(\phi)\left[1-S_{BB}({\rm i}\pi+\phi)\right]\sinh\phi.
\eeq
The rate $\tau_{BB}$ is obtained using $S_{BB}(\mathrm{i}\pi+\phi)^*=S_{BB}(-\phi)^*=S_{BB}(\phi)$. 
Together with the complex conjugate term we can write
\beq
D_{0,2+1;0,2}+D_{0,2;0,2+1}=-2\,f_{B}^{\beta/2}\,{\rm Re}\left[\frac{g}{2}\,e^{-\Delta_B{\rm i}t}\frac{t}{\tau_{BB}}\right].
\eeq

\subsection{Order $\boldsymbol{K^4}$, $\boldsymbol{gK^3}$ and $\boldsymbol{g^2K^2}$}
The leading contributions at long times at this order of $K$ are
\beq
D_{4,0;4,0}&=&C_{4,0;4,0}-Z_{4,0}C_{0,0;0,0}+Z_{2,0}^2C_{0,0;0,0}-Z_{2,0}C_{2,0;2,0},\\
D_{0,4;0,4}&=&C_{0,4;0,4}-Z_{0,4}C_{0,0;0,0}+Z_{0,2}^2C_{0,0;0,0}-Z_{0,2}C_{0,2;0,2},\\
D_{2,2;2,2}&=&C_{2,2;2,2}-Z_{2,2}C_{0,0;0,0}+2Z_{2,0}Z_{0,2}C_{0,0;0,0}-Z_{2,0}C_{0,2;0,2}-Z_{0,2}C_{2,0;2,0},\\
D_{2,0+1;2,0+1}&=&C_{2,0+1;2,0+1}-Z_{2,0+1}C_{0,0;0,0}+2Z_{2,0}Z_{0,0+1}C_{0,0;0,0}-Z_{2,0}C_{0,0+1;0,0+1}-Z_{0,0+1}C_{2,0;2,0},\\
D_{0,2+1;0,2+1}&=&C_{0,2+1;0,2+1}-Z_{0,2+1}C_{0,0;0,0}+2Z_{0,2}Z_{0,0+1}C_{0,0;0,0}-Z_{0,2}C_{0,0+1;0,0+1}-Z_{0,0+1}C_{0,2;0,2}.
\eeq
The first of these terms has been computed in Ref.~\cite{repulsive}, he we simply state the result
\beq
D_{4,0;4,0}=\frac{\mathcal{G}_{\beta/2}}{2}\left(\frac{t}{\tau}\right)^2+\dots
\eeq
This result, together with $D_{2,0;2,0}$, suggest that one may be able to resum all the terms $D_{n,0;n,0}$ as an exponential,
\beq
D_{2,0;2,0}+D_{4,0;4,0}+\dots=\mathcal{G}_{\beta/2}\left[1-\frac{t}{\tau}+\frac{1}{2}\left(\frac{t}{\tau}\right)^2+\dots\right]=\mathcal{G}_{\beta/2}e^{-t/\tau}(1+\dots).
\eeq
These leading contributions of the terms $D_{n,0;n,0}$ come from the most divergent parts of the form factors. One can use the annihilation-pole axiom to obtain a relation between the $2n$-soliton form factor and the $(2n-2)$-soliton form factor, continuing this procedure until one reaches the $0$-particle form factor, $\mathcal{G}_{\beta/2}$, while picking up $n$ annihilation poles. 

We now turn our attention to the term $D_{0,4;0,4}$. We will not calculate this term explicitly, but simply point out that the contributions from this term are sub-leading at long times. The argument for this is the same as the argument we provided for the term $D_{0,2;0,2}$. The long-time leading contributions come from the most divergent parts of the form factors.  The two-breather form factor does not have an annihilation pole. This means that most divergent part of the term $D_{0,n;0,n}$ has $n-1$ poles, hence it contains one pole less than $D_{n,0;n,0}$, which implies that it is sub-leading at long times.

The same argument also applies to the term $D_{2,2;2,2}$, which contains one pole less, and is therefore sub-leading at long times, in comparison to $D_{4,0;4,0}$. We expect the terms $D_{0,4;0,4}$ and $D_{2,2;2,2}$ to yield a linear behaviour in $t$ at late times, which is expected to give the $\mathcal{O}(K^4)$ contributions to the rates (\ref{decayrates1})--(\ref{decayrates3}).

\subsubsection{The term $\boldsymbol{D_{2,0+1;2,0+1}}$}
We first consider
\begin{equation}
C_{2,0+1;2,0+1}=\frac{\vert g\vert^2}{4}\int_0^\infty\frac{d\theta^\prime d\theta}{(2\pi)^2}\left(K^{ab}(\theta^\prime)\right)^*K^{cd}(\theta) \,_{ba}\langle \theta^\prime,-\theta^\prime,0_B\vert e^{{\rm i}\beta\Phi/2}\vert0_B,-\theta,\theta\rangle_{cd}\,e^{2\Delta{\rm i}t(\cosh\theta^\prime-\cosh\theta)}.
\end{equation}
The form factor can be written introducing two regulators $\kappa$ and $\kappa_B$ as
\beq
&&\,_{ba}\langle \theta^\prime,-\theta^\prime,0_B\vert e^{{\rm i}\beta\Phi/2}\vert 0_B,-\theta,\theta\rangle_{cd}=\,_{ba}\langle \theta^\prime,-\theta^\prime\vert B(0)\,e^{{\rm i}\beta\Phi/2}\,B^\dagger(\kappa_B)\vert-\theta+\kappa,\theta+\kappa\rangle_{cd}\nonumber\\
&&\qquad=2\pi\delta(-\kappa_B)\left[(2\pi)^2\delta_a^c\delta_b^d\delta(-2\kappa)\delta(\theta^\prime-\theta+\kappa)+2\pi S_{ba}^{ef}(2\theta-2\kappa)S_{cd}^{fh}(-2\theta)\delta(\theta^\prime-\theta+\kappa)\,_e\langle\theta^\prime+{\rm i}0\vert e^{{\rm i}\beta\Phi/2}\vert \theta+\kappa\rangle_h\right.\nonumber\\
&&\qquad\qquad\left.+2\pi\delta_b^d\delta(\theta^\prime-\theta-\kappa)\,_a\langle-\theta^\prime+{\rm i}0\vert e^{{\rm i}\beta\Phi/2}\vert -\theta+\kappa\rangle_c+\,_{ba}\langle \theta^\prime+{\rm i}0,-\theta^\prime+{\rm i}0\vert e^{{\rm i}\beta\Phi/2}\vert -\theta+\kappa,\theta+\kappa\rangle_{cd}\right]\nonumber\\
&&\qquad+(2\pi)^2\delta_a^c\delta_b^d\delta(-2\kappa)\delta(\theta^\prime-\theta+\kappa)\langle 0_B+\mathrm{i}0\vert e^{{\rm i}\beta\Phi/2}\vert 0_B+\kappa_B\rangle\nonumber\\
&&\qquad+2\pi S_{ba}^{ef}(2\theta-2\kappa)S_{cd}^{fh}(-2\theta)\delta(\theta^\prime-\theta+\kappa)\,_e\langle\theta^\prime+{\rm i}0, 0_B+\mathrm{i}0\vert e^{{\rm i}\beta\Phi/2}\vert0_B+\kappa_B, \theta+\kappa\rangle_h\nonumber\\
&&\qquad+2\pi\delta_b^d\delta(\theta^\prime-\theta-\kappa)\,_a\langle-\theta^\prime+{\rm i}0,0_B+\mathrm{i}0\vert e^{{\rm i}\beta\Phi/2}\vert0_B+\kappa_B, -\theta+\kappa\rangle_c\nonumber\\
&&\qquad+\,_{ba}\langle \theta^\prime+{\rm i}0,-\theta^\prime+{\rm i}0,0_B+\mathrm{i}0\vert e^{{\rm i}\beta\Phi/2}\vert0_B+\kappa_B, -\theta+\kappa,\theta+\kappa\rangle_{cd}.\label{foursolitontwobreather}
\eeq
The first four terms (the ones proportional to $\delta(-\kappa_B)$) yield contributions that cancel exactly with the terms $Z_{2,0+1}$, $Z_{0,0+1}C_{2,0;2,0}$ and one of the terms $Z_{2,0}Z_{0,0+1}C_{0,0;0,0}$. The fifth term gives a contribution that cancels exactly with $Z_{2,0}C_{0,0+1;0,0+1}$ and the second  $Z_{2,0}Z_{0,0+1}C_{0,0;0,0}$ term. The last three terms give contributions which we can denote as semi-connected and fully connected.

The semi-connected piece is
\beq
C_{2,0+1;2,0+1}^1&=&-\frac{\vert g\vert^2}{4}\int_0^\infty\frac{d\theta}{2\pi}\left(K^{ab}(\theta)\right)^*K^{cd}(\theta+\kappa)\,f_{\bar{e}BBh}^{\beta/2}({\rm i}\pi+{\rm i}0,{\rm i}\pi+\mathrm{i}0-\theta+\kappa,\kappa_B-\theta+\kappa,2\kappa)\nonumber\\
&&\,\,\,\,\,\,\,\,\,\,\,\,\,\,\,\,\,\,\,\,\,\,\,\,\,\,\,\,\,\,\,\,\,\,\,\,\,\,\,\,\,\,\,\,\,\,\,\,\,\,\,\,\,\,\,\,\,\,\,\,\,\,\,\,\,\,\,\,\,\,\,\,\,\times S_{ba}^{ef}(2\theta)S_{cd}^{fh}(-2\theta-2\kappa)e^{2\Delta{\rm i}t\left[\cosh\theta-\cosh(\theta+\kappa)\right]}\nonumber\\
&&-\frac{\vert g\vert^2}{4}\int_0^\infty\frac{d\theta}{2\pi}\left(K^{ab}(\theta)\right)^*K^{cb}(\theta-\kappa)\, f_{\bar{a}BBc}^{\beta/2}({\rm i}\pi+{\rm i}0,{\rm i}\pi+\mathrm{i}0+\theta+\kappa,\kappa_B+\theta+\kappa,2\kappa)\nonumber\\
&&\,\,\,\,\,\,\,\,\,\,\,\,\,\,\,\,\,\,\,\,\,\,\,\,\,\,\,\,\,\,\,\,\,\,\,\,\,\,\,\,\,\,\,\,\,\,\,\,\,\,\,\,\,\,\,\,\,\,\,\,\,\,\,\,\,\,\,\,\,\,\,\,\,\times e^{2\Delta{\rm i}t\left[\cosh\theta-\cosh(\theta-\kappa)\right]}.
\eeq
Using the annihilation pole axiom, we can write (in the limit $\kappa_B\to0$)
\beq
f_{\bar{e}BBh}^{\beta/2}({\rm i}\pi+{\rm i}0,{\rm i}\pi+\mathrm{i}0-\theta+\kappa,\kappa_B-\theta+\kappa,2\kappa)=\frac{-2{\rm i}C_{ab}}{2\kappa-{\rm i}0} f_{BB}^{\beta/2}({\rm i}\pi,0)+{\rm F.\,P.},
\eeq
where F.P. denote finite terms in the $\kappa\to 0$ limit and we have used the property $S_B({\rm i}\pi+\theta)S_B(\theta)=1$ as well as that the two-breather form factor does not contain an annihiliation pole.

The computation from this point onwards is completely equivalent to the computation of $C_{2,0;2,0}$ but replacing $\mathcal{G}_{\beta/2}$ with $\frac{\vert g\vert^2}{4}f_{BB}^{\beta/2}({\rm i}\pi,0)$. We will then just write the final result,
\beq
D_{2,0+1;2,0+1}=-\frac{\vert g\vert^2}{4}f_{BB}^{\beta/2}({\rm i}\pi,0)\frac{t}{\tau}.
\eeq

\subsubsection{The term $\boldsymbol{D_{0,2+1;0,2+1}}$}
Performing a calculation similar to that of the previous subsection, but with all the particles being breathers, we find that the most divergent contributions from the annihilation poles of four-breather form factors have vanishing residue. Thus the term $D_{0,2+1;0,2+1}$ does not contribute to the late-time behaviour at leading order, analogous to the results of the last subsection. Instead we expect it to contribute to the $\mathcal{O}(K^4)$ contributions to the rates (\ref{decayrates1})--(\ref{decayrates3}).

\subsection{Order $\boldsymbol{gK^4}$}
The leading long-times contributions at this order of $K$ are
\beq
D_{4,0+1;4,0}&=&C_{4,0+1;4,0}-Z_{4,0}C_{0,0+1;0,0}+Z_{2,0}^2C_{0,0+1;0,0}-Z_{2,0}C_{2,0+1;2,0},\\
D_{0,4+1;0,4}&=&C_{0,4+1;0,4}-Z_{0,4}C_{0,0+1;0,0}+Z_{0,2}^2C_{0,0+1;0,0}-Z_{0,2}C_{0,2+1,0,2},\\
D_{2,2+1;2,2}&=&C_{2,2+1;2,2}-Z_{2,2}C_{0,0+1;0,0}+2Z_{2,0}Z_{0,2}C_{0,0+1;0,0},
\eeq
as well as the similar terms $D_{4,0;4,0+1}$, $D_{0,4;0,4+1}$ and $D_{2,2;2,2+1}$.

\subsubsection{The term $\boldsymbol{D_{4,0+1,4,0}}$}
The leading long-time contribution for this term is given by the connected term $C_{4,0+1;4,0}$.  To simplify the calculation, we  will only compute the leading terms, which diverge as $t^2$ for long times. We also use the simplifying assumption that the initial state is of ``Dirichlet type", so $K^{aa}(\theta)=0$. We then consider the term
\beq
C_{4,0+1;4,0}&=&\frac{1}{4}\frac{g^*}{2}e^{{\rm i}\Delta_Bt}\int_0^\infty\frac{d\theta_1^\prime d\theta_2^\prime d\theta_1 d\theta_2}{(2\pi)^4}\left(K^{a\bar{a}}(\theta_1^\prime)\right)^*\left(K^{b\bar{b}}(\theta_2^\prime)\right)^* K^{c\bar{c}}(\theta_1)K^{d\bar{d}}(\theta_2) e^{2\Delta{\rm i}t\sum_i(\cosh\theta_i^\prime-\cosh\theta_i)}\nonumber\\
&&\times\,_{\bar{a}a\bar{b}b}\langle \theta_1^\prime,-\theta_1^\prime,\theta_2^\prime,-\theta_2^\prime,0_B\vert e^{{\rm i}\beta\Phi/2}\vert-\theta_2+\kappa_2,\theta_2+\kappa_2,-\theta_1+\kappa_1,\theta_1+\kappa_1\rangle_{d\bar{d}c\bar{c}}.\label{foursoliton}
\eeq

We can write the form factor in $(\ref{foursoliton})$ as
\beq
&&\!\!\!\!\!\!\!\,_{\bar{a}a\bar{b}b}\langle \theta_1^\prime,-\theta_1^\prime,\theta_2^\prime,-\theta_2^\prime,0_B\vert e^{{\rm i}\beta\Phi/2}\vert-\theta_2+\kappa_2,\theta_2+\kappa_2,-\theta_1+\kappa_1,\theta_1+\kappa_1\rangle_{d\bar{d}c\bar{c}}\nonumber\\
&&\!\!\!\!\!=f_{a\bar{a}b\bar{b}Bd\bar{d}c\bar{c}}^{\beta/2}(\theta_1^\prime+{\rm i}\pi+{\rm i}0,-\theta_1^\prime+{\rm i}\pi+{\rm i}0,\theta_2^\prime+{\rm i}\pi+{\rm i}0,-\theta_2^\prime+{\rm i}\pi+{\rm i}0,{\rm i}\pi,-\theta_2+\kappa_2,\theta_2+\kappa_2,-\theta_1+\kappa_1,\theta_1+\kappa_1)\label{fone}\quad\\
&&\!\!\!\!-2\pi\delta_a^c\delta(\theta_1^\prime-\theta_1-\kappa_1)f_{\bar{a}b\bar{b}Bd\bar{d}c}^{\beta/2}(-\theta_1^\prime+{\rm i}\pi+{\rm i}0,\theta_2^\prime+{\rm i}\pi+{\rm i}0,-\theta_2^\prime+{\rm i}\pi+{\rm i}0,{\rm i}\pi,-\theta_2+\kappa_2,\theta_2+\kappa_2,-\theta_1+\kappa_1)\label{ftwo}\\
&&\!\!\!\!\!-2\pi\delta_{\bar{a}}^g\delta(\theta_1^\prime-\theta_2-\kappa_2)S_{\bar{d}c}^{ef}(\theta_1+\theta_2)S_{e\bar{c}}^{gh}(\theta_2-\theta_1)\nonumber\\*
&&\times f_{\bar{a}b\bar{b}Bdfh}^{\beta/2}(-\theta_1^\prime+{\rm i}\pi+{\rm i}0,\theta_2^\prime+{\rm i}\pi+{\rm i}0,-\theta_2^\prime+{\rm i}\pi+{\rm i}0,{\rm i}\pi,-\theta_2+\kappa_2,-\theta_1+\kappa_1,-\theta_1+\kappa_1)\label{fthree}\\
&&\!\!\!\!\!-2\pi\delta_h^{\bar{c}}\delta(\theta_2^\prime-\theta_1-\kappa)S_{a\bar{b}}^{ef}(-\theta_1^\prime-\theta_2^\prime)S_{\bar{a}f}^{gh}(\theta_1^\prime-\theta_2^\prime)\nonumber\\*
&&\times f_{\bar{g}\bar{e}\bar{d}Bd\bar{d}c}^{\beta/2}(\theta_1^\prime+{\rm i}\pi+{\rm i}0,-\theta_1^\prime+{\rm i}\pi+{\rm i}0,-\theta_2^\prime+{\rm i}\pi+{\rm i}0,{\rm i}\pi,-\theta_2+\kappa_2,\theta_2+\kappa_2,-\theta_1+\kappa_1)\label{ffour}\\
&&\!\!\!\!\!-2\pi\delta_h^m\delta(\theta_2^\prime-\theta_2-\kappa_2)S_{a\bar{b}}^{ef}(-\theta_1^\prime-\theta_2^\prime)S_{\bar{a}f}^{gh}(\theta_1^\prime-\theta_2^\prime)S_{\bar{d}c}^{ij}(\theta_1+\theta_2)S_{i\bar{c}}^{mn}(\theta_2-\theta_1)\nonumber\\*
&&\times f_{\bar{g}\bar{e}\bar{b}Bdjn}^{\beta/2}(\theta_1^\prime+{\rm i}\pi+{\rm i}0,-\theta_1^\prime+{\rm i}\pi+{\rm i}0,-\theta_2^\prime+{\rm i}\pi+{\rm i}0,{\rm i}\pi,-\theta_2+\kappa_2,-\theta_1+\kappa_1,\theta_1+\kappa_1)\label{ffive}\\
&&\!\!\!\!\!-2\pi\delta_e^f\delta(-\theta_1^\prime+\delta_1-\kappa_1)S_{\bar{a}a}^{\bar{e}e}(2\theta_1^\prime)S_{c\bar{c}}^{f\bar{f}}(-2\theta_1)\nonumber\\*
&&\times f_{eb\bar{b}Bd\bar{d}f}^{\beta/2}(\theta_1^\prime+{\rm i}\pi+{\rm i}0,\theta_2^\prime+{\rm i}\pi+{\rm i}0,-\theta_2^\prime+{\rm i}\pi+{\rm i}0,{\rm i}\pi,-\theta_2+\kappa_2,\theta_2+\kappa_2,\theta_1+\kappa_1)\label{fsix}\\
&&\!\!\!\!\!-2\pi\delta_i^e\delta(-\theta_1^\prime+\theta_2-\kappa_2)S_{a\bar{a}}^{\bar{e}e}(2\theta_1^\prime)S_{d\bar{d}}^{f\bar{f}}(-2\theta_2)S_{fc}^{gh}(\theta_1-\theta_2)S_{g\bar{c}}^{ij}(-\theta_1-\theta_2)\nonumber\\*
&&\times f_{eb\bar{b}B\bar{f}hj}^{\beta/2}(\theta_1^\prime+{\rm i}\pi+{\rm i}0,\theta_2^\prime+{\rm i}\pi+{\rm i}0,-\theta_2^\prime+{\rm i}\theta+{\rm i}0,{\rm i}\pi,\theta_2+\kappa_2,-\theta_+\kappa_1,\theta_1+\kappa_1)\label{fseven}\\
&&\!\!\!\!\!-2\pi\delta_j^f\delta(-\theta_2^\prime+\theta_1-\kappa_1)S_{\bar{b}b}^{\bar{e}e}(2\theta_2^\prime)S_{ae}^{gh}(\theta_2^\prime-\theta_1^\prime)S_{\bar{a}h}^{ij}(\theta_1^\prime+\theta_2^\prime)S_{c\bar{c}}^{f\bar{f}}(-2\theta_1)\nonumber\\*
&&\times f_{\bar{i}\bar{g}eBd\bar{d}f}^{\beta/2}(\theta_1^\prime+{\rm i}\pi+{\rm i}0,-\theta_1^\prime+{\rm i}\pi+{\rm i}0,\theta_2^\prime+{\rm i}\pi+{\rm i}0,{\rm i}\pi,-\theta_2+\kappa_2,\theta_2+\kappa_2,\theta_1+\kappa_1)\label{feight}\\
&&\!\!\!\!\!-2\pi\delta_j^p\delta(-\theta_2^\prime+\theta_2-\kappa_2)S_{\bar{b}b}^{\bar{e}e}(2\theta_2^\prime)S_{ae}^{gh}(\theta_2^\prime-\theta_1^\prime)S_{\bar{a}h}^{ij}(\theta_1^\prime+\theta_2^\prime)S_{d\bar{d}}^{f\bar{f}}(-2\theta_2)S_{fc}^{mn}(\theta_1-\theta_2)S_{m\bar{c}}^{pq}(-\theta_1-\theta_2)\nonumber\\*
&&\times f_{\bar{i}\bar{g}eB\bar{f}nq}(\theta_1^\prime+{\rm i}\pi+{\rm i}0,-\theta_1^\prime+{\rm i}\pi+{\rm i},\theta_2^\prime+{\rm i}\pi+{\rm i}0,{\rm i}\pi,\theta_2+\kappa_2,-\theta_1+\kappa_1,\theta_1+\kappa_1).\label{fnine}\\
&&\!\!\!\!\!+(2\pi)^2S_{a\bar{b}}^{ef}(-\theta_1^\prime-\theta_2^\prime)S_{\bar{d}c}^{gh}(\theta_1+\theta_2)f_{\bar{e}\bar{b}Bdh}^{\beta/2}(-\theta_1^\prime+{\rm i}\pi+{\rm i}0,-\theta_2^\prime+{\rm i}\pi+{\rm i}0,{\rm i}\pi,-\theta_2+\kappa_2,-\theta_1+\kappa_1)\nonumber\\*
&&\times\left[\delta_a^c\delta_f^g\delta(\theta_1^\prime-\theta_1-\kappa_1\delta(\theta_2^\prime-\theta_2-\kappa_2)+\delta_a^{\bar{j}}\delta_c^{\bar{i}}\delta(\theta_1^\prime-\theta_2-\kappa_2)\delta(\theta_2^\prime-\theta_1-\kappa_1)S_{gi}^{if}(\theta_2-\theta_2^\prime)\right]\label{ften}\\
&&\!\!\!\!\!+(2\pi)^2S_{\bar{a}a}^{\bar{e}e}(2\theta_1^\prime)S_{\bar{b}b}^{\bar{f}f}(2\theta_2^\prime)S_{\bar{e}f}^{ij}(\theta_1^\prime+\theta_2^\prime)S_{c\bar{c}}^{g\bar{g}}(-2\theta_1)S_{d\bar{d}}^{h\bar{h}}(-2\theta_2)S_{h\bar{g}}^{mn}(-\theta_1-\theta_2)\nonumber\\*
&&\times f_{\bar{i}fB\bar{h}n}^{\beta/2}(\theta_1^\prime+{\rm i}\pi+{\rm i}0,\theta_2^\prime+{\rm i}\pi+{\rm i}0,{\rm i}\pi,\theta_2+\kappa_2,\theta_1+\kappa_1)\nonumber\\*
&&\times\left[\delta_e^g\delta_j^m\delta(-\theta_1^\prime+\theta_1-\kappa_1)+\delta_e^q\delta_g^p\delta(-\theta_1^\prime+\theta_2-\kappa_2)\delta(-\theta_2^\prime+\theta_1-\kappa_1)S_{mp}^{qj}(\theta_2^\prime-\theta_2)\right]\label{feleven}\\
&&\!\!\!\!\!+(2\pi)^2\delta_a^c\delta_h^i\delta(\theta_1^\prime-\theta_1-\kappa_1)\delta(-\theta_2^\prime+\theta_2-\kappa_2)S_{\bar{b}b}^{\bar{e}e}(2\theta_2^\prime)S_{ae}^{gh}(\theta_2^\prime-\theta_1^\prime)S_{d\bar{d}}^{f\bar{f}}(-2\theta_2)S_{fc}^{ij}(\theta_1-\theta_2)\nonumber\\*
&&\times f_{\bar{g}eB\bar{f}j}^{\beta/2}(-\theta_1^\prime+{\rm i}\pi+{\rm i}0,\theta_2^\prime+{\rm i}\pi+{\rm i}0,{\rm i}\pi,\theta_2+\kappa_2,-\theta_1+\kappa_1)\label{ftwelve}\\
&&\!\!\!\!\!+(2\pi)^2\delta_{\bar{a}}^m\delta_h^n\delta(\theta_1^\prime-\theta_2-\kappa_2)\delta(-\theta_2^\prime+\theta_1-\kappa_1)S_{\bar{b}b}^{\bar{e}e}(2\theta_2^\prime)S_{ae}^{gh}(\theta_2^\prime-\theta_1^\prime)S_{c\bar{c}}^{f\bar{f}}(-2\theta_1)\nonumber\\*
&&\times S_{\bar{d}\bar{f}}^{ij}(\theta_2-\theta_1)S_{if}^{mn}(\theta_1+\theta_2)f_{\bar{g}eBdj}^{\beta/2}(-\theta_1^\prime+{\rm i}\pi+{\rm i}0,\theta_2^\prime+{\rm i}\pi+{\rm i}0,{\rm i}\pi,-\theta_2+\kappa_2,\theta_1+\kappa_1)\label{fthirteen}\\
&&\!\!\!\!\!+(2\pi)^2\delta_j^{\bar{c}}\delta_i^m\delta(\theta_2^\prime-\theta_1-\kappa_1)\delta(-\theta_1^\prime+\theta_2-\kappa_2)S_{\bar{a}a}^{\bar{e}e}(2\theta_1^\prime)S_{\bar{e}\bar{b}}^{gh}(\theta_1^\prime-\theta_2^\prime)S_{eh}^{ij}(-\theta_1^\prime-\theta_2^\prime)S_{d\bar{d}}^{f\bar{f}}(-2\theta_2)\nonumber\\*
&&\times S_{fc}^{mn}(\theta_1-\theta_2)f_{\bar{g}\bar{b}B\bar{f}n}^{\beta/2}(\theta_1^\prime+{\rm i}\pi+{\rm i}0,-\theta_2^\prime+{\rm i}\pi+{\rm i}0,{\rm i}\pi,\theta_2+\kappa_2,-\theta_1+\kappa_1)\label{ffourteen}\\
&&\!\!\!\!\!+(2\pi)^2\delta_e^f\delta_h^i\delta(-\theta_1^\prime+\theta_1-\kappa_1)\delta(\theta_2^\prime-\theta_2-\kappa_2)S_{\bar{a}a}^{\bar{e}e}(2\theta_1^\prime)S_{e\bar{b}}^{gh}(\theta_1^\prime-\theta_2^\prime)S_{c\bar{c}}^{f\bar{f}}(-2\theta_1)S_{\bar{d}\bar{f}}^{ij}(\theta_2-\theta_1)\nonumber\\*
&&\times f_{\bar{g}\bar{b}Bdj}^{\beta/2}(\theta_1^\prime+{\rm i}\pi+{\rm i}0,-\theta_2^\prime+{\rm i}\pi+{\rm i}0,{\rm i}\pi,-\theta_2+\kappa_2,\theta_1+\kappa_1)\label{ffifteen}\\
&&\!\!\!\!\!+\dots,\nonumber
\eeq
where the dots here and below represent contributions that do not lead to terms growing as $t^2$. The term (\ref{fone}), when inserted back into (\ref{foursoliton}), gives the fully connected contribution, which we will denote as $C_{4,0+1;4,0}^4$. We denote the semi-connected contributions coming from the terms (\ref{ftwo})--(\ref{fnine}) by $C_{4,0+1;4,0}^3$, and those coming from the terms (\ref{ften})--(\ref{ffifteen}) by $C^2_{4,0+1,4,0}$.

We now extract the leading contributions from $C^4_{4,0+1,4,0}$. The leading contributions at long times come from the the poles at $\theta_1=\theta_1^\prime+\kappa_1-{\rm i}0$ and $\theta_2=\theta_2^\prime+\kappa_2-{\rm i}0$, or $\theta_1=\theta_2^\prime+\kappa_1-{\rm i}0$ and $\theta_2=\theta_1^\prime+\kappa_2-{\rm i}0$, such that
\beq
C_{4,0+1;4,0}^4&=&-\frac{1}{4}\frac{g^*}{2}e^{{\rm i}\Delta_Bt}\int_0^\infty\frac{d\theta_1^\prime d\theta_2^\prime}{(2\pi)^2}\left(K^{a\bar{a}}(\theta_1^\prime)\right)^*\left(K^{b\bar{b}}(\theta_2^\prime)\right)^*e^{2\Delta{\rm i}t(\cosh\theta_1^\prime+\cosh\theta_2^\prime)}\nonumber\\
&&\quad\times\left\{K^{c\bar{c}}(\theta_2^\prime)K^{d\bar{d}}(\theta_1^\prime)e^{-2\Delta{\rm i}t(\cosh(\theta_1^\prime+\kappa_1)+\cosh(\theta_2^\prime+\kappa_2))}\right.\nonumber\\*
&&\qquad\times{\rm Res}\left[{\rm Res}\left[f_{a\bar{a}b\bar{b}Bc\bar{c}d\bar{d}}^{\beta/2}(\theta_1^\prime+{\rm i}\pi+{\rm i}0,-\theta_1^\prime+{\rm i}\pi+{\rm i}0,\theta_2^\prime+{\rm i}\pi+{\rm i}0,-\theta_2^\prime+{\rm i}\pi+{\rm i}0,\right.\right.\nonumber\\*
&&\qquad\quad\left.\left.{\rm i}\pi,-\theta_2+\kappa_2,\theta_2+\kappa_2,-\theta_1+\kappa_1,\theta_1+\kappa_1),\theta_1=\theta_1^\prime+\kappa_1-{\rm i}0\right],\theta_2=\theta_2^\prime+\kappa_2-{\rm i}0\right]\nonumber\\
&&\quad+K^{c\bar{c}}(\theta_1^\prime)K^{d\bar{d}}(\theta_2^\prime)e^{-2\Delta{\rm i}t(\cosh(\theta_1^\prime+\kappa_2)+\cosh(\theta_2^\prime+\kappa_1))}\nonumber\\
&&\qquad\times{\rm Res}\left[{\rm Res}\left[f_{a\bar{a}b\bar{b}Bc\bar{c}d\bar{d}}^{\beta/2}(\theta_1^\prime+{\rm i}\pi+{\rm i}0,-\theta_1^\prime+{\rm i}\pi+{\rm i}0,\theta_2^\prime+{\rm i}\pi+{\rm i}0,-\theta_2^\prime+{\rm i}\pi+{\rm i}0,\right.\right.\nonumber\\
&&\qquad\quad\left.\left.\left.{\rm i}\pi,-\theta_2+\kappa_2,\theta_2+\kappa_2,-\theta_1+\kappa_1,\theta_1+\kappa_1),\theta_1=\theta_2^\prime+\kappa_1-{\rm i}0\right],\theta_2=\theta_1^\prime+\kappa_2-{\rm i}0\right]\right\}\nonumber\\
&&+\dots\,.
\eeq
Extracting only the $\propto t^2$ terms, we find
\beq
C_{4,0+1;4,0}^4=C_{4,0+1;4,0}^{4,1}+C_{4,0+1;4,0}^{4,2}+\dots,\nonumber
\eeq
where
\beq
C_{4,0+1;4,0}^{4,1}&=&\frac{g^*}{2}e^{{\rm i}\Delta_Bt}f_{B}^{\beta/2}\frac{\Delta^2t^2}{16\pi^2}\int_0^\infty d\theta_1^\prime d\theta_2^\prime \vert K^{a\bar{a}}(\theta_1^\prime)\vert^2\vert K^{b\bar{b}}(\theta_2^\prime)\vert^2\sinh\theta_1^\prime\sinh\theta_2^\prime\nonumber\\
&&\,\,\,\,\,\times\left\{\left[1+S_B({\rm i}\pi+\theta_1^\prime)\right]\left[1+S_B({\rm i}\pi+\theta_2^\prime)\right]\right.\label{cfouronea}\\
&&\,\,\,\,\,\qquad+\left[1+S_B({\rm i}\pi+\theta_1^\prime)\right]\left[1+S_B({\rm i}\pi-\theta_2^\prime)\right]\label{cfouroneb}\\
&&\,\,\,\,\,\qquad+\left[1+S_B({\rm i}\pi-\theta_1^\prime)\right]\left[1+S_B({\rm i}\pi+\theta_2^\prime)\right]\label{cfouronec}\\
&&\,\,\,\,\,\qquad+\left.\left[1+S_B({\rm i}\pi-\theta_1^\prime)\right]\left[1+S_B({\rm i}\pi-\theta_2^\prime)\right]\right\},\label{cfouroned}\\
&\equiv&C_{4,0+1;4,0}^{4,1,a}+C_{4,0+1;4,0}^{4,1,b}+C_{4,0+1;4,0}^{4,1,c}+C_{4,0+1;4,0}^{4,1,d}
\eeq
and
\beq
C_{4,0+1;4,0}^{4,2}&=&\frac{g^*}{2}e^{{\rm i}\Delta_Bt}f_{B}^{\beta/2}\frac{\Delta^2t^2}{16\pi^2}\int_0^\infty d\theta_1^\prime d\theta_2^\prime\left(K^{a\bar{a}}(\theta_1^\prime)\right)^*\left(K^{b\bar{b}}(\theta_2^\prime)\right)^*K_s^{c\bar{c}}(\theta_1^\prime)K_s^{d\bar{d}}(\theta_2^\prime)\sinh\theta_1^\prime\sinh\theta_2^\prime\nonumber\\
&&\,\,\,\,\,\times\left\{S_{gf}^{\bar{d}c}(\theta_1^\prime+\theta_2^\prime)S_{\bar{a}b}^{\bar{e}k}(-\theta_1^\prime-\theta_2^\prime)S_{be}^{\bar{g}c}(\theta_2^\prime-\theta_1^\prime)S_{fd}^{ak}(\theta_1^\prime-\theta_2^\prime)\right.\nonumber\\
&&\,\,\,\,\,\,\,\,\,\,\,\,\,\,\,\,\,\,\,\,\times\left[1+S_B({\rm i}\pi+\theta_1^\prime)\right]\left[1+S_B({\rm i}\pi+\theta_2^\prime)\right]\label{cfourtwoa}\\
&&\,\,\,\,\,\,\,\,\,\,+S_{a\bar{a}}^{m\bar{m}}(2\theta_1^\prime)S_{ef}^{pq}(-2\theta_1^\prime)S_{gf}^{\bar{d}c}(\theta_1^\prime+\theta_2^\prime)S_{be}^{\bar{g}c}(\theta_2^\prime-\theta_1^\prime)S_{p\bar{d}}^{mk}(-\theta_1^\prime-\theta_2^\prime)S_{qk}^{m\bar{b}}(\theta_1^\prime-\theta_2^\prime)\nonumber\\
&&\,\,\,\,\,\,\,\,\,\,\,\,\,\,\,\,\,\,\,\,\times\left[1+S_B({\rm i}\pi-\theta_1^\prime)\right]\left[1+S_B({\rm i}\pi+\theta_2^\prime)\right]\label{cfourtwob}\\
&&\,\,\,\,\,\,\,\,\,\,S_{b\bar{b}}^{e\bar{e}}(2\theta_2^\prime)S_{d\bar{d}}^{f\bar{f}}(-2\theta_2^\prime)S_{\bar{a}\bar{e}}^{gh}(\theta_2^\prime-\theta_1^\prime)S_{ah}^{i\bar{f}}(\theta_1^\prime+\theta_2^\prime)S_{ge}^{\bar{c}k}(-\theta_1^\prime-\theta_2^\prime)S_{cf}^{ik}(\theta_1^\prime-\theta_2^\prime)\nonumber\\
&&\,\,\,\,\,\,\,\,\,\,\,\,\,\,\,\,\,\,\,\,\times\left[1+S_B({\rm i}\pi+\theta_1^\prime)\right]\left[1+S_B({\rm i}\pi-\theta_2^\prime)\right]\label{cfourtwoc}\\
&&\,\,\,\,\,\,\,\,\,\,+S_{b\bar{b}}^{e\bar{e}}(2\theta_2^\prime)S_{d\bar{d}}^{f\bar{f}}(-2\theta_2^\prime)S_{ig}^{mn}(2\theta_1^\prime)S_{c\bar{c}}^{k\bar{k}}(-2\theta_1^\prime)S_{\bar{a}\bar{e}}^{gh}(\theta_2^\prime-\theta_1^\prime)S_{ah}^{i\bar{f}}(\theta_1^\prime+\theta_2^\prime)S_{k\bar{f}}^{\bar{n}p}(-\theta_1^\prime-\theta_2^\prime)S_{k\bar{p}}^{me}(\theta_1^\prime-\theta_2^\prime)\nonumber\\
&&\,\,\,\,\,\,\,\,\,\,\,\,\,\,\,\,\,\,\,\,\left.\times\left[1+S_B({\rm i}\pi-\theta_1^\prime)\right]\left[1+S_B({\rm i}\pi-\theta_2^\prime)\right]\right\}\label{cfourtwod}\\
&\equiv&C_{4,0+1;4,0}^{4,2,a}+C_{4,0+1;4,0}^{4,2,b}+C_{4,0+1;4,0}^{4,2,c}+C_{4,0+1;4,0}^{4,2,d}.
\eeq

We now consider the semi-connected terms, (\ref{ftwo})--(\ref{fnine}). We label the corresponding contributions to $C_{4,0+1;4,0}$ coming from these terms consecutively as $C_{4,0+1;4,0}^{3,1}$ to $C_{4,0+1;4,0}^{3,8}$. We can then find the relations
\beq
C_{4,0+1;4,0}^{3,1}&=&C_{4,0+1;4,0}^{4,1,a}+C_{4,0+1;4,0}^{4,1,b},\\
C_{4,0+1;4,0}^{3,2}&=&C_{4,0+1;4,0}^{4,2,a}+C_{4,0+1;4,0}^{4,2,b},\\
C_{4,0+1;4,0}^{3,3}&=&C_{4,0+1;4,0}^{4,2,a}+C_{4,0+1;4,0}^{4,2,c},\\
C_{4,0+1;4,0}^{3,4}&=&C_{4,0+1;4,0}^{4,1,a}+C_{4,0+1;4,0}^{4,1,c},\\
C_{4,0+1;4,0}^{3,5}&=&-C_{4,0+1;4,0}^{4,1,b}-C_{4,0+1;4,0}^{4,1,d},\\
C_{4,0+1;4,0}^{3,6}&=&-C_{4,0+1;4,0}^{4,2,b}-C_{4,0+1;4,0}^{4,2,d},\\
C_{4,0+1;4,0}^{3,7}&=&-C_{4,0+1;4,0}^{4,2,c}-C_{4,0+1;4,0}^{4,2,d},\\
C_{4,0+1;4,0}^{3,8}&=&-C_{4,0+1;4,0}^{4,1,c}-C_{4,0+1;4,0}^{4,1,d}.
\eeq
The total contribution from these terms is then
\beq
C_{4,0+1;4,0}^{3}=\sum_{i=1}^{8}C_{4,0+1;4,0}^{3,i}=2C_{4,0+1;4,0}^{4,1,a}+2C_{4,0+1;4,0}^{4,2,a}-2C_{4,0+1;4,0}^{4,1,d}-2C_{4,0+1;4,0}^{4,2,d}.
\eeq

Now we turn to the semi-connected terms (\ref{ften})--(\ref{ffifteen}), whose corresponding contributions to $C_{4,0+1;4,0}$, we label $C_{4,0+1;4,0}^{2,1}$ to $C_{4,0+1;4,0}^{2,6}$, respectively. These contributions are given by
\beq
C_{4,0+1;4,0}^{2,1}&=&C_{4,0+1;4,0}^{4,1,a}+C_{4,0+1;4,0}^{4,2,a},\\
C_{4,0+1;4,0}^{2,2}&=&C_{4,0+1;4,0}^{4,1,d}+C_{4,0+1;4,0}^{4,2,d},\\
C_{4,0+1;4,0}^{2,3}&=&-C_{4,0+1;4,0}^{4,1,b},\\
C_{4,0+1;4,0}^{2,4}&=&-C_{4,0+1;4,0}^{4,2,c},\\
C_{4,0+1;4,0}^{2,5}&=&-C_{4,0+1;4,0}^{4,2,b},\\
C_{4,0+1;4,0}^{2,6}&=&-C_{4,0+1;4,0}^{4,1,c}.
\eeq
Adding these terms we find
\beq
C_{4,0+1;4,0}^{2}&=&\sum_{i=1}^{6}C_{4,0+1;4,0}^{2,i}\nonumber\\
&&\!\!\!\!\!\!\!\!\!\!\!\!\!\!\!\!\!\!\!\!\!\!\!\!\!\!\!=\;C_{4,0+1;4,0}^{4,1,a}-C_{4,0+1;4,0}^{4,1,b}-C_{4,0+1;4,0}^{4,1,c}+C_{4,0+1;4,0}^{4,1,d}+C_{4,0+1;4,0}^{4,2,a}-C_{4,0+1;4,0}^{4,2,b}-C_{4,0+1;4,0}^{4,2,c}+C_{4,0+1;4,0}^{4,2,d}.
\eeq
We are now ready to combine all the $\propto t^2$ contributions,
\begin{equation}
C_{4,0+1;4,0}=C_{4,0+1;4,0}^{4}+C_{4,0+1;4,0}^{3}+C_{4,0+1;4,0}^{2}+\dots=C_{4,0+1;4,0}^{4,1,a}+4C_{4,0+1;4,0}^{4,2,a}+\dots .
\end{equation}

It was discussed in Ref.~\cite{repulsive}, that if the boundary state satisfies the boundary Yang-Baxter equation, the terms $C_{4,0+1;4,0}^{4,1,a}$ and $C_{4,0+1;4,0}^{4,2,a}$ can be shown to be equivalent, by virtue of the relationship
\beq
K^{a_1\bar{a}_1}(\theta_1)K^{e_2\bar{e}_2}(\theta_2)\delta_{\bar{a}_1}^{e_1}\delta_{\bar{e}_2}^{d_2}=K^{c_1\bar{c}_1}(\theta_1)K^{c_2\bar{c}_2}(\theta_2)S_{\bar{c}_2c_1}^{b_2c_4}(\theta_1+\theta_2)S_{c_2c_4}^{a_2a_1}(\theta_1-\theta_2)S_{b_2\bar{c}_1}^{d_2d_1}(\theta_2-\theta_1)S_{a_2d_1}^{e_2e_1}(-\theta_1-\theta_2).\label{BYB}
\eeq
We can then write the main result
\beq
C_{4,0+1;4,0}&=&\frac{g^*}{2}e^{{\rm i}\Delta_Bt}f_{B}^{\beta/2}\frac{\Delta^2t^2}{2\pi^2}\int_0^\infty d\theta_1^\prime d\theta_2^\prime \vert K^{a\bar{a}}(\theta_1^\prime)\vert^2\vert K^{b\bar{b}}(\theta_2^\prime)\vert^2\sinh\theta_1^\prime\sinh\theta_2^\prime\nonumber\\
&&\,\,\,\,\,\,\,\,\,\,\,\,\,\,\,\,\,\,\,\,\,\,\,\,\,\,\,\,\,\,\,\,\,\,\,\,\,\,\,\,\,\,\,\,\,\,\,\,\,\,\times\left[1+S_B({\rm i}\pi+\theta_1^\prime)\right]\left[1+S_B({\rm i}\pi+\theta_2^\prime)\right]\nonumber\\
&=&\frac{1}{2}\frac{g^*}{2}e^{{\rm i}\Delta_Bt}f_{B}^{\beta/2}\left(\frac{t}{\tau_B^*}\right)^2+\dots\,.
\eeq

\subsubsection{The term $\boldsymbol{D_{0,4+1;0,4}}$}
The leading contribution for this term at long times is again given by the connected term, $C_{0,4+1;0,4}$. The computation of this leading contribution is very similar to the computation presented in the previous section, so we will simply present the necessary results, while omitting the details of the calculation. We consider the term
\beq
C_{0,4+1;0,4}&=&\frac{1}{4}\frac{g^*}{2}e^{{\rm i}\Delta_B t}\int_0^\infty\frac{d\phi_1^\prime d\phi_2^\prime d\phi_1 d\theta_2}{(2\pi)^4}\left(K_B(\phi_1^\prime)\right)^*\left(K_B(\phi_2)\right)^*K_B(\phi_1)K_B(\phi_2)e^{2\Delta_B{\rm i}t\sum_i(\cosh\phi_i^\prime-\cosh\phi_i)}\nonumber\\
&&\,\,\,\,\,\,\,\,\,\,\,\,\,\,\,\,\,\,\,\,\times\langle\phi_1^\prime,-\phi_1^\prime,\phi_2^\prime,-\phi_2^\prime,0_B\vert e^{{\rm i}\beta\Phi/2}\vert-\phi_2+\kappa_2,\phi_2+\kappa_2,-\phi_1+\kappa_1,\phi_1+\kappa_1\rangle.\label{fourplusonebreather}
\eeq
The form factor in (\ref{fourplusonebreather}) can be expanded in connected and disconnected pieces analogous to (\ref{fone})--(\ref{ffifteen}).

We first consider the fully connected term, which includes nine-breather form factor, which we denote $C_{0,4+1;0,4}^4$. The leading contribution at long times can be computed by extracting the residues at the poles $\phi_1=\phi_1^\prime+\kappa_1-{\rm i}0$ and $\phi_2=\phi_2^\prime+\kappa_2-{\rm i}0$, or $\phi_1=\phi_2^\prime+\kappa_1-{\rm i}0$ and $\phi_2=\phi_1^\prime+\kappa_2-{\rm i}0$. The result can be written as
\beq
C_{0,4+1;0,4}^4&=&g^*e^{{\rm i}\Delta_Bt}f_B^{\beta/2}\frac{\Delta_B^2t^2}{16\pi^2}\int_0^\infty d\phi_1^\prime d\phi_2^\prime\vert K_B(\phi_1^\prime)\vert^2\vert K_B(\phi_2^\prime)\vert^2\sinh\phi_1^\prime\sinh\phi_2^\prime\nonumber\\
&&\,\,\,\,\,\,\,\,\,\,\,\,\,\,\,\,\,\,\,\,\times\left\{\left[1-S_{BB}({\rm i}\pi+\phi_1^\prime)\right]\left[1-S_{BB}({\rm i}\pi+\phi_2^\prime)\right]\right.\nonumber\\
&&\,\,\,\,\,\,\,\,\,\,\,\,\,\,\,\,\,\,\,\,+\left[1-S_{BB}({\rm i}\pi+\phi_1^\prime\right]\left[1-S_{BB}({\rm i}\pi-\phi_2^\prime)\right]\nonumber\\
&&\,\,\,\,\,\,\,\,\,\,\,\,\,\,\,\,\,\,\,\,+\left[1-S_{BB}({\rm i}\pi-\phi_1^\prime)\right]\left[1-S_{BB}({\rm i}\pi+\phi_2^\prime)\right]\nonumber\\
&&\,\,\,\,\,\,\,\,\,\,\,\,\,\,\,\,\,\,\,\,+\left.\left[1-S_{BB}({\rm i}\pi-\phi_1^\prime)\right]\left[1-S_{BB}({\rm i}\pi-\phi_2^\prime)\right]\right\}+\dots\nonumber\\
&\equiv&2C_{0,4+1;0,4}^{4,a}+2C_{0,4+1;0,4}^{4,b}+2C_{0,4+1;0,4}^{4,c}+2C_{0,4+1;0,4}^{4,d}+\dots\,.
\eeq

We now consider the semi-connected terms which involve seven-breather form factors, which we collectively call $C_{0,4+1;0,4}^{3}$ (which are analogous to the terms (\ref{ftwo})--(\ref{fnine})). We find up to sub-leading corrections
\beq
C_{0,4+1;0,4}^{3}=4C_{0,4+1;0,4}^{4,a}-4C_{0,4+1;0,4}^{4,d}.
\eeq
We then proceed to the semi-connected terms which are analogous to (\ref{ften})--(\ref{ffifteen}), which involve five-breather form factors, and we label collectively as $C_{0,4+1;0,4}^{2}$, for which we find
\beq
C_{0,4+1;0,4}^{2}=2C_{0,4+1;0,4}^{4,a}-2C_{0,4+1;0,4}^{4,b}-2C_{0,4+1;0,4}^{4,c}+2C_{0,4+1;0,4}^{4,d}.
\eeq
Combining all the $\propto t^2$ terms, find
\beq
C_{0,4+1;0,4}=8C_{0,4+1;0,4}^{4,a}=\frac{1}{2}\frac{g^*}{2}e^{{\rm i}\Delta_Bt}f_B^{\beta/2}\left(\frac{t}{\tau_{BB}^*}\right)^2+\dots\,.
\eeq

\subsubsection{The term $\boldsymbol{D_{2,2+1;2,2}}$}
The leading contribution at long times for this terms is given only by $C_{2,2+1;2,2}$. This term is given by
\beq
C_{2,2+1;2,2}&=&\frac{g^*}{2}e^{{\rm i}\Delta_Bt}\int_0^\infty\frac{d\theta^\prime d\phi^\prime d\theta d\phi}{(2\pi)^4}\left(K^{a\bar{a}}(\theta^\prime)\right)^*\left(K_B(\phi^\prime)\right)^*K^{c\bar{c}}(\theta)K_B(\phi)e^{2\Delta{\rm i}t\left(\cosh\theta^\prime-\cosh\theta\right)}\nonumber\\
&&\,\,\,\,\,\times e^{2\Delta_B{\rm i}t\left(\cosh\phi^\prime-\cosh\phi\right)}\,_{\bar{a}a}\langle\theta^\prime,-\theta^\prime,\phi^\prime,-\phi^\prime,0_B\vert e^{{\rm i}\beta\Phi/2}\vert -\phi+\kappa_2,\phi+\kappa_2,-\theta+\kappa_1,\theta+\kappa_1\rangle_{c\bar{c}}.\label{twosolitontwobreather}
\eeq
We can expand the form factor analogously to the expansion (\ref{fone})--(\ref{ffifteen}).

The fully connected term, which we denote $C_{2,2+1;2,2}^4$, involves a four-soliton, five-breather form factor. The leading contributions at long times come from taking the residues at the poles $\theta=\theta^\prime+\kappa_1-{\rm i}0$ and $\phi=\phi^\prime+\kappa_2-{\rm i}0$, such that
\beq
C_{2,2+1;2,2}^4&=&-\frac{g^*}{2}e^{{\rm i}\Delta_B t}\int_0^\infty\frac{d\theta^\prime d\phi^\prime}{(2\pi)^2}\left(K^{a\bar{a}}(\theta^\prime)\right)^*\left(K_B(\phi^\prime)\right)^*K^{c\bar{c}}(\theta)K_B(\phi)e^{2\Delta{\rm i}t\left(\cosh\theta^\prime-\cosh(\theta'+\kappa_1)\right)}\\
&&\,\,\,\,\,\times e^{2\Delta_B{\rm i}t\left(\cosh\phi^\prime-\cosh(\phi'+\kappa_2)\right)}\,
{\rm Res}\left[{\rm Res}\left[f_{a\bar{a}BBBBBc\bar{c}}^{\beta/2}(\theta^\prime+{\rm i}\pi+{\rm i}0,-\theta^\prime+{\rm i}\pi+{\rm i}0,\phi^\prime+{\rm i}\pi+{\rm i}0,\right.\right.\nonumber\\
&&\qquad\qquad\left.\left.-\phi^\prime+{\rm i}\pi+{\rm i}0,{\rm i}\pi,-\phi+\kappa_2,\phi+\kappa_2,-\theta+\kappa_1,\theta+\kappa_1),\theta=\theta^\prime+\kappa_1-{\rm i}0\right],\phi=\phi^\prime+\kappa_2-{\rm i}0\right]\nonumber
\eeq
Extracting the $\propto t^2$ terms, we find
\beq
C_{2,2+1;2,2}^4&=&4 g^*f_B^{\beta/2}\frac{\Delta_B\Delta t^2}{16\pi^2}\int_0^\infty d\theta^\prime d\phi^\prime \vert K^{a\bar{a}}(\theta^\prime)\vert^2\vert K_B(\phi^\prime)\vert^2\sinh\theta^\prime\sinh\phi\nonumber\\
&&\,\,\,\,\,\,\,\,\,\,\,\,\,\,\,\,\,\,\,\,\times\left\{\left[1+S_B({\rm i}\pi+\theta^\prime)\right]\left[1-S_{BB}({\rm i}\pi+\phi^\prime)\right]\right.\nonumber\\
&&\,\,\,\,\,\,\,\,\,\,\,\,\,\,\,\,\,\,\,\,\quad+\left[1+S_B({\rm i}\pi+\theta^\prime)\right]\left[1-S_{BB}({\rm i}\pi-\phi^\prime)\right]\nonumber\\
&&\,\,\,\,\,\,\,\,\,\,\,\,\,\,\,\,\,\,\,\,\quad+\left[1+S_B({\rm i}\pi-\theta^\prime)\right]\left[1-S_{BB}({\rm i}\pi+\phi^\prime)\right]\nonumber\\
&&\,\,\,\,\,\,\,\,\,\,\,\,\,\,\,\,\,\,\,\,\quad+\left.\left[1+S_B({\rm i}\pi-\theta^\prime)\right]\left[1-S_{BB}({\rm i}\pi-\phi^\prime)\right]\right\}+\dots\nonumber\\
&\equiv& C_{2,2+1;2,2}^{4,a}+C_{2,2+1;2,2}^{4,b}+C_{2,2+1;2,2}^{4,c}+C_{2,2+1;2,2}^{4,d}+\dots\,.
\eeq

We now consider semi-connected terms analogous to the contributions (\ref{ftwo})--(\ref{fnine}), which we denote as $C_{2,2+1;2,2}^3$, and in this case involve either two-soliton, five-breather form factors, or four-soliton, three-breather form factors. From these therms we find
\beq
C_{2,2+1;2,2}^3=2C_{2,2+1;2,2}^{4,a}-2C_{2,2+1;2,2}^{4,d}.
\eeq

Similarly to the previous sections, we now consider semi-connected terms which we denote as $C_{2,2+1;2,2}^2$. The only $\propto t^2$ contributions to this term come from terms involving two-soliton, three-breather form factors.  These contributions are
\beq
C_{2,2+1;2,2}^{2}=C_{2,2+1;2,2}^{4,a}-C_{2,2+1;2,2}^{4,c}-C_{2,2+1;2,2}^{4,c}+C_{2,2+1;2,2}^{4,d}
\eeq

Combining all the contributions, we find
\beq
C_{2,2+1;2,2}=4C_{2,2+1;2,2}^{4,a}=\frac{g^*}{2}e^{{\rm i}\Delta_Bt}f_B^{\beta/2}\frac{t^2}{\tau_B^*\tau_{BB}^*}.
\eeq

\subsection{Order $\boldsymbol{g^2K^4}$}
The only leading contribution at this order (which is $\propto t^2$) is
\beq
D_{4,0+1;4,0+1}&=&C_{4,0+1;4,0+1}-Z_{4,0+1}C_{0,0;0,0}-Z_{4,0}C_{0,0+1;0,0+1}-Z_{0,0+1}C_{4,0;4,0}+Z_{2,0+1}^2C_{0,0;0,0}\nonumber\\
&&+Z_{2,0}^2C_{0,0+1;0,0+1}+2Z_{2,0}Z_{0,0+1}C_{0,0;0,0}-Z_{2,0}C_{2,0+1;2,0+1}-Z_{2,0+1}C_{2,0;2,0}.
\eeq
At late times, the leading contributions come only from the term $D_{4,0+1;4,0+1}\approx C_{4,0+1;4,0+1}\propto t^2$. This term can be written explicitly as
\beq
C_{4,0+1;4,0+1}&=&\frac{\vert g\vert^2}{16}\int_0^\infty\frac{d\theta_1^\prime d\theta_2^\prime d\theta_1 d\theta_2}{(2\pi)^4}\left(K^{a\bar{a}}(\theta_1^\prime)\right)^*\left(K^{b\bar{b}}(\theta_2^\prime)\right)^*K^{c\bar{c}}(\theta_1)K^{d\bar{d}}(\theta_2)e^{2\Delta{\rm i}t\sum_i(\cosh\theta_i^\prime-\cosh\theta_i)}\nonumber\\
&&\times_{\bar{a}a\bar{b}b}\langle\theta_1^\prime,-\theta_1^\prime,\theta_2^\prime,-\theta_2^\prime,0_B\vert e^{{\rm i}\beta\Phi/2}\vert 0_B+\kappa_B,-\theta_2+\kappa_2,\theta_2+\kappa_2,-\theta_1+\kappa_1,\theta_1+\kappa_1\rangle_{d\bar{d}c\bar{c}}.\label{cfouronefourone}
\eeq

The form factor in (\ref{cfouronefourone}) can be expanded similarly to (\ref{fone})--(\ref{ffifteen}), with the difference that now there will be twice as many terms, since now one can have disconnected terms proportional to $\langle0_B\vert 0_B+\kappa_B\rangle=2\pi\,\delta(-\kappa_B)$. These terms that are proportional to $\delta(-\kappa_B)$, will cancel with $-Z_{0,0+1}C_{4,0;4,0}$. The remaining terms are analogous to the terms in the expansion (\ref{fone})--(\ref{ffifteen}).

We can first consider the fully connected term, corresponding to a term analogous to (\ref{fone}), which involves an eight-soliton, two-breather form factor, and we will denote as $C_{4,0+1;4,0+1}^4$. Again, the leading contribution at long times can be extracted from taking the double residues at the poles $\theta_1=\theta_1^\prime+\kappa_1-{\rm i}0$ and $\theta_2=\theta_2^\prime+\kappa_2-{\rm i}0$, or $\theta_1=\theta_2^\prime+\kappa_1-{\rm i}0$ and $\theta_2=\theta_1^\prime+\kappa_2-{\rm i}0$, such that
\beq
C_{4,0+1;4,0+1}^4&=&-\frac{1}{4}\frac{\vert g\vert^2}{4}\int_0^\infty\frac{d\theta_1^\prime d\theta_2^\prime}{(2\pi)^2}\left(K^{a\bar{a}}(\theta_1^\prime)\right)^*\left(K^{b\bar{b}}(\theta_2^\prime)\right)^*e^{2\Delta{\rm i}t(\cosh\theta_1^\prime+\cosh\theta_2^\prime)}\nonumber\\
&&\times\left\{K^{c\bar{c}}(\theta_2^\prime)K^{d\bar{d}}(\theta_1^\prime)e^{-2\Delta{\rm i}t(\cosh(\theta_1^\prime+\kappa_1)+\cosh(\theta_2^\prime+\kappa_2))}\right.\nonumber\\
&&\,\,\,\,\,\times{\rm Res}\left[{\rm Res}\left[f_{a\bar{a}b\bar{b}BBc\bar{c}d\bar{d}}^{\beta/2}(\theta_1^\prime+{\rm i}\pi+{\rm i}0,-\theta_1^\prime+{\rm i}\pi+{\rm i}0,\theta_2^\prime+{\rm i}\pi+{\rm i}0,-\theta_2^\prime+{\rm i}\pi+{\rm i}0,\right.\right.\nonumber\\
&&\,\,\,\,\,\,\,\,\,\,\,\,\,\,\,\left.\left.{\rm i}\pi,\kappa_B,-\theta_2+\kappa_2,\theta_2+\kappa_2,-\theta_1+\kappa_1,\theta_1+\kappa_1),\theta_1=\theta_1^\prime+\kappa_1-{\rm i}0\right],\theta_2=\theta_2^\prime+\kappa_2-{\rm i}0\right]\nonumber\\
&&\,+K^{c\bar{c}}(\theta_1^\prime)K^{d\bar{d}}(\theta_2^\prime)e^{-2\Delta{\rm i}t(\cosh(\theta_1^\prime+\kappa_2)+\cosh(\theta_2^\prime+\kappa_1))}\nonumber\\
&&\,\,\,\,\,\times{\rm Res}\left[{\rm Res}\left[f_{a\bar{a}b\bar{b}BBc\bar{c}d\bar{d}}^{\beta/2}(\theta_1^\prime+{\rm i}\pi+{\rm i}0,-\theta_1^\prime+{\rm i}\pi+{\rm i}0,\theta_2^\prime+{\rm i}\pi+{\rm i}0,-\theta_2^\prime+{\rm i}\pi+{\rm i}0,\right.\right.\nonumber\\
&&\,\,\,\,\,\,\,\,\,\,\,\,\,\,\,\left.\left.\left.{\rm i}\pi,\kappa_B,-\theta_2+\kappa_2,\theta_2+\kappa_2,-\theta_1+\kappa_1,\theta_1+\kappa_1),\theta_1=\theta_2^\prime+\kappa_1-{\rm i}0\right],\theta_2=\theta_1^\prime+\kappa_2-{\rm i}0\right]\right\}\nonumber\\
&&+\dots\,.
\eeq
Extracting the $\propto t^2$, we find
\beq
C_{4,0+1;4,0+1}^4=4C_{4,0+1;4,0+1}^{4,1,a}+C_{4,0+1;4,0+1}^{4,2}+\dots\,,
\eeq
where
\beq
C_{4,0+1;4,0+1}^{4,1,a}=\vert g\vert^2f_{BB}^{\beta/2}({\rm i}\pi,0)\frac{\Delta^2 t^2}{16\pi^2}\int_0^\infty d\theta_1^\prime d\theta_2^\prime \vert K^{a\bar{a}}(\theta_1^\prime)\vert^2\vert K^{b\bar{b}}(\theta_2^\prime)\vert^2\sinh\theta_1^\prime\sinh\theta_2^\prime,
\eeq
and
\beq
C_{4,0+1;4,0+1}^{4,2}&=&\vert g\vert^2f_{BB}^{\beta/2}({\rm i}\pi,0)\frac{\Delta^2t^2}{16\pi^2}\int_0^\infty d\theta_1^\prime d\theta_2^\prime\left(K^{a\bar{a}}(\theta_1^\prime)\right)^*\left(K^{b\bar{b}}(\theta_2^\prime)\right)^*K^{c\bar{c}}(\theta_1^\prime)K^{d\bar{d}}(\theta_2^\prime)\sinh\theta_1^\prime\sinh\theta_2^\prime\nonumber\\
&&\,\,\,\,\,\times\left\{S_{gf}^{\bar{d}c}(\theta_1^\prime+\theta_2^\prime)S_{\bar{a}b}^{\bar{e}k}(-\theta_1^\prime-\theta_2^\prime)S_{be}^{\bar{g}c}(\theta_2^\prime-\theta_1^\prime)S_{fd}^{ak}(\theta_1^\prime-\theta_2^\prime)\right.\nonumber\\
&&\,\,\,\,\,\,\,\,\,\,+S_{a\bar{a}}^{m\bar{m}}(2\theta_1^\prime)S_{ef}^{pq}(-2\theta_1^\prime)S_{gf}^{\bar{d}c}(\theta_1^\prime+\theta_2^\prime)S_{be}^{\bar{g}c}(\theta_2^\prime-\theta_1^\prime)S_{p\bar{d}}^{mk}(-\theta_1^\prime-\theta_2^\prime)S_{qk}^{m\bar{b}}(\theta_1^\prime-\theta_2^\prime)\nonumber\\
&&\,\,\,\,\,\,\,\,\,\,+S_{b\bar{b}}^{e\bar{e}}(2\theta_2^\prime)S_{d\bar{d}}^{f\bar{f}}(-2\theta_2^\prime)S_{\bar{a}\bar{e}}^{gh}(\theta_2^\prime-\theta_1^\prime)S_{ah}^{i\bar{f}}(\theta_1^\prime+\theta_2^\prime)S_{ge}^{\bar{c}k}(-\theta_1^\prime-\theta_2^\prime)S_{cf}^{ik}(\theta_1^\prime-\theta_2^\prime)\nonumber\\
&&\,\,\,\,\,\,\,\,\,\,\left.+S_{b\bar{b}}^{e\bar{e}}(2\theta_2^\prime)S_{d\bar{d}}^{f\bar{f}}(-2\theta_2^\prime)S_{ig}^{mn}(2\theta_1^\prime)S_{c\bar{c}}^{k\bar{k}}(-2\theta_1^\prime)S_{\bar{a}\bar{e}}^{gh}(\theta_2^\prime-\theta_1^\prime)S_{ah}^{i\bar{f}}(\theta_1^\prime+\theta_2^\prime)S_{k\bar{f}}^{\bar{n}p}(-\theta_1^\prime-\theta_2^\prime)S_{k\bar{p}}^{me}(\theta_1^\prime-\theta_2^\prime)\right\}\nonumber\\
&\equiv&C_{4,0+1;4,0+1}^{4,2,a}+C_{4,0+1;4,0+1}^{4,2,b}+C_{4,0+1;4,0+1}^{4,2,c}+C_{4,0+1;4,0+1}^{4,2,d}.
\eeq

We now examine the semi-connected terms which we denote collectively as $C_{4,0+1;4,0+1}^3$, which arise from terms analogous to (\ref{ftwo})--(\ref{fnine}), and involve six-soliton, two-breather form factors. We find
\beq
C_{4,0+1;4,0+1}^3=2C_{4,0+1;4,0+1}^{4,2,a}-2C_{4,0+1;4,0+1}^{4,2,d}.
\eeq

Similarly, we examine the terms $C_{4,0+1;4,0+1}^2$, which arise from terms analogous to (\ref{ften})--(\ref{ffifteen}), and involve four-soliton, two-breather form factors. From these, we find
\beq
C_{4,0+1;4,0+1}^2=C_{4,0+1;4,0+1}^{4,2,a}-C_{4,0+1;4,0+1}^{4,2,b}-C_{4,0+1;4,0+1}^{4,2,c}+C_{4,0+1;4,0+1}^{4,2,d}.
\eeq

Combining $C_{4,0+1;4,0+1}^{4}$, $C_{4,0+1;4,0+1}^{3}$ and $C_{4,0+1;4,0+1}^{2}$, we find
\beq
C_{4,0+1;4,0+1}=4C_{4,0+1;4,0+1}^{4,1,a}+4C_{4,0+1;4,0+1}^{4,2,a}.
\eeq
We can use the relation (\ref{BYB}), to show that $C_{4,0+1;4,0+1}^{4,2,a}=C_{4,0+1;4,0+1}^{4,1,a}$, such that
\beq
C_{4,0+1;4,0+1}=8C_{4,0+1;4,0+1}^{4,1,a}=\frac{\vert g\vert^*}{4}f_{BB}^{\beta/2}({\rm i}\pi,0)\frac{t^2}{2\tau^2}.
\eeq

\section{Infinite-volume regularisation}\label{app:regularisation}
We now review briefly the infinite-volume regularisation which was introduced first for studying finite-temperature correlation functions in Ref.~\cite{infinitereg}. This has been generalised for the study of quantum quenches in \cite{SE} and shown to give identical results to a finite-volume regularisation~\cite{finiteL}. The terms of the linked-cluster expansion appear to be divergent due to the intertwining of particles with rapidities $\theta_i$ and $-\theta_i$ in the initial state. These divergences, however, are an artefact of working in infinite volume, and when the theory is properly regularised, these divergent terms cancel with each other. The regularisation procedure we follow consists firstly on shifting the rapidities of each pair of particles, $\{-\theta_i,\theta_i\}$ in the ket states by a real parameter, $\kappa_i$, which shifts the rapidities away from singularities. The resulting expressions are understood as generalised functions of the variables $\kappa_i$. The divergencies are then explicitly exhibited by introducing a smooth function $P(\kappa)$ (for each $\kappa_i$) which is strongly peaked around $\kappa=0$ and satisfies
\beq
P(0)=L,\quad\int d\kappa\,P(\kappa)=1,
\eeq
where $L$ is a finite system size that is introduced as a regularisation. Possible choice of the regularisation function are $P(\kappa)=L\,e^{-\pi L^2\kappa^2}$ or $P(\kappa)=L\Delta/[1+(\pi L\Delta\kappa)^2]$.


\end{document}